\documentclass[a4paper,11pt]{article}
\pdfoutput=1 

\usepackage{jheppub} 
		
\usepackage{graphicx}				
\usepackage{amssymb}
\usepackage{array}
\usepackage{tabularx}
\usepackage{longtable}

\usepackage{xcolor}
\usepackage{soul}

\usepackage{doi}
\usepackage{bm}
\usepackage{textcomp} 
\usepackage{feynmp-auto}
\usepackage{slashed}
\usepackage{amsmath}
\usepackage{tikz}
\usetikzlibrary{positioning}
\usepackage{multirow}
\usepackage{multicol}
\usepackage{booktabs}

 \usepackage{caption}
 
 \usepackage{float}

\newcommand{\be}{\begin{equation}}
\newcommand{\ee}{\end{equation}}
\def\ba{\begin{eqnarray}}
\def\ea{\end{eqnarray}}
\newcommand{\nl}{\nonumber \\ }
\newcommand{\nn}{\nonumber }

\newcommand{\bfp}{{\bf p} }

\setstcolor{red}

\allowdisplaybreaks

\preprint{{\small USTC-ICTS/PCFT-25-50}}

\title{Effective Theory for Light Portal Dark Matter Detection}

\author[a]{Qing Chen}
\author[b,c]{and Shuang-Yong Zhou}

\affiliation[a]{Center for Fundamental Physics, School of Mechanics and Physics, Anhui University of Science and Technology, Huainan, Anhui 232001, China}
\affiliation[b]{Interdisciplinary Center for Theoretical Study, University of Science and Technology of China, Hefei, Anhui 230026, China}
\affiliation[c]{Peng Huanwu Center for Fundamental Theory, Hefei, Anhui 230026, China}

\emailAdd{qingchen@aust.edu.cn}
\emailAdd{zhoushy@ustc.edu.cn}

\abstract{
We develop a general framework for the calculation of direct-detection signals from light-portal dark matter, incorporating a consistent treatment of finite momentum transfer.
In this framework, dark matter interacts with Standard Model matter through a light mediator, which simultaneously serves as the force carrier for dark matter self-interactions, potentially with a distinct coupling strength. We systematically construct the corresponding effective theory relevant to the detection of this class of dark matter.
Our analysis focuses on light (semi)relativistic dark matter, which may be produced through cosmic-ray boosting and can be probed in high-threshold experiments such as large-volume neutrino detectors.
In this context, the nucleon matrix elements of the effective operators at finite momentum transfer are required and have become available through recent advances in lattice QCD and related nonperturbative methods.
The relativistic Fermi gas model is used to translate the nucleon-level description to the nuclear level, thereby incorporating nuclear effects relevant to experiments with heavy targets.
To demonstrate the utility of the framework, we present ultraviolet-complete examples featuring spin-1 and spin-2 portal dark matter.
For these models, we compute the differential cross sections with respect to momentum transfer, using parameter choices that address the so-called ``core--cusp'' problem in astrophysical observations via dark matter self-interactionss.
}

\begin{document}

\begin{fmffile}{feynmffile} 
\fmfcmd{
vardef middir(expr p,ang) = dir(angle direction length(p)/2 of p + ang) enddef;
style_def arrow_left expr p = shrink(.7); cfill(arrow p shifted(4thick*middir(p,90))); endshrink enddef;
style_def arrow_left_more expr p = shrink(.7); cfill(arrow p shifted(6thick*middir(p,90))); endshrink enddef;
style_def arrow_right expr p = shrink(.7); cfill(arrow p shifted(4thick*middir(p,-90))); endshrink enddef;}

\fmfset{arrow_len}{2.5mm}
\fmfset{arrow_ang}{12}
\fmfset{wiggly_len}{3mm}
\fmfset{wiggly_slope}{75}
\fmfset{curly_len}{2mm}
\fmfset{zigzag_len}{2.5mm}
\fmfset{zigzag_width}{1mm}
\fmfset{dash_len}{1.5mm}

\fmfcmd{
    path quadrant, q[], otimes;
    quadrant = (0, 0) -- (0.5, 0) & quartercircle & (0, 0.5) -- (0, 0);
    for i=1 upto 4: q[i] = quadrant rotated (45 + 90*i); endfor
    otimes = q[1] & q[2] & q[3] & q[4] -- cycle;
}
\fmfwizard

\maketitle

\flushbottom

\section{Introduction}
Weakly Interacting Massive Particles (WIMPs), typically with masses ranging from 10 GeV to the TeV scale, have long been well-motivated dark matter (DM) candidates (see the reviews \cite{Jungman:1995df,Bertone:2004pz}), yet they remain undetected. 
Heavy TeV-scale electroweak WIMPs naturally elude current experiments due to their tiny cross sections \cite{Hill:2013hoa,Chen:2018uqz,Chen:2019gtm,Chen:2023bwg}, which are severely suppressed by cancellations between different types of operators. 
While TeV electroweak WIMPs remain well motivated, the null results from recent experiments \cite{XENON:2025vwd,PandaX:2024qfu,LZ:2022lsv} make it worthwhile to explore other possibilities. 
Without departing from the WIMP paradigm, attention has been drawn to light dark matter with sub-GeV masses \cite{Battaglieri:2017aum,Ema:2018bih,Song:2023jqm,Dutta:2024kuj}. 

In typical direct detection experiments, sensitivity to cold dark matter with velocities on the order of $10^{-3}$ times the speed of light deteriorates rapidly for masses below 1 GeV. 
Because the nuclear recoil energy is proportional to the square of the dark matter particle's three-momentum, a sufficiently light and cold dark matter particle cannot impart enough energy to a nucleus to be registered by the detector. 
However, if these sub-GeV dark matter particles are boosted by cosmic rays and become (semi-)relativistic \cite{Emken:2017hnp,Bringmann:2018cvk}, they can deposit considerable energy via nuclear recoils, making them potentially detectable \cite{PandaX-II:2021kai,Super-Kamiokande:2022ncz,LZ:2025iaw}. 
{In addition to cosmic-ray boosting, another mechanism---known as fermionic dark matter absorption---can impart a detectable signal on nuclei \cite{Elor:2021swj,Dror:2019dib}.} 
Other ways to circumvent the poor sensitivity to sub-GeV dark matter include detecting light DM-electron scattering \cite{Essig:2011nj,Super-Kamiokande:2017dch,Liang:2024ecw,Krnjaic:2024bdd,PandaX:2024syk}, which can trigger inelastic atomic processes such as ionization and excitation. 
This mechanism is beyond the scope of this paper; instead, we will focus on cosmic-ray-accelerated DM-nucleon/nucleus scattering.

Specifically, we focus on light-portal dark matter, motivated in part by the ``core-cusp'' problem \cite{Moore:1994yx,Flores:1994gz,Navarro:1995iw,Navarro:1996gj} arising from astronomical dark matter observations. 
This problem refers to the discrepancy between the steeply peaked central density (the cusp) of dark matter halos predicted by collisionless cold dark matter (CDM) simulations and the flatter density profiles (the cores) actually observed. 
A leading solution for resolving this small-scale anomaly is to introduce dark matter self-interactions \cite{Spergel:1999mh,Tulin:2017ara}. 
This scenario relies on self-interacting DM collisions with a mean free path on the order of $1$~kpc to heat up the low-entropy central material. These collisions prevent collapse into a dense cusp and instead produce a shallower, core-like density profile. 

{However, beyond the core-cusp anomaly, observations also reveal a ``diversity problem'' in the rotation curves of galaxies. Addressing this problem while satisfying the combined constraints from cluster, galactic, and dwarf scales, generally requires strongly velocity-dependent self-interactions. A solution across all astrophysical scales requires cross sections of $\sim 0.1-0.5$~cm$^2$/g on cluster scales, $\sim 1-10$~cm$^2$/g on galaxy scales, and even larger values for dwarf galaxies \cite{Oman:2015xda, Kaplinghat:2015aga, Tulin:2017ara}.}

In the weakly interacting regime, the light mediator acts as a short-range force carrier relative to the macroscopic dark matter interparticle spacing (roughly $1$~cm in the local halo), yet its range remains much longer than that of the standard weak force. Depending on the interaction strength and the mean free path, the requisite dark matter mass typically spans $1$~MeV to $10$~GeV \cite{Spergel:1999mh, Tulin:2017ara}, and thus largely encompasses the sub-GeV regime. 
{This specific mass range serves as an illustrative example capable of explaining the astronomical data; however, the framework developed in this paper is completely general and applies to any light-portal process involving finite momentum transfer, without assuming a specific dark matter mass.}

When the mediator of the interaction between dark matter and Standard Model (SM) matter is light, the momentum transfer in a single scattering event involving boosted dark matter can be comparable to the mediator mass. 
The zero-momentum transfer approximation---typically employed in analyses of dark matter scattering with single nucleons~\cite{PandaX-II:2021kai,Super-Kamiokande:2022ncz} or nuclei~\cite{LZ:2025iaw}---is valid only for contact interactions or heavy-mediator scenarios. 
Consequently, this approximation becomes invalid in the light-portal situation. 
In this work, we address such light-portal scenarios by constructing effective non-local operators with finite momentum transfer. 
We construct both DM and QCD bilinear operators, which are connected via the light mediator propagator.

The nucleon matrix elements of QCD operators with finite momentum transfer are inherently nonperturbative objects at low energies. 
In recent years, significant effort and progress have been made in computing them using lattice QCD~\cite{Cates:2011pz,Alexandrou:2019ali,Alexandrou:2021wzv,Alexandrou:2022dtc,Alexandrou:2023qbg,Hackett:2023rif,Wang:2024lrm,Bhattacharya:2024wtg,Bhattacharya:2025yba}, within the QCD instanton vacuum~\cite{Liu:2024rdm}, via light-front methods~\cite{Nair:2025sfr}, and through dispersive analyses~\cite{Hoferichter:2012wf,Cao:2024zlf,Cao:2025dkv}, evolving from spin-0 scalar to spin-2 gravitational form factors.
In addition to light-portal dark matter detection, these matrix elements are crucial for calculations in other nucleon-target experiments, such as muon-to-electron conversion experiments~\cite{Haxton:2024lyc}.
We will also translate these nucleon matrix elements with finite momentum transfer to the nuclear level, utilizing the relativistic Fermi gas model~\cite{Smith:1972xh} to account for nuclear effects.

{Existing analyses of cosmic-ray-boosted dark matter typically treat DM-nucleon or DM-nucleus scattering using simplified models, primarily assuming contact or vector/axial-vector interactions. These studies generally incorporate finite momentum transfer effects through phenomenological form factors, such as a dipole form factor with an empirical mass parameter or the Helm form factor~\cite{Bringmann:2018cvk, Bondarenko:2019vrb,Dent:2019krz,Ge:2020yuf, Bell:2021xff, Alvey:2022pad,Bell:2023sdq, Cappiello:2024acu, Diurba:2025lky}.}
{Our framework goes beyond these existing approaches by treating the momentum transfer microscopically at the quark and gluon levels. This is achieved by building the finite momentum transfer into non-local quark and gluon interactions, before convoluting the results to the nuclear level. This microscopic nuclear treatment also yields quantitatively different results from those obtained with existing methods.}

Compared to signals from cold WIMPs or light DM electron scattering, signals from cosmic-ray-boosted, light-portal sub-GeV dark matter are more readily detectable in high-threshold fixed-target experiments. 
Prominent examples include large-volume neutrino detectors---such as DUNE~\cite{DUNE:2020lwj}, JUNO~\cite{JUNO:2021vlw}, Super-Kamiokande~\cite{Super-Kamiokande:2002weg}, and Hyper-Kamiokande~\cite{Hyper-Kamiokande:2018ofw} which are sensitive to recoil energies in the 10--100~MeV range~\cite{Dutta:2024kuj}.

The remainder of this paper is organized as follows: Section~\ref{sec:EFT} constructs an effective field theory for interactions of light-portal dark matter with quarks and gluons; Section~\ref{sec:nme} parametrizes the nucleon matrix elements with form factors at finite momentum transfer; Section~\ref{sec:rfg} revisits the relativistic Fermi gas model and translates nucleon-level hadronic tensors to those at the nucleus level;
Section~\ref{sec:exam} showcases examples of spin-1 and spin-2 portal calculations of dark-matter--nucleus scattering cross sections using the framework developed in previous sections;
we summarize in Section~\ref{sec:sum};
Appendix~\ref{app:nid} lists on-shell nucleon-level identities in momentum space; 
Appendix~\ref{app:RFG} presents details of the relativistic Fermi-gas model nuclear treatment;
Appendix~\ref{app:trans} details the translation of tensors from the nucleon-level to the nuclear level in the relativistic Fermi gas model;
Appendix~\ref{app:ABrelation} derives the relations between the two integral basis functions;
Appendix~\ref{app:Bfunc} lists the explicit forms of integral basis functions.
Appendix~\ref{app: fmrules} provides the relevant Feynman rules for spin-2 portal dark matter.

\section{Light portal dark matter effective theory \label{sec:EFT}}
Dark matter scattering on a nuclear target involves interactions with nucleons composed of quarks and gluons~\cite{Hill:2014yxa}. 
We consider a scenario involving energetic dark matter and a light mediator, where the momentum transfer in the scattering is comparable to the mediator mass. 
This differs from the conventional four-Fermi regime, where the interaction is treated as pointlike and momentum transfer is negligible compared to a heavy mediator mass. 
Conversely, this light-portal scenario neither resembles deep inelastic scattering, where momentum transfer dominates and the interaction can be expanded in inverse powers of the transfer.

To describe this intermediate regime, we shall construct effective DM-quark and DM-gluon interaction operators with the mediator propagator explicitly built in. 
We emphasize that in our scenario, the light-portal particle (i.e., the mediator) is not integrated out; instead, it provides a propagator within the effective operator. 
The resulting (non-local) Lagrangian can still be viewed as an effective theory in the sense that it allows for all possible effective interactions with unknown, weakly coupled coefficients. 
Indeed, this approach conceptually resembles the use of form factors.

Since this new force remains unobserved, we assume it is significantly weaker than all Standard Model (SM) forces and focus our analysis on the leading-order processes. 
We begin by enumerating the dark matter (DM), quark, and gluon bilinear operators, which are subsequently connected via the mediator propagator. 
Given our focus on leading tree-level interactions, we classify these processes by the spin of the mediator. 
For simplicity, we restrict the mediator spin to at most two for simplicity. Higher-spin exchanges may arise in theories with an infinite tower of higher-spin states, as in string-theory constructions. However, in general, for a higher-spin mediator, the effective operator must contain more derivatives to contract the extra Lorentz indices in the mediator's propagator, and thus is further suppressed compared to lower-spin mediators.

To construct the bilinears, the building blocks include Dirac matrices
$$\left\{1,\, i\gamma^5,\, \gamma^\mu,\, \gamma^\mu\gamma^5, \sigma^{\mu\nu}\right\}$$
and (covariant) derivatives
$$\left\{i\partial_-^\mu,\, iD_-^\mu,\, \partial_+^\mu\right\}$$
where DM is assumed to be a SM gauge singlet and only ordinary derivatives appear in the DM bilinear operators.
Covariant derivatives are associated with QCD $SU(3)$ color gauge and only appear in the QCD operators.
Explicitly, $\partial_+^\mu\equiv \overrightarrow{\partial}^\mu+\overleftarrow{\partial}^\mu$, 
$\partial_-^\mu\equiv (\overrightarrow{\partial}^\mu-\overleftarrow{\partial}^\mu)/2$,
$D_-^\mu\equiv (\overrightarrow{D}^\mu-\overleftarrow{D}^\mu)/2$
and $\overrightarrow{D}_\mu=\overrightarrow{\partial}_\mu-ig_s A_\mu^A T^A$ and $\overleftarrow{D}_\mu=\overleftarrow{\partial}_\mu+ig_s A_\mu^A T^A$, \
where $g_s$ is the strong coupling constant, $A_\mu^A$ is the gluon field, and $T^A$ is an $SU(3)$ generator. 

 To derive a basis that is non-redundant, we will make use of relations that can be derived from the QCD equations of motion (EOMs) which involve one or two derivatives for our purpose. As usual, the use of these EOMs is equivalent to field redefinitions, as they are leading order in terms of the weak couplings, which we focus on in this paper. Together with integration by parts, this allows us to convert operators with higher derivatives to ones with fewer derivatives.
 Whenever a total derivative $\partial_+^\mu$ appears on the QCD side, we move it to the dark matter side. We focus on bilinear operators, which are the lowest-order contributions in perturbative QCD and dominate the interaction. Therefore, we use these identities to eliminate operators that are not quark bilinears—such as those involving gluon field strengths or quark-mixing terms—and retain only the bilinear ones, even if they contain higher derivatives.

Explicitly, QCD EOMs involving one derivative that we use are
\ba
&&\bar{q}i\slashed{D}_-q=m_q\bar{q}q\,,
\\
&&\bar{q}\slashed{\partial}_+q=0\,,
\\
&&\bar{q}i\slashed{D}_-\gamma^5q=0\,,
\\
&&\bar{q}\slashed{\partial}_+\gamma^5 q=-2m_q\bar{q}i\gamma^5q\,,
\\
&&\bar{q}iD_{-\,}^{\nu}q-\frac{1}{2}\bar{q}\sigma^{\mu\nu}{\partial}_{+\mu}q=m_q\bar{q}\gamma^\nu q\,,
\\
&&\bar{q}D_{-\,}^{\nu}\gamma^5 q=-\frac{1}{2}\bar{q}\sigma^{\mu\nu}{\partial}_{+\mu}i\gamma^5 q\,,
\\
&&-\bar{q}{\partial}_{+\,}^{\nu}q=2\bar{q}\sigma^{\mu\nu}i{D}_{-\mu} q\,,
\\
&&\bar{q}i\gamma^5{\partial}_{+\,}^{\nu}q-2\bar{q}\sigma^{\mu\nu}\gamma^5{D}_{-\mu} q=2m_q\bar{q}\gamma^5\gamma^\nu q\,,
\\
&&\bar{q}\gamma^{[\alpha}iD_{-}^{\nu]}q+\frac{1}{4}\epsilon^{\mu\nu\alpha\beta}\bar{q}{\partial}_{+\,\mu}\gamma_\beta\gamma^5q=0\,,
\\
&&\bar{q}\gamma^{[\alpha}\partial_{+}^{\nu]}q-\epsilon^{\mu\nu\alpha\beta}\bar{q}{iD}_{-\,\mu}\gamma_\beta\gamma^5q=-m_q\bar{q}\sigma^{\alpha\nu} q\,,
\ea
where ${}^{[\,...\,]}$ denotes antisymmetrization $A^{[\mu}B^{\nu]}=(A^\mu B^\nu-A^\nu B^\mu)/2$. QCD EOMs involving two derivatives that we use are
\ba
&&\bar{q}\gamma_\mu iD_{-}^{\{\mu} iD_{-}^{\nu\}} q=m_q\bar{q}iD_{-}^\nu q\,,
\\
&&\bar{q}D_{-}^2q+\frac{1}{4}\bar{q}\partial_{+}^2 q-\frac{g_s}{2}\bar{q}\sigma^{\mu\nu}G_{\mu\nu}q=-m_q^2\bar{q} q\,,\label{cmnm}
\\
&&\bar{q} iD_{-}^{\{\mu} iD_{-}^{\nu\}}\gamma_\mu\gamma^5 q=0\,,
\\
&&\bar{q}D_{-}^2i\gamma^5 q+\frac{1}{4}\bar{q}\partial_{+}^2 i\gamma^5  q-\frac{g_s}{2}\bar{q}i\gamma^5\sigma^{\mu\nu}G_{\mu\nu}q=-m_q^2\bar{q}i\gamma^5 q\,,\label{cedm}
\\
&&\bar{q}D_{-\{\nu}D_{-\alpha\}}q+\frac{1}{2}\bar{q}\partial_{+}^{\mu}\sigma_{\mu\{\nu}iD_{-\alpha\}}q+\frac{g_s}{2}\bar{q}\sigma_{\mu\{\nu}G_{\alpha\}}^{~~\mu}q=-m_q\bar{q}\gamma_{\{\nu}iD_{-\alpha\}} q\,,
\\
&&\bar{q}D_{-\{\nu}D_{-\alpha\}}i\gamma^5q+\frac{1}{2}\bar{q}\partial_{+}^{\mu}\sigma_{\mu\{\nu}iD_{-\alpha\}}i\gamma^5q-\frac{g_s}{2}\bar{q}\sigma_{\mu\{\nu}G_{\alpha\}}^{~~\mu}i\gamma^5q=0\,,
\\
&&\bar{q}\partial_{+\nu}iD_{-\alpha}q-2\bar{q}D_{-}^{\mu}\sigma_{\mu\nu}D_{-\alpha}q+i{g_s}\bar{q}\sigma_{\mu\nu}G_{\alpha}^{~~\mu}q+{g_s}\bar{q}G_{\alpha\nu}q=0\,,\label{Gterm}
\\
&&\bar{q}\partial_{+\nu}D_{-\alpha}\gamma^5q+2\bar{q}D_{-}^{\mu}i\sigma_{\mu\nu}D_{-\alpha}\gamma^5q+{g_s}\bar{q}\sigma_{\mu\nu}G_{\alpha}^{~~\mu}\gamma^5q
-i{g_s}\bar{q}G_{\alpha\nu}\gamma^5q
\nn\\
&&=m_q\bar{q}\gamma_{\nu}\gamma^5iD_{-\alpha} q\,,
\\
&&\bar{q}\sigma^{\alpha\beta}D_{-}^2q+\frac{1}{4}\bar{q}\sigma^{\alpha\beta}\partial_{+}^2 q-\frac{g_s}{2}\bar{q}\sigma^{\alpha\beta}\sigma^{\mu\nu}G_{\mu\nu}q=-m_q^2\bar{q}\sigma^{\alpha\beta}q\,,
\\
&&\bar{q}\gamma^5\sigma^{\alpha\beta}D_{-}^2q+\frac{1}{4}\bar{q}i\gamma^5\sigma^{\alpha\beta}\partial_{+}^2 q-\frac{g_s}{2}\bar{q}\gamma^5\sigma^{\alpha\beta}\sigma^{\mu\nu}G_{\mu\nu}q=-m_q^2\bar{q}\gamma^5\sigma^{\alpha\beta}q\,.
\ea
where ${}^{\{\,...\,\}}$ denotes $A^{\{\mu}B^{\nu\}}=(A^\mu B^\nu+A^\nu B^\mu)/2$ and $D_{-[\mu}D_{-\nu]}=-ig_sG_{\mu\nu}/2$ has been used. By integrating by part to move the total derivative $\partial_{+}$ to the dark matter side, we obtain the following replacement rules
\ba
&&|\phi|^2\bar{q}\partial_{+}^\mu\partial_{+}^\nu q\to\phi^*\partial_{+}^{\mu}\partial_{+}^\nu\phi\bar{q}q
\label{r1}\\
&&|\phi|^2\bar{q}\partial_{+}^\mu\partial_{+}^\nu i\gamma^5 q\to\phi^*\partial_{+}^{\mu}\partial_{+}^\nu\phi\bar{q}i\gamma^5q
\label{r2}\\
&&|\phi|^2\bar{q}iD_{-}^\mu\partial_{+}^\nu q\to \frac{\epsilon^{\alpha\mu\rho\sigma}}{2}\left(\phi^*\partial_{+\,\alpha}\partial_{+}^\nu\phi\right)\left(\bar{q}i\sigma_{\rho\sigma}\gamma^5 q\right)
-m_q\left(\phi^* \partial_{+}^\nu\phi\right)\left(\bar{q}\gamma^\mu q\right)
\label{r3}\\
&&|\phi|^2\bar{q}D_{-}^\mu\partial_{+}^\nu \gamma^5 q\to -\left(\phi^*\partial_{+\,\alpha}\partial_{+}^\mu\phi\right)\left(\bar{q}\sigma^{\alpha\nu}i\gamma^5 q\right)
\label{r4}
\ea

After these procedures, the QCD bilinear operators up to dimension five for quarks are tabulated in Table~\ref{table:qops} and for gluons in Table~\ref{table:gops}.
Together with the DM bilinear operators, and without counting the dimension of the propagator, the total dimension of the effective interactions we consider is up to eight.\footnote{Neglecting the propagator dimension allows comparison with traditional four-fermion operators, where the propagator is heavy and integrated out. Including the propagator dimension instead would make the total dimension up to six.} This is because, for spin-2 processes, QCD contains dimension-4 operators at lowest order, and for fermionic DM, a spin-2 operator is also dimension-4 at lowest order, leading to an overall dimension-8 interaction. The scalar and fermion DM bilinear operators are constructed up to dimension five in Tables~\ref{table:sdm} and~\ref{table:fdm}, respectively. We keep the low-dimensional operators as the leading contributions, since the theory is weakly coupled and higher-dimensional operators correspond to higher-loop processes when the mediator’s spin is no greater than two.

\begin{table}[tbp]
    \centering
    \begin{tabular}{|c|c c|}
\hline  
Dimension & \hspace{2.5cm} Quark operators   &    \\
\hline  
\multirow{1}*{3}&  $ \bar{q}\gamma^\mu q$ &  $\bar{q}\gamma^\mu \gamma^5 q$
 \\
\hline  
\multirow{3}*{4}  & $m_q \bar{q}{q}$ &  $m_q \bar{q}i\gamma^5{q}$
\\
&  $ \bar{q}(\gamma^{\{\mu}iD_{-}^{\nu\}}-\frac{g^{\mu \nu}}{4}i\slashed{D}_{-})q $ & 
$\bar{q}\gamma^{\{\mu}iD_{-}^{\nu\}}\gamma^5 q $
\\
& $im_q\bar{q}\sigma^{\mu\nu}\gamma^5 q $  &
\\
\hline  
\multirow{4}*{5}  & $\bar{q}D_{-}^2q$ & $\bar{q}D_{-}^2i\gamma^5 q$
\\
&  $\bar{q}(D_{-}^{\{\mu}D_{-}^{\nu\}}-\frac{g^{\mu \nu}}{4}{D}^2_{-})q $ & $ \bar{q}D_{-}^{\{\mu}D_{-}^{\nu\}}i\gamma^5q $
\\
&  $ \bar{q}D_{-}^{\{\lambda}D_{-}^{[\mu\}}\sigma^{\nu]}{}_\lambda q $ & $ \bar{q}D_{-}^2\sigma^{\mu\nu}q $
\\
\hline  
    \end{tabular}
            \caption{Quark bilinear operators. The operator  $ \bar{q}D_{-}^{\{\lambda}D_{-}^{[\mu\}}\sigma^{\nu]}{}_\lambda q \equiv-2\bar{q}D_{-}^{\rho}\sigma_\rho{}^{[\nu}D^{-\mu]}q+i{g_s}\bar{q}\sigma_{\rho}{}^{[\nu}G^{\mu]\rho}q$ and can alternatively be replaced by $\bar{q}G_{\alpha\nu}q$ upon moving the total derivative term to the dark matter side in Eq.~(\ref{Gterm}). However, we still keep it this way to allow only bilinear operators in the basis.
            } \label{table:qops}
\end{table}

\begin{table}[tbp]
    \centering
    \begin{tabular}{|c|c c|}
\hline
Dimension & \hspace{2.5cm} Gluon operators & \\
\hline
\multirow{2}*{4}  & $G^{A\, \mu\nu}G^A_{\mu\nu}$ &$G^{A\, \mu\nu}\tilde{G}^A_{\mu\nu}$
\\
&$ -G^{A\mu\lambda}G^{A\nu}_{~~~\lambda}+\frac{1}{d}g^{\mu\nu}(G^{A}_{\alpha\beta})^{2}$&
\\
\hline
5  & $G^{A\, \alpha\beta}iD_-^\mu G^A_{\alpha\beta}$
&$G^{A\, \alpha\beta}iD_-^\mu \tilde{G}^A_{\alpha\beta}$
\\
\hline
    \end{tabular}
            \caption{Gluon bilinear operators. We define $\tilde{G}^A_{\mu\nu}=\epsilon^{\mu\nu\alpha\beta}G^{A}_{\alpha\beta}/2$ with totally antisymmetric $\epsilon^{0123}=+1$ notation.}       \label{table:gops}
\end{table}

\begin{table}[tbp]
    \centering
    \begin{tabular}{|c|c c|}
\hline
Dimension & \hspace{3cm} Scalar DM &    \\
\hline
\multirow{1}*{2}&  $|\phi|^2 $ & 
 \\
\hline
\multirow{1}*{3}  & $\phi^*i\partial_{-}^\mu\phi $ &  $\phi^*\partial_{+}^\mu\phi $
\\
\hline
\multirow{3}*{4}  & $ \phi^*\partial_{+}^2\phi $ &  $ \phi^*\partial_{-}^2\phi$
\\
&$\phi^* (\partial^\mu_{-}\partial^\nu_{-}-\frac{g^{\mu\nu}}{4}\partial_{-}^2)\phi $ & $\phi^* (\partial^\mu_{+}\partial^\nu_{+}-\frac{g^{\mu\nu}}{4}\partial_{+}^2)\phi $
\\
&$\phi^* i\partial^{\{\mu}_{+}\partial^{\nu\}}_{-}\phi $&$\phi^* i\partial^{[\mu}_{+}\partial^{\nu]}_{-}\phi  $
\\
\hline
\multirow{2}*{5}  & $\phi^*i\partial^2_{-}\partial_{-}^\mu\phi$ &  $\phi^*\partial^2_{-}\partial_{+}^\mu\phi $
\\
&$\phi^*i\partial^2_{+}\partial_{-}^\mu\phi $&$\phi^*\partial^2_{+}\partial_{+}^\mu\phi$
\\
\hline
    \end{tabular}
            \caption{Scalar dark matter bilinear operators.}       \label{table:sdm}
\end{table}

\begin{table}[tbp]
    \centering
    \begin{tabular}{|c|c c|}
\hline
Dimension & \hspace{3cm} Fermion DM   &  \\
\hline
\multirow{3}*{3}  &   $\bar{\psi}\psi$&   $\bar{\psi}i\gamma^5 \psi $ 
\\
&   $\bar{\psi}\gamma^\mu\psi$ &  $\bar{\psi}\gamma^\mu\gamma^5\psi$
\\
&$ \bar{\psi}\sigma^{\mu\nu}\psi $ &$ \epsilon^{\mu\nu\rho\sigma} \bar{\psi}\sigma_{\rho\sigma}\psi $
\\
\hline
\multirow{12}*{4}  & $\bar{\psi}i\slashed{\partial}_{-}\psi $ & $\bar{\psi}\slashed{\partial}_{-}\gamma^5\psi $
\\
&$\bar{\psi}\slashed{\partial}_{+}\psi $ & $\bar{\psi}i\slashed{\partial}_{+}\gamma^5\psi $
\\
&$\bar{\psi}i\partial_{-}^\mu\psi$ &$\bar{\psi}i\partial_{-}^\mu i\gamma^5\psi$
\\
&  $\bar{\psi}\partial_{+}^\mu\psi$  & $\bar{\psi}\partial_{+}^\mu i\gamma^5\psi$
\\
 & $\bar{\psi}i\partial_{-}^\alpha\sigma_{\alpha\mu}\psi$& $\epsilon^{\alpha\beta\rho\mu}\bar{\psi}i\partial_{-\alpha}\sigma_{\beta\rho}\psi$
\\
   & $\bar{\psi}\partial_{+}^\alpha\sigma_{\alpha\mu}\psi$ & $\epsilon^{\alpha\beta\rho\mu}\bar{\psi}\partial_{+\alpha}\sigma_{\beta\rho}\psi$
\\
 & $\bar{\psi}(\gamma^{\{\mu}i\partial_{-}^{\nu\}}-\frac{g^{\mu\nu}}{4}i\slashed{\partial}_{-}) \psi $& $\bar{\psi}\gamma^{\{\mu}i\partial_{-}^{\nu\}} \gamma^5\psi$
\\
 & $ \bar{\psi}(\gamma^{\{\mu}\partial_{+}^{\nu\}}-\frac{g^{\mu\nu}}{4}\slashed{\partial}_{+}) \psi $& $ \bar{\psi}\gamma^{\{\mu}\partial_{+}^{\nu\}} \gamma^5\psi$
\\
&$ \bar{\psi}i\gamma^{[\mu}\partial_-^{\nu]} \psi $&$\bar{\psi}\gamma^{[\mu}i\partial_-^{\nu]} \gamma^5\psi $
\\
&$ \bar{\psi}\gamma^{[\mu}\partial_+^{\nu]} \psi $&$\bar{\psi}\gamma^{[\mu}\partial_+^{\nu]} \gamma^5\psi $
\\
&$\epsilon^{\mu\nu\rho\sigma} \bar{\psi}i\gamma_{[\rho}\partial_{-\sigma]} \psi$&$\epsilon^{\mu\nu\rho\sigma} \bar{\psi}\gamma_{[\rho}i\partial_{-\sigma]}\gamma^5 \psi$
\\
&$\epsilon^{\mu\nu\rho\sigma} \bar{\psi}\gamma_{[\rho}\partial_{+\sigma]} \psi $&$\epsilon^{\mu\nu\rho\sigma} \bar{\psi}\gamma_{[\rho}\partial_{+\sigma]}\gamma^5 \psi$
\\
\hline
\multirow{15}*{5}  &$\bar{\psi}{\partial}_{-}^2\psi$ &$\bar{\psi}{\partial}_{-}^2i\gamma^5\psi$
\\
&$\bar{\psi}{\partial}_{+}^2\psi$ &$\bar{\psi}{\partial}_{+}^2i\gamma^5\psi$
\\
&$\bar{\psi}\sigma_{\mu\nu}i{\partial}_{-}^{[\mu}{\partial}_{+}^{\nu]}\psi$ &$\epsilon_{\alpha\beta\mu\nu}\bar{\psi}\sigma^{\alpha\beta}i{\partial}_{-}^{[\mu}{\partial}_{+}^{\nu]}\psi$
\\
& $\bar{\psi}i\slashed{\partial}_{-}i\partial_{-}^\mu\psi$ &$\bar{\psi}i\slashed{\partial}_{-}i\partial_{-}^\mu\gamma^5\psi$
\\
&$\bar{\psi}i\slashed{\partial}_{-}\partial_{+}^\mu\psi$&$\bar{\psi}i\slashed{\partial}_{-}\partial_{+}^\mu\gamma^5\psi$
\\
&$\bar{\psi}\slashed{\partial}_{+}i\partial_{-}^\mu\psi$&$\bar{\psi}\slashed{\partial}_{+}i\partial_{-}^\mu\gamma^5\psi$
\\
&$\bar{\psi}\slashed{\partial}_{+}\partial_{+}^\mu\psi$ &$\bar{\psi}\slashed{\partial}_{+}\partial_{+}^\mu\gamma^5\psi$
\\
&$\bar{\psi}{\partial}^\mu_{-}\partial_{-}^\nu\psi$ &$\bar{\psi}{\partial}^\mu_{-}\partial_{-}^\nu i\gamma^5\psi$
\\
&$\bar{\psi}{\partial}^{\{\mu}_{+}i\partial_{-}^{\nu\}}\psi$ &$\bar{\psi}{\partial}^{\{\mu}_{+}\partial_{-}^{\nu\}} \gamma^5\psi$
\\
&$\bar{\psi}{\partial}^{[\mu}_{+}i\partial_{-}^{\nu]}\psi$ &$\bar{\psi}{\partial}^{[\mu}_{+}\partial_{-}^{\nu]} \gamma^5\psi$
\\
&$\bar{\psi}{\partial}^\mu_{+}\partial_{+}^\nu\psi$ &$\bar{\psi}{\partial}^\mu_{+}\partial_{+}^\nu i\gamma^5\psi$
\\
&$\bar{\psi}\sigma_{\mu\alpha}i\partial_{-}^\alpha i\partial_{-\nu}\psi$ &$\bar{\psi}i\gamma^5\sigma_{\mu\alpha}i\partial_{-}^\alpha i\partial_{-\nu}\psi$ 
\\
&$\bar{\psi}\sigma_{\mu\alpha}\partial_{+}^\alpha i\partial_{-\nu}\psi$ &$\bar{\psi}i\gamma^5\sigma_{\mu\alpha}\partial_{+}^\alpha i\partial_{-\nu}\psi$ 
\\
&$\bar{\psi}\sigma_{\mu\alpha}i\partial_{-}^\alpha \partial_{+\nu}\psi$ &$\bar{\psi}i\gamma^5\sigma_{\mu\alpha}i\partial_{-}^\alpha \partial_{+\nu}\psi$ 
\\
&$\bar{\psi}\sigma_{\mu\alpha}\partial_{+}^\alpha \partial_{+\nu}\psi$ &$\bar{\psi}i\gamma^5\sigma_{\mu\alpha}\partial_{+}^\alpha \partial_{+\nu}\psi$ 
\\
\hline       
    \end{tabular}     
\caption{Fermion dark matter bilinear operators.}       \label{table:fdm}
\end{table}

\begin{figure}[h]
        \centering
        \includegraphics[width=.65\linewidth]{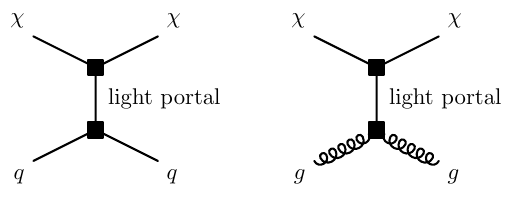}
\caption{\label{fig:tchannel}
Interaction of dark matter with quarks/gluons via a light mediator in the $t$ channel, where the black squares denote effective vertices that may represent spin-0, spin-1, and spin-2 currents.
}
\end{figure}

Then, {by applying EOM and integration-by-parts at the QCD Lagrangian level for the operator basis  of the QCD sector and constructing all possible Lorentz structures that contract with the QCD sector operators from the DM sector,} together with the propagators, we obtain a complete set of non-local effective operators that account for the leading contributions up to dimension 8 and spin-2 propagators. We denote the DM bilinear operators by $O_\chi$, and the quark/gluon bilinear operators by $O_{q/g}$, as shown in Figure~\ref{fig:tchannel}. The spin-0, spin-1, and spin-2 effective operators take the form
\ba
O_\chi\frac{\Xi }{\partial^2-m^2}O_q
\\
O_\chi^\mu\frac{\Xi _{\mu\nu}}{\partial^2-m^2} O_q^\nu
\\
O_\chi^{\mu\nu}\frac{\Xi_{\mu\nu\alpha\beta} }{\partial^2-m^2}O_q^{\alpha\beta}
\ea
where $m$ is the mass of the mediator, the $\Xi$'s are dimensionless tensors, specifically, $\Xi=i$, $\Xi_{\mu\nu}=-ig_{\mu\nu}$ and $\Xi_{\mu\nu\alpha\beta}=-i\left(\frac{1}{2}\tilde{\eta}_{\mu\alpha}\tilde{\eta}_{\nu\beta}+\frac{1}{2}\tilde{\eta}_{\mu\beta}\tilde{\eta}_{\nu\alpha}
-\frac{1}{3}\tilde{\eta}_{\mu\nu}\tilde{\eta}_{\alpha\beta}\right)$  with $\tilde{\eta}_{\mu\nu}=-{\eta}_{\mu\nu}+{\partial_{\mu}\partial_\nu}/{m^2}$. The propagators will be abbreviated as $P^{(0)}$, $P^{(1)}_{\mu\nu}$ and $P^{(2)}_{\mu\nu\alpha\beta}$ below. 

\section{Nucleon matrix elements and hadronic tensors \label{sec:nme}}

With the effective operators systematically enumerated in the previous section, a generic amplitude for the dark matter and nucleon scattering $\chi(k)+N(p)\to \chi(k^\prime)+N(p^\prime)$ is then given by
\ba
\mathcal{M}=\sum_i\sum_{q,\,g} c_i \langle \chi (k^\prime) |\mathcal{O}^{i}_\chi|\chi(k)\rangle P^{i}
\langle N(p^\prime) |\mathcal{O}^{i}_{q/\,g}|N(p)\rangle\,,
\ea
where the index $i$ labels a combination of dark matter effective operator and quark/gluon operator via a propagator from all possible combinations and the $c_i$ are the corresponding Wilson coefficients.
Here, $\mathcal{O}^i_\chi$ represents the dark matter bilinear operator, $\mathcal{O}^i_{q/g}$ represents the quark or gluon bilinear operator, and $P^i$ denotes the propagator. Microscopically, DM interacts with the nucleon through quarks and gluons. The nucleon is a composite system of quarks and gluons, and the nucleon matrix elements of the quark/gluon operators $\langle N(p^\prime) |\mathcal{O}^{i}_{q/\,g}|N(p)\rangle$ are nonperturbative at low energies. Their forms are fixed by the Lorentz structures of the operators, with momentum-transfer-dependent coefficients known as form factors. These form factors reflect the distribution of quarks and gluons inside a nucleon.

The squared amplitude, summed over the spins of the final states, is
\ba
\sum_{\rm spins}{|\mathcal{M}|^2}=\sum_{i,\,j} c_ic_j^* T^\chi_{ij} P_i P_j^* T^N_{ij}/\left(1+\delta_{i\,j}\right)\,,
\ea
where $$T^\chi_{ij}=\mathrm{Tr}\left[\langle \chi (k^\prime) |\mathcal{O}^{j}_\chi|\chi(k)\rangle^*\langle \chi (k^\prime) |\mathcal{O}^{i}_\chi|\chi(k)\rangle\right]$$ is the dark matter tensor, 
and $$T^N_{ij}=\mathrm{Tr}\left[\langle N(p^\prime) |\mathcal{O}^{j}_{q/\,g}|N(p)\rangle^*\langle N(p^\prime) |\mathcal{O}^{i}_{q/\,g}|N(p)\rangle\right]$$ is the hadronic tensor.

The quark and gluon bilinear operators $\mathcal{O}^{j}_{q/g}$ are listed in Tables~\ref{table:qops} and~\ref{table:gops}.
We will evaluate their matrix elements between nucleon states and express the results in terms of form factors at finite momentum transfer. 

(1) The quark scalar matrix element may be written as
\ba
m_q\langle N(p^\prime) |\bar{q}q|N(p)\rangle=m_N F_s^q(Q^2)\bar{u}(p^\prime)u(p)\,,
\ea
where $Q^2\equiv -q^2$ and $m_N$ is the nucleon mass. At zero momentum transfer, the form factor $F_s^q(0)\equiv g^q_s$ is just the flavor-diagonal scalar charge, intimately related to the usual quantities
\ba
\sigma_{\pi N}=\frac{m_u+m_d}{2}\langle N(p)|\bar{u}u+\bar{d}d|N(p)\rangle\,,\quad \sigma_s=m_s\langle N(p)|\bar{s}s|N(p)\rangle\,.
\ea
At finite momentum transfer, the scalar form factor $F_s^q(Q^2)$ can be expressed in terms of the nucleon gravitational form factors $A(t),\, J(t)$ and $D(t)$  \cite{Hackett:2023rif} with uncertainties approaching $\sim 10\%-15\%$, which are defined as
\ba
\langle N(p^\prime) |T_i^{\mu\nu}|N(p)\rangle=\frac{\bar{u}\left(p^\prime\right)}{m_N}\!\left[P^\mu P^\nu A_i(t)+i P^{\{\mu}\sigma^{\nu\}\rho} q_\rho J_i(t)
+\!\frac{q^\mu q^\nu\!-\!g^{\mu\nu}q^2}{4}D_i(t)\right]\! u\left(p\right),
\ea
where $i=q,\,g$, $t\equiv q^2$, and $T_q^{\mu\nu}=\bar{q}\gamma^{\{\mu}iD_{-}^{\nu\}}q$
and $T_g^{\mu\nu}= -G^{A\mu\lambda}G^{A\nu}_{~~~\lambda}+\frac{1}{4}g^{\mu\nu}(G^{A}_{\alpha\beta})^{2}$
are the quark and gluon parts of the QCD energy momentum tensor $T^{\mu\nu}$.
The form factors $A_i(t),\, J_i(t)$ and $D_i(t)$ have been computed in lattice QCD \cite{Hackett:2023rif}. The trace of the energy momentum tensor is 
\ba
\langle N(p^\prime) |T^{\mu}_\mu|N(p)\rangle=\frac{1}{m_N}\bar{u}\left(p^\prime\right)\left[\left(m_N^2-\frac{t}{4}\right)A(t)-\frac{t}{2}J(t)+\frac{3}{4}tD(t)\right]u\left(p\right)\,,
\label{trace}
\ea
with $A(t)=\sum_{i=q,\,g}A_i(t),\,J(t)=\sum_{i=q,\,g}J_i(t),\,D(t)=\sum_{i=q,\,g}D_i(t)$. For each quark flavor, we have
\begin{align}
m_q\langle N(p^\prime) |\bar{q}q|N(p)\rangle
&=\langle N(p^\prime) |T^{\mu}_{q,\,\mu}|N(p)\rangle  \nonumber
\\
&=
\frac{1}{m_N}\bar{u}\left(p^\prime\right)\left[\left(m_N^2-\frac{t}{4}\right)A_q(t)-\frac{t}{2}J_q(t)+\frac{3}{4}tD_q(t)\right]u\left(p\right)\,,
\end{align}
which leads to
\ba
F_s^{q}(Q^2)=\left(1-\frac{t}{4m_N^2}\right)A_q(t)-\frac{t}{2m_N^2}J_q(t)+\frac{3t}{4m_N^2}D_q(t)\,.
\ea

At next-to-leading order, we have the QCD sum rule
\ba
\sum_q^{n_f}(1-\gamma_m)\langle N(p^\prime) m_q\bar{q}q|N(p)\rangle+\langle N(p^\prime)|\frac{\beta}{2g}(G^{A}_{\alpha\beta})^{2}|N(p)\rangle
=\langle N(p^\prime) |T^{\mu}_\mu|N(p)\rangle\,,
\label{qcd_sum}
\ea
where the quark mass anomalous dimension is given by $\gamma_m={d\log{m_q}}/{d\log{\mu}}=-{2\alpha_s}/{\pi}+O(\alpha_s^2)$, and the QCD beta function is $(\beta/g){d\log{g}}/{d\log{\mu}}=-(11-2n_f/3){\alpha_s}/{4\pi}+O(\alpha_s^2)$, with $\alpha_s$ being the strong coupling and $n_f$ being the number of flavors. The gluon contribution in Eq.~(\ref{qcd_sum}) has been computed in lattice QCD  \cite{Wang:2024lrm}.
Combining Eq.~(\ref{trace}) and Eq.~(\ref{qcd_sum}), we then see that the scalar form factor $F_s^q(Q^2)$ can be written as
\ba
F_s^{u+d+s}(Q^2)=\frac{1}{1-\gamma_m}\left[\left(1-\frac{t}{4m_N^2}\right)A(t)-\frac{t}{2m_N^2}J(t)+\frac{3t}{4m_N^2}D(t)-G_N(Q^2)\right]\,,
\ea
where $G_N(Q^2)$ is defined by $\langle N(p^\prime)|\frac{\beta}{2g}(G^{A}_{\alpha\beta})^{2}|N(p)\rangle=m_NG_N(Q^2)\bar{u}(p^\prime)u(p)$ \cite{Wang:2024lrm}, and the scalar form factor of each flavor is approximately
\ba
F_s^{q}(Q^2)=\frac{1}{1-\gamma_m}\left[\left(1-\frac{t}{4m_N^2}\right)A_q(t)-\frac{t}{2m_N^2}J_q(t)+\frac{3t}{4m_N^2}D_q(t)-\frac{1}{3}G_N(Q^2)\right]\,.\label{qsff}
\ea

(2) The quark pseudoscalar matrix element
 \ba
 m_q\langle N(p^\prime) |\bar{q}i\gamma^5 q|N(p)\rangle=m_NF_5^q(Q^2)\bar{u}(p^\prime)i\gamma^5 u(p)
 \ea
 can be derived from the experimentally more accessible matrix element
  \ba
 \bar{m}\langle N(p^\prime) |\bar{Q}i\gamma^5 \tau^a Q|N(p)\rangle=m_NF_5^a(Q^2)\bar{u}(p^\prime)i\gamma^5 u(p) .
 \ea
To see this, note that for the flavor singlet, triplet and octet, we have axial-vector currents
 \ba
 &&A_\mu^0=\frac{1}{3}\bar{Q}\gamma_\mu\gamma_5Q=\frac{1}{3}\left(\bar{u}\gamma_\mu\gamma_5u+\bar{d}\gamma_\mu\gamma_5d+\bar{s}\gamma_\mu\gamma_5s\right)\,,
 \\
&& A_\mu^3=\bar{Q}\gamma_\mu\gamma_5\tau^3Q=\frac{1}{2}\left(\bar{u}\gamma_\mu\gamma_5u-\bar{d}\gamma_\mu\gamma_5d\right)\,,
\\
&& A_\mu^8=\bar{Q}\gamma_\mu\gamma_5\tau^8Q=\frac{1}{2\sqrt{3}}\left(\bar{u}\gamma_\mu\gamma_5u+\bar{d}\gamma_\mu\gamma_5d-2\bar{s}\gamma_\mu\gamma_5s\right)\,,
\ea
where $\tau^a$ are the Gell-mann matrices (generators of $SU(3)$ flavor symmetry). The partial conservation of the axial-vector current (PCAC) gives rise to
\ba
\partial^\mu A_\mu^a =2i\bar{m}^a\bar{Q}\tau^a\gamma^5 Q\,,
\ea
where $\bar{m}^a$ is the PCAC mass and we take $\bar{m}^3=(m_u+m_d)/2$, $\bar{m}^8=(m_u+m_d+m_s)/3$.
The PCAC relation relates the axial-vector form factors and the pseudoscalar form factors,
 \ba
 F^a_A(Q^2)-\frac{Q^2}{4m_N^2}F^a_P(Q^2)=F_5^a(Q^2)\,,
 \label{pcac_form}
 \ea
 where $a=3,\,8$. Along with the axial anomaly identity for the flavor singlet
  \ba
\sum_q \partial_\mu\left( \bar{q}\gamma_\mu\gamma^5 q\right)=-\frac{n_f\alpha_s}{8\pi}G^{A\,\alpha\beta}\tilde{G}_{\alpha\beta}^A+\sum_q 2im_q\bar{q}\gamma^5q\,,
 \ea
where $n_f=3$ is the flavor number, it yields
\ba
\sum_q \left[F^q_A(Q^2)-\frac{Q^2}{4m_N^2}F^q_P(Q^2)\right]=-\frac{n_f\alpha_s}{16\pi}F^g_5(Q^2)+\sum_q F_5^q(Q^2)\,.
\label{axial_anomaly}
\ea
For notational convenience, we define $\kappa(Q^2)\equiv\sum_q \left[F^q_A(Q^2)-\frac{Q^2}{4m_N^2}F^q_P(Q^2)\right]+\frac{n_f\alpha_s}{16\pi}F^g_5(Q^2)$. With these established, the single-flavor quark pseudoscalar form factors are then solved by combining Eq.~(\ref{pcac_form}) and Eq.~(\ref{axial_anomaly}), which gives
\ba
&&\frac{F_5^u}{m_u}-\frac{F_5^d}{m_d}=\frac{4}{m_u+m_d}F_5^3(Q^2)\,,
\\
&&\frac{F_5^u}{m_u}+\frac{F_5^d}{m_d}-2\frac{F_5^s}{m_s}=\frac{6\sqrt{3}}{m_u+m_d+m_s}F_5^8(Q^2)\,,
\\
&&F_5^u+F_5^d+F_5^s=\kappa(Q^2)\,,
\ea
leading to
\begin{align}
F_5^u(Q^2)&=\frac{R_{ud}}{1+R_{ud}+R_{sd}}\left[\frac{2F_5^3(Q^2)\left(2+R_{sd}\right)}{1+R_{ud}}
+\frac{3\sqrt{3}F_5^8(Q^2)R_{sd}}{\left(1+R_{ud}+R_{sd}\right)}+\kappa(Q^2)\right]\,,
\\
F_5^d(Q^2)&=\frac{-1}{1+R_{ud}+R_{sd}}\left[\frac{2F_5^3(Q^2)\left(2R_{ud}\!+\!R_{sd}\right)}{1+R_{ud}}
-\frac{3\sqrt{3}F_5^8(Q^2)R_{sd}}{1+R_{ud}+R_{sd}}-\kappa(Q^2) \right]\,,
\\
F_5^s(Q^2)&=\frac{R_{sd}}{1+R_{ud}+R_{sd}}\!\left[\frac{2F_5^3(Q^2)\left(1\!-\!R_{ud}\right)}{1+R_{ud}}
-\frac{3\sqrt{3}F_5^8(Q^2)\left(1\!+\!R_{ud}\right)}{1+R_{ud}+R_{sd}}+\kappa(Q^2)\right].
\end{align}
where we have defined $R_{q_1q_2}=m_{q_1}/m_{q_2}$.
The flavor-combined form factors $F_5^a(Q^2)$ can be determined by $ F^a_A(Q^2) $ and $F^a_P(Q^2)$ via Eq.~(\ref{pcac_form}). For the isovector form factor $F_5^3(Q^2)$, we can directly get from the $z$-expansion fit \cite{Alexandrou:2023qbg}
\ba
F_5^3(Q^2)=\frac{m_\pi^2}{4m_N^2}\sum_{k=0}^3 a_k z^k(Q^2)\,,
\ea
where the pion mass is $m_\pi=0.135$ GeV, the nucleon mass is $m_N=0.938$ GeV, $t_0=0$ GeV$^2$, and 
\ba
a_0=4.62(33)\,,\quad a_1=-2.2(2.5)\,,\quad a_2=-2.9(3.7)\,,\quad a_3=-1.2(2.4)
\ea
We can derive the octet form factor $F_5^8(Q^2)$ by $2\sqrt{3}F_A^8(Q^2)$ and $2\sqrt{3}F^8_P(Q^2)$ from  \cite{Alexandrou:2021wzv} with their dipole parametrization
\ba
&&g_A^{u+d-2s}=0.46(5)\,,\quad m_A^{u+d-2s}=0.898(134)(22)(256) \,\rm{GeV}\,,
\\
&& F_{P}^{u+d-2s}(0)=6.621(618)(597)(1.966)\,,\quad m_P^{u+d-2s}=0.484(20)(53)(118) \,\rm{GeV}\,.
\ea

 (3) The quark vector matrix element can be written as
 \ba
 \langle N(p^\prime) |\bar{q}\gamma^\mu q|N(p)\rangle=\bar{u}(p^\prime)\left[F_{V1}^q(Q^2)\gamma^\mu+\frac{i}{2m_N}F_{V2}^q(Q^2)\sigma^{\mu\nu}q_\nu\right]u(p)\,,
\ea
where the Dirac form factor $F_{V1}^{p(n)}(Q^2)$ and the Pauli form factor $F_{V2}^{p(n)}(Q^2)$ for the neutron and proton can be extracted experimentally, 
and the single-flavor form factors can be derived by the electromagnetic current decomposition of the quark vector currents \cite{Cates:2011pz,Qattan:2011ke}
\ba
&&F_{V1}^{p(n)}(Q^2)=\frac{2}{3}F_{V1}^{u,p(n)}(Q^2)-\frac{1}{3}F_{V1}^{d,p(n)}(Q^2)-\frac{1}{3}F_{V1}^{s,p(n)}(Q^2)\,,
\\
&&F_{V2}^{p(n)}(Q^2)=\frac{2}{3}F_{V2}^{u,p(n)}(Q^2)-\frac{1}{3}F_{V2}^{d,p(n)}(Q^2)-\frac{1}{3}F_{V2}^{s,p(n)}(Q^2)\,.
\ea
The numerical values of the proton $u$ and $d$ flavor form factors $F_{V1}^u(Q^2)$, $F_{V1}^d(Q^2)$, $F_{V2}^u(Q^2)$, $F_{V2}^d(Q^2)$ at different $Q^2$ are tabulated in  \cite{Cates:2011pz}.
 Neglecting the strange flavor contributions, the neutron form factors can be obtained by interchanging the $u$ and $d$ flavors.

(4) The quark axial-vector matrix element can be parametrized as
 \ba
 \langle N(p^\prime) |\bar{q}\gamma^\mu\gamma^5 q|N(p)\rangle=\bar{u}(p^\prime)\left[F_{A}^q(Q^2)\gamma^\mu\gamma^5+\frac{q_\mu}{2m_N}F^q_P(Q^2)\gamma^5\right]u(p)
\ea
The single-flavor axial form factors have been computed in lattice QCD \cite{Alexandrou:2021wzv} with a dipole model,
\ba
F_{A}^q(Q^2)=\frac{F_{A}^q(0)}{\left(1+{Q^2}/{m_A^{q\,2}}\right)^2}\,,\quad F_{P}^q(Q^2)=\frac{F_{P}^q(0)}{\left(1+{Q^2}/{m_P^{q\,2}}\right)^2}\,,
\ea
with $q=u,\,d,\,s$ and $F_{A}^q(0)\equiv g_A^q$, as well as by the $z$-expansion  \cite{Hill:2010yb, Bhattacharya:2011ah}
\ba
F_{A}^q(Q^2)=\sum_{k=0}^{k_{\rm max}} a^{A,\,q}_k z^k(Q^2)\,,\quad F_{P}^q(Q^2)=\sum_{k=0}^{k_{\rm max}} a^{P,\,q}_k z^k(Q^2)\,,
\ea
with
\ba
z(Q^2)=\frac{\sqrt{t_{\rm cut}+Q^2}-\sqrt{t_{\rm cut}}}{\sqrt{t_{\rm cut}+Q^2}+\sqrt{t_{\rm cut}}}\,,\quad t_{\rm cut}=9m_\pi^2\,.
\ea
The strange flavor axial form factor can also be extracted from experiments  \cite{Pate:2024acz}.

(5) The quark $C$-even, spin-2 matrix element can be parametrized as
\ba
&&\langle N(p^\prime) | \bar{q}\left(\gamma^{\{\mu}iD_{-}^{\nu\}}-\frac{g^{\mu \nu}}{4}i\slashed{D}_{-}\right)q |N(p)\rangle
\nn\\
&&=\frac{1}{m_N}\bar{u}\left(p^\prime\right)\left[P^\mu P^\nu A_q(t)+i P^{\{\mu}\sigma^{\nu\}\rho} q_\rho J_q(t)
+\frac{1}{4}\left(q^\mu q^\nu-g^{\mu\nu}q^2\right)D_q(t)\right]u\left(p\right)
\nn\\
&&-\frac{g^{\mu \nu}}{4}m_N F_s^q(Q^2)\bar{u}(p^\prime)u(p)\,,
\ea
where the gravitational form factors for $u$, $d$ and $s$ flavor have been computed in lattice QCD~ \cite{Hackett:2023rif} and $F_s^q(Q^2)$ is the quark scalar form factor in Eq.~(\ref{qsff})

(6) The quark $C$-odd, spin-2 matrix element can be parametrized as
\ba
&&\langle N(p^\prime) |\bar{q}\gamma^{\{\alpha}iD_{-}^{\beta\}}\gamma^5 q|N(p)\rangle
\nn\\
&&=\bar{u}(p^\prime)\bigg[F^{5q,(2)}_A(Q^2)\gamma^{\{\alpha}P^{\beta\}}\gamma^5+F^{5q,(2)}_B(Q^2)\frac{q^{\{\alpha}P^{\beta\}}\gamma^5}{m_N}\bigg]u(p)
\ea
where the form factors are related to the generalized parton distributions by
\ba
F^{5q,(2)}_A(Q^2)=\tilde{A}^q_{20}(q^2),\quad F^{5q,(2)}_B(Q^2)=\tilde{B}^q_{20}(q^2) .
\ea
To date, both the isoscalar and isovector combinations $\tilde{A}^{u+d}_{20}$ and  $\tilde{A}^{u-d}_{20}$  \cite{Bhattacharya:2024wtg,Alexandrou:2019ali}, and the isovector combination $\tilde{B}^{u-d}_{20}$ have been computed in lattice QCD \cite{Alexandrou:2019ali}.

(7) The quark antisymmetric tensor matrix element can be parametrized as
\ba
&&~~~~ m_q\langle N(p^\prime) |\bar{q}\sigma^{\alpha\beta}i\gamma^5 q|N(p)\rangle
\nn\\
&&=m_q\bar{u}(p^\prime)\bigg[\sigma^{\alpha\beta}i\gamma^5F^{t,q}_1(Q^2)+\frac{P^{[\alpha}\gamma^{\beta]}\gamma^5}{m_N}F^{t,q}_2(Q^2)
+\frac{P^{[\alpha}q^{\beta]}\gamma^5}{m_N^2}F^{t,q}_3(Q^2)\bigg]u(p)\,,
\ea
where the form factors are related to the generalized parton distributions via
\ba
F^{t,q}_1(Q^2)=A_{T10}^q(q^2),\quad F^{t,q}_2(Q^2)=A_{T4}^q(q^2),\quad F^{t,q}_3(Q^2)=A_{T2}^q(q^2),
\ea
To date, only the isovector combinations $A_{T10}^{u-d}$, $A_{T4}^{u-d}$ and $A_{T2}^{u-d}$ have been computed in lattice QCD  \cite{Bhattacharya:2025yba}.

(8) The chiral-odd double-derivative scalar matrix element can be parametrized as
\ba
\langle N(p^\prime) |\bar{q}D_{-}^2q|N(p)\rangle=m_N^2F_{2s}^q(Q^2)\bar{u}(p^\prime)u(p)\,,
\ea
where form factor $F_{2s}^q(Q^2)$ is related to the quark scalar form factor given in Eq.~(\ref{qsff}) and the quark chromo-magnetic dipole moment form factor $\langle N(p^\prime) |\bar{q}\sigma_{\mu\nu}G^{\mu\nu}q|N(p)\rangle$ via
Eq.~(\ref{cmnm}), as analyzed in  \cite{Bhattacharya:2025blb}.

(9) The chiral-odd double-derivative pseudoscalar matrix element can be parametrized as
\ba
\langle N(p^\prime) |\bar{q}D_{-}^2i\gamma^5 q|N(p)\rangle=m_N^2F_{2a}^q(Q^2)\bar{u}(p^\prime)i\gamma^5u(p)\,,
\ea
where form factor $F_{2a}^q(Q^2)$ is related to the quark pseudoscalar form factor given in Eq.~(\ref{qsff}) and quark chromo-electric dipole moment form factor $\langle N(p^\prime) |\bar{q}\sigma_{\mu\nu}G^{\mu\nu}i\gamma^5 q|N(p)\rangle$ via Eq.~(\ref{cedm}).

(10) Chiral-odd double-derivative spin-2 matrix element can be parametrized as  \cite{Diehl:2005jf}
\ba
&&\langle N(p^\prime) |\bar{q}\left(iD_{-}^{\{\alpha} iD_{-}^{\beta\}}+\frac{g^{\alpha\beta}}{4}D_{-}^2\right)q|N(p)\rangle
\nn\\
&&=\bar{u}(p^\prime)\bigg[m_N P^{\{\alpha}\gamma^{\beta\}}A_{S20}(Q^2)+\frac{1}{2}i\sigma^{\{\alpha\rho}P^{\beta\}}q_{\rho}B_{S20}(Q^2)
+\frac{q^{\alpha}q^{\beta}}{m_N^2}C_{S2}(Q^2)\bigg]u(p)\,,
\ea
where $A_{S20}$, $B_{S20}$ and $C_{S2}$ are twist-3 form factors.

(11) Chiral-odd pseudo double-derivative spin-2 matrix element can be parametrized as  \cite{Diehl:2005jf}
\ba
&&\langle N(p^\prime) |\bar{q}iD_{-}^{\{\alpha} iD_{-}^{\beta\}}i\gamma^5 q|N(p)\rangle
\nn\\
&&=\bar{u}(p^\prime)\bigg[m_N i\gamma^{\{\alpha}q^{\beta\}}\gamma^5 \tilde{A}_{P21}(Q^2)+\frac{1}{2}P^{\alpha}P^{\beta}i\gamma^5\tilde{B}_{P20}(Q^2)
+\frac{1}{2}q^{\alpha}q^{\beta}i\gamma^5 \tilde{B}_{P22}(Q^2)\bigg]u(p)\,,\nn\\
\ea
where $ \tilde{A}_{P21}$, $\tilde{B}_{P20}$ and $ \tilde{B}_{P22}$ are the twist-3 form factors.

(12) The chiral-odd double-derivative symmetric tensor matrix element can be parametrized as
\ba
&&\langle N(p^\prime)  | \bar{q}D_{-}^{\{\lambda}D_{-}^{[\alpha\}}\sigma^{\beta]}{}_\lambda q |N(p)\rangle
\nn\\
&&=\bar{u}(p^\prime)\bigg[\sigma^{\alpha\beta}F^{2D,q}_1(Q^2)+i\frac{q^{[\alpha}\gamma^{\beta]}}{m_N}F^{2D,q}_2(Q^2)
+i\frac{P^{[\alpha}q^{\beta]}}{m_N^2}F^{2D,q}_3(Q^2)\bigg]u(p)\,,
\ea

(13) The chiral-odd double-derivative  antisymmetric tensor matrix element can be parametrized as
\ba
&&\langle N(p^\prime)  | \bar{q}D_{-}^2\sigma^{\alpha\beta}q  |N(p)\rangle
\nn\\
&&=m_q\bar{u}(p^\prime)\bigg[\sigma^{\alpha\beta}F^{2t,q}_1(Q^2)+i\frac{q^{[\alpha}\gamma^{\beta]}}{m_N}F^{2t,q}_2(Q^2)
+i\frac{P^{[\alpha}q^{\beta]}}{m_N^2}F^{2t,q}_3(Q^2)\bigg]u(p)\,,
\ea

(14) The gluon scalar matrix element can be parametrized as
\ba
\langle N(p^\prime) |G^{A\,\mu\nu}G_{\mu\nu}^A|N(p)\rangle=m_NF_s^g(Q^2)\bar{u}(p^\prime)u(p)\,,
\ea
where $F_s^g(Q^2)=\frac{2g}{\beta}G_N(Q^2)$, and $G_N(Q^2)$ has been computed in lattice QCD  \cite{Wang:2024lrm}.

(15) The gluon pseudoscalar matrix element can be parametrized as
\ba
\langle N(p^\prime) |G^{A\,\mu\nu}\tilde{G}_{\mu\nu}^A|N(p)\rangle=m_NF_5^g(Q^2)\bar{u}(p^\prime)i\gamma^5 u(p)\,,
\ea
which has been computed in the QCD instanton vacuum  \cite{Liu:2024rdm}.

(16) The gluon spin-2 matrix element can be parametrized as
\ba
&&\langle N(p^\prime) |-G^{A\mu\lambda}G^{A\nu}_{~~~\lambda}+\frac{1}{4}g^{\mu\nu}(G^{A}_{\alpha\beta})^{2}|N(p)\rangle
\nn\\
&&=\frac{1}{m_N}\bar{u}\left(p^\prime\right)\left[P^\mu P^\nu A_g(t)+i P^{\{\mu}\sigma^{\nu\}\rho} q_\rho J_g(t)
+\frac{1}{4}\left(q^\mu q^\nu-g^{\mu\nu}q^2\right)D_g(t)\right]u\left(p\right)\,,
\ea
where $P^\mu=\left(p^\mu+p^{\prime \mu}\right)/2$, $q^\mu=p^{\prime \mu}-p^\mu$. 
The gravitational form factors $A_g(t),\,J_g(t),\,D_g(t)$ have been computed in lattice QCD  \cite{Hackett:2023rif}.

(17) The gluon vector matrix element can be parametrized as
\ba
&&\langle N(p^\prime) |G^{A\, \alpha\beta}iD_-^\mu G^A_{\alpha\beta}|N(p)\rangle
\nn\\
&&=m_N^2\bar{u}(p^\prime)\left[F_{V1}^g(Q^2)\gamma^\mu+\frac{i}{2m_N}F_{V2}^g(Q^2)\sigma^{\mu\nu}q_\nu\right]u(p)\,.
\ea

(18) The gluon axial-vector matrix element can be parametrized as
\ba
&&\langle N(p^\prime) |G^{A\, \alpha\beta}iD_-^\mu \tilde{G}^A_{\alpha\beta}|N(p)\rangle
\nn\\
&&=m_N^2\bar{u}(p^\prime)\left[F_{A}^g(Q^2)\gamma^\mu\gamma^5+\frac{q_\mu}{2m_N}F^g_P(Q^2)\gamma^5\right]u(p)\,.
\ea

Note that the form factor structures in this section are derived from all possible products of Dirac matrices and derivatives (momenta), with the nucleon-level equations of motion (EOMs) in Appendix~\ref{app:nid} used to eliminate redundant terms and to ensure time-reversal (CP-even) symmetry. We follow the conventional choices adopted in most of the literature, but one can easily convert to a preferred basis using the EOMs provided in Appendix~\ref{app:nid}.

With the form factors obtained, the nucleon matrix elements above $\langle N(p^\prime) |\mathcal{O}^{i}_{q/\,g}|N(p)\rangle\equiv \sum_k{F}^{ik}(Q^2)\bar{u}(p^\prime)\Gamma^k(P,\,q) u(p)$
can be used to infer the tensor structures $\Gamma^k(P,\,q)$, with which we can write the hadronic tensor as follows
\ba
T^N_{ij}&&=\mathrm{Tr}\left[\langle N(p^\prime) |\mathcal{O}^{j}_{q/\,g}|N(p)\rangle^\dagger\langle N(p^\prime) |\mathcal{O}^{i}_{q/\,g}|N(p)\rangle\right]
\nn\\
&&={F}^{ik}{F}^{jl}\mathrm{Tr} \left[\left(\bar{u}(p^\prime)\Gamma^l(P,\,q) u(p)\right)^\dagger\bar{u}(p^\prime)\Gamma^k (P,\,q)u(p)\right]
\nn\\
&&={F}^{ik}{F}^{jl}\mathrm{Tr}\left[\left(\slashed{p}+m_N\right)\Gamma^{l}(P,\,-q)\left(\slashed{p}^\prime+m_N\right)\Gamma^k(P,\,q)\right]\,,
\label{hadronTensor}
\ea
 where ${F}^{ik}$ is the $k$-th form factor in the nucleon matrix element of an operator $O^i$. Notice that all structures satisfy $\gamma^0\Gamma^{k\,\dagger}(P,\,q)\gamma^0=\Gamma^{k}(P,\,-q)$, as required by unitarity.

\section{Dark matter scattering on nucleus \label{sec:rfg}}

 Dark matter direct detection experiments take place on compound nuclei versus isolated nucleons. With finite momentum transfer situation, the nuclear effect becomes non-negligible. To put it simple, the momentum transfer is registered by a nucleus, then how do we translate the information that we have, i.e. the nucleon-level amplitude at a certain momentum transfer to that of the nucleus level?
 In this section, we systematically treat the nuclear effect by incorporating the relativistic Fermi gas model~\cite{Moniz:1969sr} and convert the cross section from the nucleon-level to that at the nucleus level at finite momentum transfer. 
{While we employ the relativistic Fermi gas model for its computational efficiency, we acknowledge that it lacks the sophisticated treatment of its sharp cutoff (binding energy choice/Pauli blocking) \cite{Ankowski:2014yfa}
 and nucleon correlations (e.g., 2p2h/meson exchange currents) \cite{Nieves:2011pp}. These effects primarily introduce uncertainties, which we account for by applying a 15-20\% systematic uncertainty comparing with more realistic spectral functions.}

\subsection{Relativistic Fermi gas model revisited}
 The cross section of dark matter and a single nucleon scattering $\chi(k)+N(p)\to \chi(k^\prime)+N(p^\prime)$ is given by
 \be
 \sigma_{\rm free}= \frac{1}{4|k\cdot p|}\int\!\frac{d^3\mathbf{p}^\prime}{(2\pi)^3 2E_{p^\prime}}\int\!\frac{d^3\mathbf{k}^\prime}{(2\pi)^3 2E_{k^\prime}}
(2\pi)^4\delta^{(4)}(p^\prime+k^\prime-p-k)\frac{1}{\rm{d.o.f}_\chi}\frac{1}{2}\sum_{\rm spins}{|\mathcal{M}|^2}\,,
\label{freecs}
\ee
where we have denoted the degree of freedom of dark matter by $\rm{d.o.f}_\chi$, and $\sum{|\mathcal{M}|^2}$ is the squared amplitude summed over final states spins.

One can use the relativistic Fermi gas model~\cite{Moniz:1969sr, Smith:1972xh, Bhattacharya:2011ah,Bhattacharya:2024win} to account for nuclear effects and compute dark matter–nucleus scattering at finite momentum transfer.
Following Ref.~\cite{Bhattacharya:2011ah}, the cross section for scattering off a nucleus can be related to that for scattering off a free nucleon by
\begin{align}
\sigma_{\rm nuclear}&=2V\int\frac{d^3\mathbf{p}}{\left(2\pi\right)^3}n_i(\mathbf{p})\sigma_{\rm free}\left[1-n_f(\mathbf{p}^\prime)\right]
\\
&=\int\frac{d^3\mathbf{k}^\prime}{(2\pi)^3 2E_{k^\prime}}\int\frac{d^3\mathbf{p}}{\left(2\pi\right)^3 2E_{p^\prime}}n_i(\mathbf{p})\left[1-n_f(\mathbf{p}^\prime)\right]
\frac{V}{4|k\cdot p|}(2\pi)\delta(p^0-p^{\prime\,0}+q^0)
\nn\\
&~~~~~~~~~~\times\frac{1}{\rm{d.o.f}_\chi}\sum_{\rm spins}{|\mathcal{M}|^2}\,,\label{cs_nuclear}
\end{align}
where the free nucleon cross section $\sigma_{\rm free}$ is given by Eq.~(\ref{freecs}), and $n_i(\mathbf{p})$ and $n_f(\mathbf{p}^\prime)$ denote the distribution functions of the initial- and final-state nucleons with three-momenta $\mathbf{p}$ and $\mathbf{p}^\prime$, respectively. The normalization $V = 3 \pi^2 A / (2 p_F^3)$ is fixed by requiring that half of the nucleons, for a nucleus with nucleon number $A$, lie below the Fermi surface, where $p_F$ is a parameter of the model.
A key step in this approach is to introduce the binding energy $\epsilon_b$ such that $p^0=E_p-\epsilon_b$, and $p^{\prime\,0}=E_{p^\prime}$, with $E_p=({m_N^2+|\mathbf{p}|^2})^{1/2}$ and $E_{p^\prime}=({m_N^2+|\mathbf{p}^\prime|^2})^{1/2}$ \cite{Bhattacharya:2011ah}.
{The integral in Eq.~(\ref{cs_nuclear}) involves hadronic tensor reduction from nucleon-level to nucleus level. The detailed derivation are given in Appendix~\ref{app:RFG}}.

\subsection{Cross section}

{After conversion of hadronic tensor from nucleon-level to nucleus level using relativistic Fermi gas model, we obtain the cross section}
\ba
\sigma_{\rm nuclear}&&=\int\frac{d^3\mathbf{k}^\prime}{(2\pi)^3 2E_{k^\prime}}\int d^3\mathbf{p} f(\mathbf{p},\,q^0,\,\mathbf{q})
\frac{1}{\rm{d.o.f}_\chi}\sum_{\rm spins}{|\mathcal{M}|^2}\,,\label{nuclearcs}
\ea
{where $f(\mathbf{p},\,q^0,\,\mathbf{q})$ are given in Appendix~\ref{app:RFG}.}
For fixed incoming dark matter momentum $|\bf k|$, we replace the integral $\int{d^3\mathbf{k}^\prime}$ by $\int{d^3\mathbf{q}}$ and obtain
\ba
\sigma_{\rm nuclear}&&=\int\frac{d^3\mathbf{q}}{(2\pi)^3 2E_{k^\prime}}\int d^3\mathbf{p}f(\mathbf{p},\,q^0,\,\mathbf{q})
\frac{1}{\rm{d.o.f}_\chi}\sum_{\rm spins}{|\mathcal{M}|^2}
\nn\\
&&=\int\frac{|\mathbf{q}|^2 d|\mathbf{q}|d\cos{\alpha}}{(2\pi)^2 2\sqrt{\mathbf{k}^2-2|\mathbf{k}||\mathbf{q}|\cos{\alpha}+|\mathbf{q}|^2+M^2}}\int d^3\mathbf{p} f(\mathbf{p},\,q^0,\,\mathbf{q})
\frac{1}{\rm{d.o.f}_\chi}\sum_{\rm spins}{|\mathcal{M}|^2}\,,
\nn\\
\label{nucs}
\ea
where $\alpha$ is the angle between $\mathbf{q}$ and $\mathbf{k}$. Next, we make use of the derivative
\ba
\frac{d|\mathbf{q}|}{dt}=\frac{|\mathbf{k}|\cos{\alpha}}{2\left(E_k^2-\cos^2{\alpha}|\mathbf{k}|^2\right)}
+\frac{E_k\left(-2E_k^2+2\cos^2{\alpha}|\mathbf{k}|^2+t\right)}{2\left(E_k^2-\cos^2{\alpha}|\mathbf{k}|^2\right)\sqrt{t^2+4\cos^2{\alpha}|\mathbf{k}|^2t-4E_k^2t}}\,,
\ea
to replace $|\mathbf{q}|$ in terms of $t$, and further integrate over $d\cos{\alpha}$ to obtain the differential cross section $d\sigma_{\rm{nuclear}}/dQ^2$, with $Q^2=-t$. 

Note that the maximum of momentum transfer $Q^2$ is  \cite{Bringmann:2018cvk}
\ba
Q^2_{\rm max}=2m_N\frac{T_{\chi}^2+2MT_{\chi}}{T_\chi+(M+m_N)^2/(2m_N)}\,,
\ea
where $T_\chi=E_k-M$ is the kinetic energy of dark matter. 
On the other hand, the minimum momentum transfer $Q^2_{\rm min}$ is determined by the experimental energy resolution. For high-threshold experiments such as DUNE~\cite{DUNE:2020lwj}, the resolution $\Delta E$ is on the order of MeV.
We therefore take the minimum momentum transfer to be $Q^2_{\rm min} = 2 m_N \Delta E \sim 0.001~\mathrm{GeV}^2$.

The event rate can be computed with given dark matter flux for different DM kinetic energy and three-momentum direction, which we leave for future work.

\section{Examples \label{sec:exam}}
As an application of the effective theory framework for light-portal dark matter and dark matter–nucleus scattering developed in the previous sections, we perform some example calculations here.
We focus on spin-1 and spin-2 portal scenarios. The relevant parameters in these models are the dark matter mass $M$, the mediator mass $m_{\mathrm{eff}}$, and the coupling constants.
We consider sub-GeV dark matter, which can be kinematically boosted and detected in high-threshold fixed-target experiments, such as large-volume neutrino detectors.
The mediator mass can also be constrained by small-scale astrophysical observations of dark matter self-interactions, which are relevant to resolving the so-called “core-cusp” problem.

\subsection{Core-cusp parametrization}

The core-cusp problem~\cite{Moore:1994yx,Flores:1994gz,Navarro:1995iw,Navarro:1996gj} refers to the discrepancy between the dark matter (DM) density profiles predicted by collisionless cold dark matter simulations and those inferred from observations. Simulations predict a steeply rising density toward the center of a DM halo, while observations of dwarf galaxies indicate a flat central density profile. This discrepancy can be explained by DM self-interactions, which can flatten the central density~\cite{Spergel:1999mh,Tulin:2017ara,Kang:2020afi}.

To address the core-cusp problem, we consider dark matter self-scattering with a reference cross section $\sigma_0 \sim 10^{-24}  (M/\mathrm{GeV})  \mathrm{cm}^2$~\cite{Spergel:1999mh}. However, there are significant variations in the inferred DM self-scattering cross sections across different astronomical systems—dwarf galaxies, larger galaxies, and galaxy clusters—deviating from {a single} value, {as referred to the ``diversity problem"} \cite{Oman:2015xda, Kaplinghat:2015aga, Tulin:2017ara}, { which generally favors strongly velocity-dependent self- interactions, with typical targets below ~0.5 cm$^2$/g on cluster scales, ~ 1-10 cm$^2$/g on galaxy scales, and much larger cross sections on dwarf-galaxy scales.} The detailed computation of the self-scattering cross section can be found in Ref.~\cite{Tulin:2013teo}. For sub-GeV dark matter, we adopt the mediator mass ranges corresponding to weak couplings from their results in Ref.~\cite{Tulin:2013teo}.

\subsection{Spin-1 portal light dark matter}

We first consider a simple effective interaction between a fermionic dark matter particle and a quark mediated by a spin-1 particle, which in momentum space can be written as,
\ba
\mathcal{L}_{\rm int}=\alpha_1\frac{\bar{\psi}\gamma^\mu\psi \bar{q}\gamma_\mu q}{q^2-m_{\mathrm{eff}}^2}+\alpha_2\frac{\bar{\psi}\gamma^\mu\gamma^5\psi \bar{q}\gamma_\mu\gamma^5 q}{q^2-m_{\mathrm{eff}}^2}+\alpha_3\frac{\bar{\psi}\gamma^\mu\gamma^5\psi \bar{q}\gamma_\mu q}{q^2-m_{\mathrm{eff}}^2}+\alpha_4\frac{\bar{\psi}\gamma^\mu\psi \bar{q}\gamma_\mu\gamma^5 q}{q^2-m_{\mathrm{eff}}^2}\,,
\ea
where the vector mediator has mass $m_{\mathrm{eff}}$. A possible UV model could be
\ba
\mathcal{L}_{\rm spin-1}&=&\bar{\psi}\left(i\slashed{\partial}+g_1\slashed{V}+g_2\slashed{V}\gamma^5-M\right)\psi+\bar{q}\left(g_1\slashed{V}+g_2\slashed{V}\gamma^5\right)q
\nn\\
&&-\frac{1}{4}\left(\partial_\mu V_\nu-\partial_\nu V_\mu\right)\left(\partial^\mu V^\nu-\partial^\nu V^\mu\right)+\frac{1}{2}m_{\mathrm{eff}}^2V_\mu V^\mu\,,
\ea
where $V^\mu$ is a spin-1 field, and $\alpha_1=g_1^2,\,\alpha_2=g_2^2,\,\alpha_3=\alpha_4=g_1g_2$. By using the effective interaction, we bypass the need to explicitly construct all possible UV models, which allows to focus on enumerating its effects at experimentally accessible energies.
With the effective interaction, the amplitude for dark-matter--nucleon scattering is given by
\ba
i\mathcal{M}&=&\sum_{q=u,d,s}\bar{u}_{\chi}(k^\prime)i\left(g_1\gamma^\mu+g_2\gamma^\mu\gamma^5\right) u_\chi(k)\frac{-i}{q^2-m_{\mathrm{eff}}^2}
\langle N\left(p^\prime\right)| \bar{q}i\left(g_1\gamma_\mu+g_2\gamma_\mu\gamma^5\right) q|N\left(p\right)\rangle
\nn\\
&=&\sum_{q=u,d,s}\bar{u}_{\chi}(k^\prime)i\left(g_1\gamma^\mu+g_2\gamma^\mu\gamma^5\right) u_\chi(k)\frac{-i}{q^2-m_{\mathrm{eff}}^2}i
\bar{u}(p^\prime)\bigg[g_1F_{V1}^q(Q^2)\gamma^\mu+g_2F_{A}^q(Q^2)\gamma^\mu\gamma^5
\nn\\
&&+\frac{ig_1}{2m_N}F_{V2}^q(Q^2)\sigma^{\mu\nu}q_\nu+g_2\frac{q_\mu}{2m_N}F^q_P(Q^2)\gamma^5\bigg]u(p)\,,\nn\\
\ea
where the complex nucleon part is parametrized by form factors discussed in Section \ref{sec:nme}. The spin-summed amplitude squared is then
\ba
\sum_{\rm spins}|\mathcal{M}|^2=\frac{1}{\left(q^2-m_{\mathrm{eff}}^2\right)^2}T^\chi_{\mu\nu}T^{N\,\mu\nu}\,,
\label{spin1ampsqrd}
\ea
with $T^\chi_{\mu\nu}=2\left[(g_1^2+g_2^2)\left(4 k_\mu k_\nu-2k_\mu q_\nu-2q_\mu k_\nu+q^2g^{\mu\nu}\right)-4g_2^2M^2g^{\mu\nu}+4ig_1g_2k_\alpha q_\beta  \epsilon^{\alpha\beta\mu\nu}\right]$ and
\ba
T^{N\,\mu\nu}=g^{\mu\nu}H_1+\frac{p^\mu p^\nu}{m_N^2}H_2+\frac{p^\mu q^\nu+p^\nu q^\mu}{2m_N^2}H_3+\frac{q^\mu q^\nu}{m_N^2}H_4
+i\epsilon^{\alpha\beta\mu\nu}\frac{p_\alpha q_\beta}{m_N^2}H_5\,,
\label{nucleonT}
\ea
where we have
\ba
&&H_1=-2\left[g_1^2\left(F_1+F_2\right)^2+g_2^2F_A^2\right] Q^2-8g_2^2F_A^2m_N^2\,,
\\
&&H_2=8\left(g_1^2F_1^2+g_2^2F_A^2\right)m_N^2+2g_1^2F_2^2 Q^2\,,
\\
&&H_3=8\left(g_1^2F_1^2+g_2^2F_A^2\right)m_N^2+2g_1^2F_2^2 Q^2\,,
\\
&&H_4=-4g_1^2F_1F_2m_N^2-g_1^2\left(2m_N^2-\frac{Q^2}{2}\right)F_2^2-g_2^2\frac{Q^2}{2}F_P^2\,, 
\\
&&H_5=8g_1 g_2 \left(F_1+F_2\right)F_A m_N^2\,,
\ea
and $F_i\equiv\sum_{q=u,d,s}F_{Vi}^q(Q^2)$, $i=1,\,2$, $F_A\equiv\sum_{q=u,d,s}F_{A}^q(Q^2)$, and $F_P\equiv \sum_{q=u,d,s}F_{P}^q(Q^2)$.

To use the relativistic Fermi gas model to treat the scattering between a dark matter particle and a nucleus, we cast the hadronic tensor Eq.\,(\ref{nucleonT}) into the similar form 
in terms of the nucleus momentum $p_T^\mu$ as follows
 \ba
T^{N\,\mu\nu}=g^{\mu\nu}W_1+\frac{p_T^\mu p_T^\nu}{m_T^2}W_2+\frac{p_T^\mu q^\nu+p_T^\nu q^\mu}{2m_T^2}W_3+\frac{q^\mu q^\nu}{m_T^2}W_4
+i\epsilon^{\alpha\beta\mu\nu}\frac{p_{T\alpha} q_\beta}{m_T^2}W_5\,,
\label{nucleusT}
\ea
where
\ba
&&W_1=A_{000}H_1+\frac{1}{2}(A_{020}-A_{001})H_2\,,
\\
&&W_2=\left[\frac{1}{2}\left(1-\frac{q_0^2}{|\mathbf{q}|^2}\right)A_{001}-\frac{1}{2}\left(1-3\frac{q_0^2}{|\mathbf{q}|^2}\right)A_{020}+
A_{200}-2\frac{q_0}{|\mathbf{q}|}A_{110}\right]H_2\,,
\\
&&W_3=\frac{m_T}{m_N}\left(A_{100}-A_{010}\frac{q_0}{|\mathbf{q}|}\right)H_3+\frac{m_T}{|\mathbf{q}|}\left[\left(A_{001}-3A_{020}\right)\frac{q_0}{|\mathbf{q}|}+2A_{110}\right]H_2\,,
\\
&&W_4=\frac{m_T^2}{m_N^2}\left[A_{000}H_4+\frac{m_N^2}{|\mathbf{q}|^2}\left(\frac{3}{2}A_{020}-\frac{1}{2}A_{001}\right)H_2+A_{010}\frac{m_N}{|\mathbf{q}|}H_3\right]\,,
\\
&&W_5=\frac{m_T}{m_N}\left[A_{010}\frac{q_0}{|\mathbf{q}|}-A_{100}\right]H_5\,,
\ea
after using Eq.~(\ref{2pterm}) and Eq.~(\ref{1pterm}) and the rest nucleus target momentum $p_T^\mu=m_T\delta^\mu_0$. 
The expressions of $A_{ijk}$ functions can be found in Eq.~(\ref{Afunc}).

\begin{figure}[h]
        \centering
        \includegraphics[width=.8\linewidth]{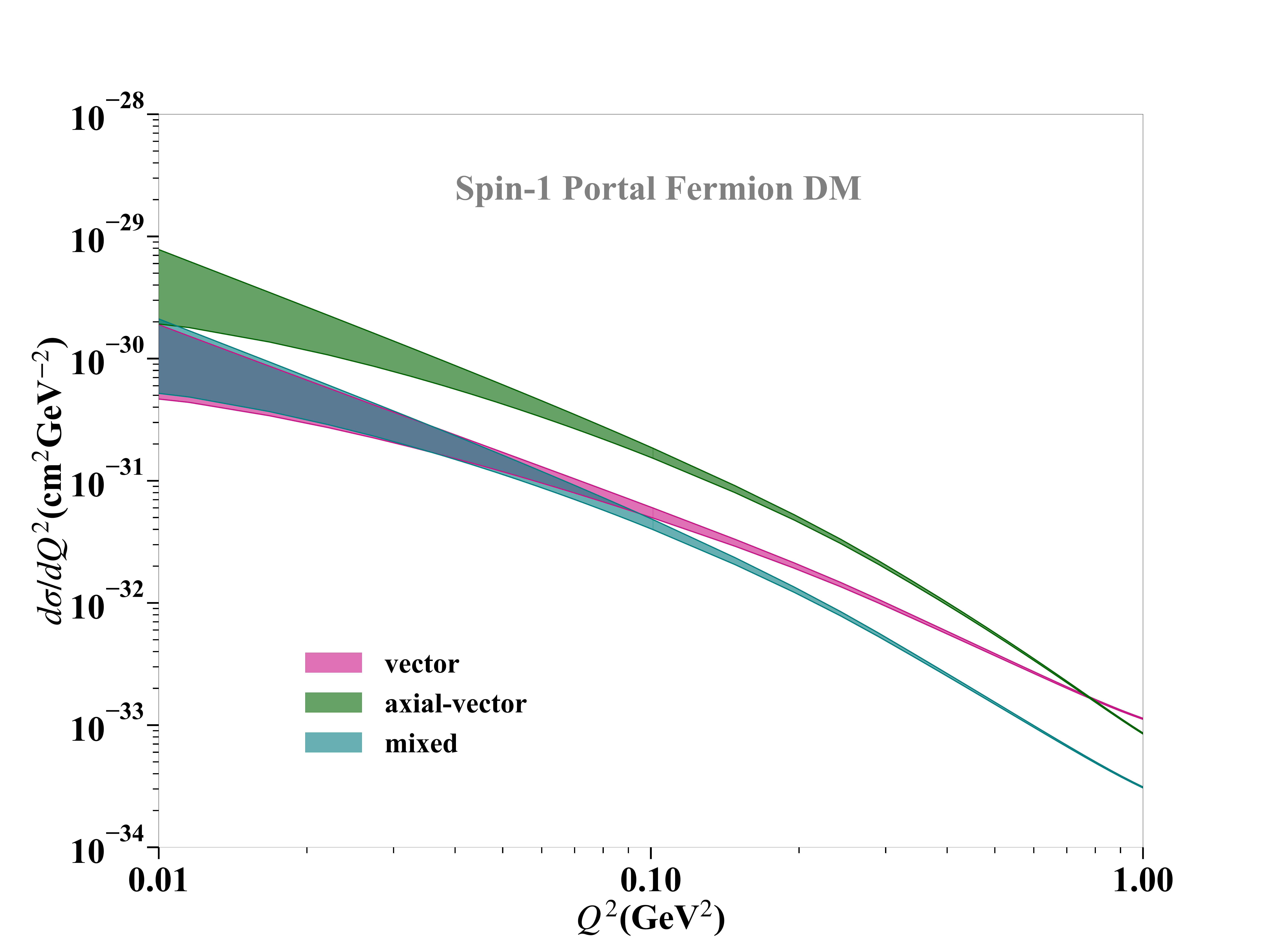}
\caption{\label{fig:spin1plot}
{The differential cross section of spin-1 portal (semi)relativistic dark matter and Argon nucleus scattering, with three types of interaction: pure vector, pure axial-vector and mixed.}
}
\end{figure}

As discussed in Section \ref{sec:nme}, the form factors can be extracted in various ways, and plugging them into the amplitude for dark matter–nucleus scattering leads to nucleon matrix elements at finite momentum transfer. The relativistic Fermi gas model further converts the cross section at finite momentum transfer from nucleon-level to the nucleus level. A numerical plot is shown in Figure.~\ref{fig:spin1plot}, where we consider sub-GeV cosmic-ray–boosted dark matter with particle mass $M=100$ MeV and kinetic energy $T=1$ GeV \cite{Bringmann:2018cvk}, scattering on Argon target ($A=40$) with three sets of coefficients: 
$\alpha_1=0.001,\,\alpha_2=\alpha_3=\alpha_4=0$ for pure vector interaction, $\alpha_2=0.001,\,\alpha_1=\alpha_3=\alpha_4=0$ for pure axial-vector interaction, and $\alpha_1=\alpha_2=\alpha_3=\alpha_4=0.001$ for vector and axial-vector mixed interaction , corresponding to magenta, dark green and teal colored bands in the figure. The Fermi momentum is taken to be $p_F = 0.27$ GeV, following Ref.~\cite{Smith:1972xh}.
The upper bound of each band corresponds to a mediator mass of $m_\mathrm{eff} = 0.001$ GeV, and the lower bound corresponds to $m_\mathrm{eff} = 0.1$ GeV, which are values taken from~\cite{Tulin:2013teo} at the coupling $\alpha\sim 0.001$ to solve the ``core-cusp" problem.
For simplicity, all form factors are taken at their central values for illustration purposes. 
For uncertainty sources, there are lattice computation uncertainties which can be found in relevant references from Section 3  and nuclear modeling of relativistic fermi gas.

Our parameter choices are illustrative, and more realistic constraints from relic abundance can be explored \cite{Elor:2021swj}.
Our parameter choices can be translated to the standard Yukawa self-interacting DM framework, with the variables $\omega$ and $\sigma_0/M$, where $\omega = m_{\mathrm{eff}}/M$ and the Born-regime zero-velocity cross section is given by $\sigma_0 = 4\pi \alpha^2 M^2 / m_{\mathrm{eff}}^4$.
We note that for the specific benchmark parameters chosen here to highlight the CRDM direct detection reach, the characteristic velocity scale is $\omega \sim 10^3 - 10^5$ km/s. At the characteristic velocities of dwarf galaxies ($v \sim 10-30$ km/s), this implies an effectively velocity-independent cross section of $\sigma_0/m_\chi \sim \mathcal{O}(0.5)$ cm$^2$/g, which is insufficient to fully resolve the dwarf-scale diversity problem. Therefore, these benchmarks should be viewed as illustrative for high-velocity comic ray boosted DM phenomenology rather than complete solutions to small-scale structure anomalies.

As the comic ray boosted DM travels through the Earth, multiple elastic scatterings cause it to lose kinetic energy. If the cross section is sufficiently large, the kinetic energy of the boosted DM flux will be degraded below the nuclear recoil or electron recoil energy thresholds of the deep underground detectors.
For sub-GeV dark matter and typical deep underground depths ($\sim 1$--$2$~km of rock equivalent), Earth attenuation generally becomes critical for DM-nucleon cross sections greater than $\sigma_{\chi N} \gtrsim 10^{-31}$ to $10^{-30}$~cm$^2$ and atmosphere attenuation becomes important for $\sigma_{\chi N} \geq 10^{-28}$ to $10^{-27}$~cm$^2$  \cite{Emken:2018run} . So for our parameter space here, the free-streaming approximation holds. However, if the parameter space of spin-1 example is changed to smaller mediator mass $ m_{\rm eff}<0.001$ GeV, stronger coupling $\alpha>0.001$, the free-streaming approximation breaks down and proper treatment of propagation effects should be incorporated.
In that case, the event rate drops as the attenuation effects. More details about earth-stopping effect can be found in \cite{Su:2022wpj}.

\subsection{Spin-2 portal dark matter}

As a second example, we now consider the effective theory for bi-gravity portal dark matter~\cite{Chen:2023oip}\footnote{Though the form of the model is the same as in~\cite{Chen:2023oip}, current paper focuses on light dark matter versus heavy dark matter in~\cite{Chen:2023oip}, which employes a different framework to compute the detection cross section.}, in which the gravitational sector is extended to a spin-2 gauge theory with two copies of diffeomorphism invariance (general covariance), $\mathrm{Diff}_L \times \mathrm{Diff}_R$. This symmetry is spontaneously broken down by the mass terms to a single diffeomorphism group.
The effective action after the symmetry breaking $\mathrm{Diff}_L \times \mathrm{Diff}_R \to \mathrm{Diff}_V$ is (see \cite{deRham:2014zqa} for a review)
\ba
&&S_{\mathrm{bi-gravity}}=\frac{1}{2}M_{\mathrm{pl}}^2\int d^4x\sqrt{|\mathrm{det}\, g|}\,R[g]+\frac{1}{2}M_f^2\int d^4x\sqrt{|\mathrm{det}\,f|}\,R[f]\nn
\\
&&\hspace{55pt}+\frac{1}{4}M_{f}^2m^2\int d^4x\sqrt{|\mathrm{det} \,f|}\,\sum_{n=0}^4\alpha_n U_n[K[g,f]]\,,
\label{bi}
\ea
where $R$ is the Ricci scalar curvature, $M_{\mathrm{pl}}$ is the Planck scale $\sim 10^{19}$ GeV, $M_f$ is another energy scale with  $M_f\ll M_{\mathrm{pl}}$ in this model and $\alpha_n$'s are real coefficients.
The mass hopping term takes the form $U_n[K]=K^{\mu_1}_{[\mu_1}K^{\mu_2}_{\mu_2}\dots K^{\mu_n}_{\mu_n]}$, 
with $K[g,f]^{\mu}_{\nu}\equiv\delta^{\mu}_{\nu}-(\sqrt{f^{-1}g})^\mu_\nu$, and ${}_{[\mu_1\,,\mu_2\,,...,\mu_n]}$ denoting total antisymmetrization of indices \cite{deRham:2010kj}. 
To reduce to the Fierz-Pauli  \cite{Fierz:1939ix} quadratic mass term at leading order, we take $\alpha_1=0$ and $\alpha_2=2$.

We can express the action $S_{\mathrm{bi-gravity}}$ in terms of perturbative mass eigenstates around flat space. To that end, we first decompose the metrics around the Minkowski metric $\eta_{\mu\nu}=\mathrm{diag}(1,\,-1,\,-1,\,-1)$,
\ba
&g_{\mu\nu}=\eta_{\mu\nu}+\frac{1}{M_{\mathrm{pl}}}\tilde{ g}_{\mu\nu}\nn
\\
&f_{\mu\nu}=\eta_{\mu\nu}+\frac{1}{M_f}\tilde{ f}_{\mu\nu}\,,
\ea
The following diagonalization 
\ba
&&H_{\mu\nu}=M_{\mathrm{eff}}\left(\frac{1}{M_f}\tilde{f}_{\mu\nu}-\frac{1}{M_{\mathrm{pl}}}\tilde{ g}_{\mu\nu}\right) ,\label{transM1}
\\
&&h_{\mu\nu}=M_{\mathrm{eff}}\left(\frac{1}{M_{\mathrm{pl}}}\tilde{f}_{\mu\nu}+\frac{1}{M_f}\tilde{ g}_{\mu\nu}\right)
\label{transM2}
\ea
with $M_{\mathrm{eff}}^2=(M_{\mathrm{pl}}^{-2}+M_f^{-2})^{-1}$ and $m_{\mathrm{eff}}^2=m^2(1+{M_f^2}{M_{\mathrm{pl}}^{-2}})$, is needed to obtain the action in terms of a massive state $H_{\mu\nu}$ and a massless state $h_{\mu\nu}$ at linear order,
\ba
S^{(2)}_{\mathrm{bi-gravity}}=\int d^4x\left[\mathcal{L}_{\mathrm{kin}}^{(2)}[H]+\mathcal{L}_{\mathrm{kin}}^{(2)}[h]
+\frac{1}{8}m_{\mathrm{eff}}^2\left(H^{\mu\nu}H_{\mu\nu}-H^2\right)\right]
\ea
where $H\equiv H^\mu_\mu$ and the linear order kinetic terms take the form,
\ba
&&\mathcal{L}_{\mathrm{kin}}^{(2)}[H]\equiv\frac{1}{4}\left(\frac{1}{2}H^{\mu\nu}\Box H_{\mu\nu}-H^{\mu\nu}\partial_\mu\partial^\alpha H_{\nu\alpha}+H\partial^\mu\partial^\nu H_{\mu\nu}-\frac{1}{2}H\Box H\right)\,,
\\
&&\mathcal{L}_{\mathrm{kin}}^{(2)}[h]\equiv\frac{1}{4}\left(\frac{1}{2}h^{\mu\nu}\Box h_{\mu\nu}-h^{\mu\nu}\partial_\mu\partial^\alpha h_{\nu\alpha}+h\partial^\mu\partial^\nu h_{\mu\nu}-\frac{1}{2}h\Box h\right)\,.
\ea
We can assume that the Standard Model and dark matter particles couple to both metrics   \cite{deRham:2014zqa, Huang:2015yga},
\ba
S=S_{\mathrm{bi-gravity}}+\int d^4x\sqrt{|\mathrm{det}\, g^{\mathrm{eff}}|}\, \mathcal{L}_{\mathrm{SM}}+\int d^4x\sqrt{|\mathrm{det}\,f^{\mathrm{eff}}|}\, \mathcal{L}_{\mathrm{DM}}\,,
\label{effL}
\ea
but with different effective metrics
\ba
&&g^{\mathrm{eff}}_{\mu\nu}=\alpha^2 f_{\mu\nu}+2\alpha\beta f_{\mu\rho} \sqrt{f^{-1}g}\big|^\rho_\nu+\beta^2 g_{\mu\nu}\,,
\\
&&f^{\mathrm{eff}}_{\mu\nu}=\alpha^{\prime2} f_{\mu\nu}+2\alpha^\prime\beta^\prime f_{\mu\rho} \sqrt{f^{-1}g}\big|^\rho_\nu+\beta^{\prime 2} g_{\mu\nu}\,,
\ea
where $\alpha$, $\beta$, $\alpha^\prime$, $\beta^\prime$ are real parameters.
In terms of the massive mode $H_{\mu\nu}$ and the massless mode $h_{\mu\nu}$, the linearized effective metrics are given by
\begin{align}
&g^{\mathrm{eff}}_{\mu\nu}=\eta_{\mu\nu}+\kappa \frac{H_{\mu\nu}}{M_{\mathrm{pl}}}
+\xi_r \frac{h_{\mu\nu}}{M_{\mathrm{pl}}}\,,\label{g_eff}
\\
&f^{\mathrm{eff}}_{\mu\nu}=\eta_{\mu\nu}+\kappa^\prime \frac{H_{\mu\nu}}{M_{\mathrm{pl}}}
+\xi_r \frac{h_{\mu\nu}}{M_{\mathrm{pl}}}\,,\label{f_eff}
\end{align}
with $\xi_r\equiv\left(1+r^2\right)^{-1/2}$, $\kappa\equiv \xi_r\left[1-(\beta/\xi_r)^2\right]/r$ and $\kappa^\prime\equiv \xi_r\left[1-(\beta^{\prime}/\xi_r)^2\right]/r$, where we have defined ratio $r\equiv M_f/M_{\mathrm{pl}}$ and chosen the normalization $\left(\alpha+\beta\right)^2 = \left(\alpha^\prime+\beta^\prime\right)^2=1$. The mass of the massive spin-2 mode is given by
$m_{\mathrm{eff}}=m/\xi_r$.

\begin{figure}[h]
        \centering
        \includegraphics[width=.65\linewidth]{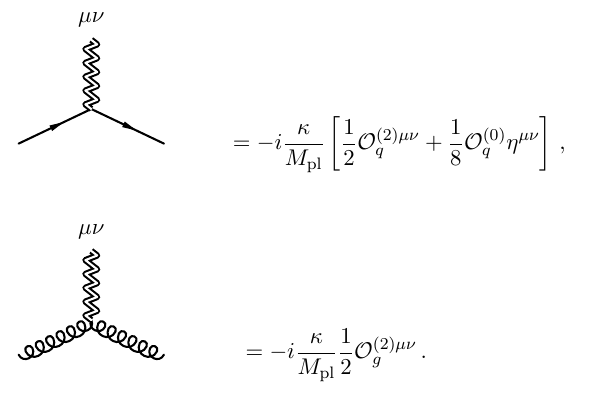}
\caption{\label{fig:spin2feyn}
Massive spin-2 currents matched onto QCD effective operators.
}
\end{figure}

We can compute the dark matter–quark and dark matter–gluon scattering amplitudes using the Feynman rules derived from the Lagrangian~Eq.\,(\ref{effL}), and then match them to the corresponding QCD operators, as illustrated in Figure.~\ref{fig:spin2feyn}.
where the relevant QCD operators are spin-0 and spin-2 quark and gluon operators:
\ba
\mathcal{O}^{(0)}_q&=&m_q \bar{q}q\,, \quad \mathcal{O}_g^{(0)}=G_{\mu\nu}^A G^{A\,\mu\nu}\,,\\
\mathcal{O}^{(2) \mu \nu }_{q} &=& \frac{1}{2}\bar{q}\left(\gamma^{\{\mu}iD_{-}^{\nu\}}-\frac{g^{\mu \nu}}{d}i\slashed{D}_{-}\right)q \,,\quad 
\mathcal{O}^{(2) \mu \nu }_{g}= -G^{A\mu\lambda}G^{A\nu}_{~~~\lambda}+\frac{1}{d}g^{\mu\nu}(G^{A}_{\alpha\beta})^{2}\,.
\ea
The amplitudes for fermion DM and nucleon scattering can be computed by using the massive spin-2 portal Feynman rules, as outlined in Appendix~\ref{app: fmrules},
\begin{align}
i\mathcal{M}=\bar{u}_{\chi}(p^\prime)\!\left(-\frac{i \kappa^\prime}{8M_{\mathrm{pl}}}\right)\left[(p^\mu+p^{\prime\mu})\gamma^\nu+(p^\nu+p^{\prime\nu})\gamma^\mu-2\eta^{\mu\nu}(\slashed{p}+\slashed{p}^\prime)
+4M \eta^{\mu\nu}\right]u_\chi(p)
\nl
\otimes P^{(2)}_{\mu\nu\alpha\beta}
\sum_{i=u,d,s,g}\left[-i\frac{\kappa}{8M_{\mathrm{pl}}}\langle N\left(p^\prime, s^\prime\right)|\mathcal{O}^{(0)}_{i}\eta^{\mu\nu}|N\left(p, s\right)\rangle-i\frac{\kappa}{2}\langle N\left(p^\prime, s^\prime\right)|\mathcal{O}^{(2) \mu \nu }_{i}|N\left(p, s\right)\rangle\right]\nn
\end{align}
where $\kappa\equiv \frac{1-(1+r^2)\beta^2}{M_{\mathrm{pl}} r\sqrt{1+r^2}}$ and  $\kappa^\prime\equiv \frac{1-(1+r^2)\beta^{\prime 2}}{M_{\mathrm{pl}} r\sqrt{1+r^2}}$,
\ba
P^{(2)}_{\mu\nu\alpha\beta}=\left(\frac{1}{2}\tilde{\eta}_{\mu\alpha}\tilde{\eta}_{\nu\beta}+\frac{1}{2}\tilde{\eta}_{\mu\beta}\tilde{\eta}_{\nu\alpha}
-\frac{1}{3}\tilde{\eta}_{\mu\nu}\tilde{\eta}_{\alpha\beta}\right)\frac{-i}{q^2-m_{\mathrm{eff}}^2+i\epsilon}\,,
\ea
with $\tilde{\eta}_{\mu\nu}=-{\eta}_{\mu\nu}+{q_{\mu}q_\nu}/{m^2_{\mathrm{eff}}}$.
Inserting the three flavor QCD operators into the nucleon states and take the lattice QCD results  \cite{Hackett:2023rif} for the form factors using the $z$-expansion fit, we get
\ba
\langle N\left(p^\prime, s^\prime\right)|\mathcal{O}^{(0)}_{i}|N\left(p, s\right)\rangle&=&m_N \mathcal{F}^{(0)}_{i, N}\left(t\right)
\bar{u}_N\left(p^\prime, s^\prime\right)u_N\left(p, s\right)\,,
\\
\langle N\left(p^\prime, s^\prime\right)|\mathcal{O}^{(2) \mu \nu }_{i}|N\left(p, s\right)\rangle
&=&\frac{1}{m_N}\bar{u}_N\left(p^\prime, s^\prime\right)\Bigg[P^\mu P^\nu A_i(t)+i P^{\{\mu}\sigma^{\nu\}\rho} q_\rho J_i(t)
\nn\\
&&+\frac{1}{4}\left(q^\mu q^\nu-g^{\mu\nu}q^2\right)D_i(t)\Bigg]u_N\left(p, s\right)\,,
\ea
where $P=\left(p+p^\prime\right)/2$, $q=p^\prime-p$, $t=q^2$.
For convenience, we define  $\sum_{i=u,d,s}\mathcal{F}^{(0)}_{i, N}\equiv F(t)$,
 $\sum_{i=u,d,s,g}A_i(t)\equiv A(t)$, $\sum_{i=u,d,s,g}J_i(t)\equiv J(t)$ and  $\sum_{i=u,d,s,g}D_i(t)\equiv D(t)$.

\begin{figure}
        \centering
        \includegraphics[width=.49\linewidth]{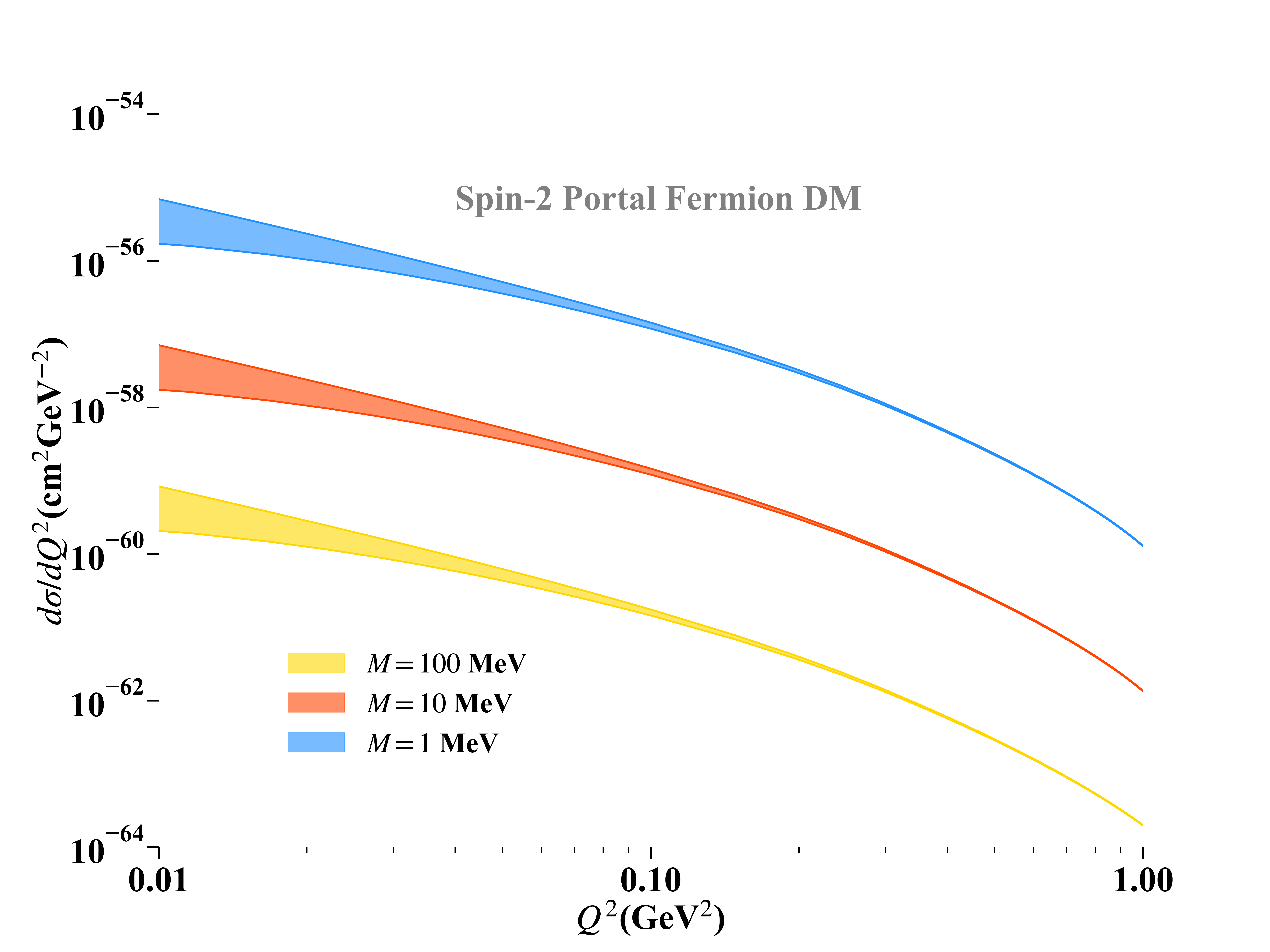}
          \includegraphics[width=.49\linewidth]{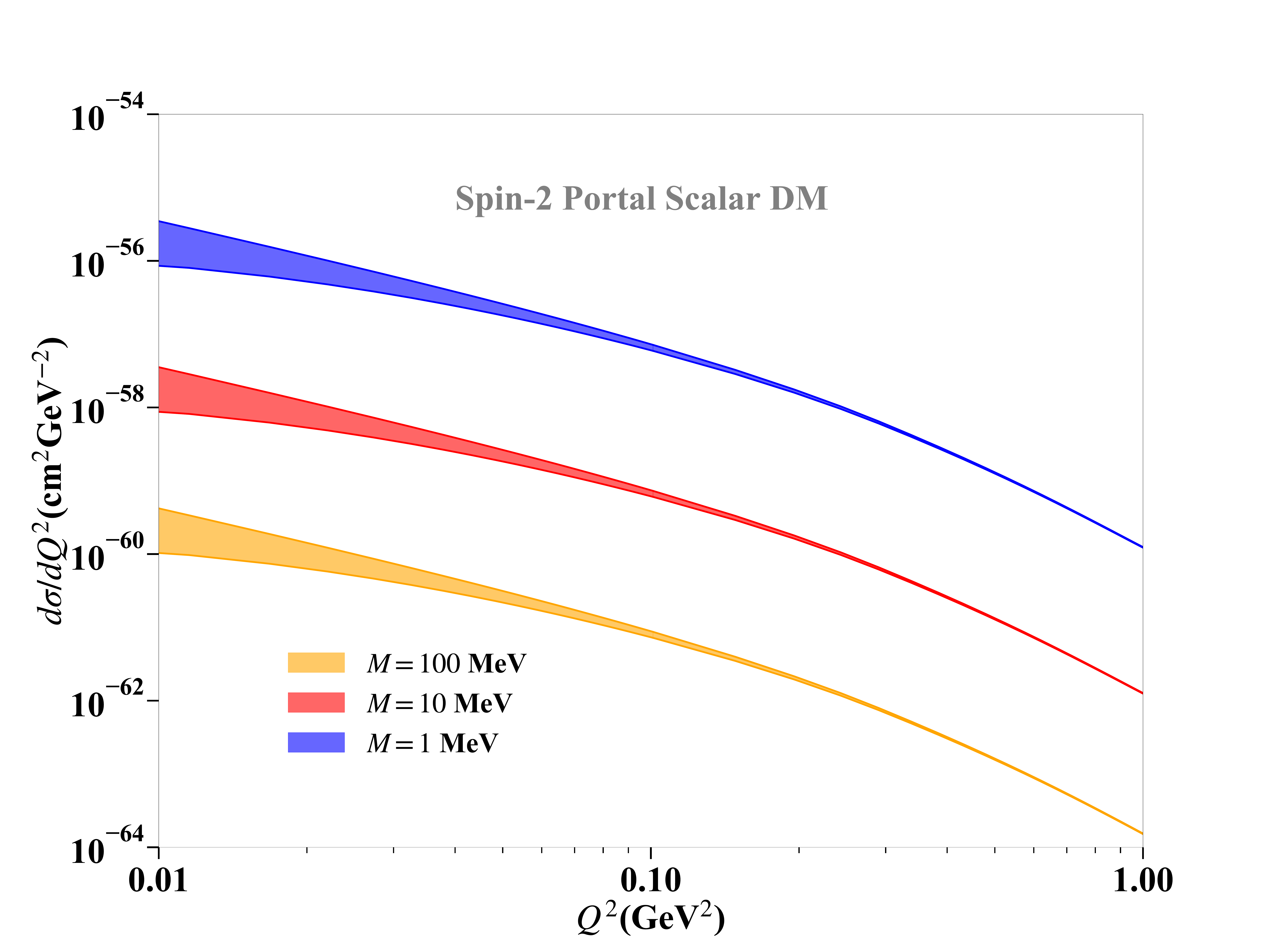}
\caption{\label{fig:spin2plot}
Differential cross sections for spin-2 portal (semi)relativistic dark matter and Argon nucleus scattering.
}
\end{figure}

Similar to the spin-1 portal case, we use the Fermi gas model to express the hadronic tensor in terms of the momentum transfer $q^\mu$ and the target momentum $p_T^\mu$, following the procedures discussed in Section~\ref{sec:rfg}.
In the spin-2 portal case, the tensor is rank-4, compared to rank-2 in the spin-1 case.
This requires decomposition formulae for the three-momentum term $p_\mu p_\nu p_\alpha$ [Eq.~(\ref{3greek})] and the four-momentum term $p_\mu p_\nu p_\alpha p_\beta$ [Eq.~(\ref{4greek})], in addition to the simpler two-momentum term $p_\mu p_\nu$ [Eq.~(\ref{2pterm})] used in the spin-1 case.
The increased complexity in the spin-2 portal case led us to write a program to perform the decomposition. We omit the lengthy intermediate expressions and present only the final numerical results in Figure.~\ref{fig:spin2plot}.

We consider both scalar and fermionic sub-GeV dark matter with masses $M = 1, 10, 100$ MeV and kinetic energy $T = 1$ GeV~\cite{Bringmann:2018cvk}, scattering off an Argon target ($A = 40$). The Fermi momentum is taken to be $p_F = 0.27$ GeV~\cite{Smith:1972xh}. The couplings are set to $\kappa = 1$~\cite{Chen:2023oip} and $\kappa^\prime = M_{\mathrm{pl}}/M$, representing the ratio between the Planck scale and the dark matter mass. 
The effective fine structure constant for nonrelativistic DM self-interaction is $\alpha_{\rm DM}=\left(\kappa^\prime M/M_{\mathrm{pl}}\right)^2/4\pi \sim 0.01$. 
According to Ref.~\cite{Tulin:2013teo}, for attractive interaction with $\alpha_{\rm DM}\sim 0.01$, the mediator mass is from below 0.01 GeV to above 0.01 GeV. 
We loosen the bounds by an order here, considering the differences among astronomical objects being used to yield the bounds.
Numerically, the upper bound of each band corresponds to a mediator mass $m_\mathrm{eff} = 0.001$ GeV, while the lower bound corresponds to $m_\mathrm{eff} = 0.1$ GeV. The numerical cross section values are small. Cross sections for other spin-2 portal models that yield higher results could be computed by this framework as well.

\subsection{Comparison with existing nuclear treatment}
We have compared our results with nuclear treatment for a vector mediator simplified-model in common existing literature. 
In literature \cite{Bringmann:2018cvk, Bondarenko:2019vrb,Dent:2019krz,Ge:2020yuf, Bell:2021xff, Alvey:2022pad,Bell:2023sdq, Cappiello:2024acu, Diurba:2025lky},
 differential cross section takes the form
\begin{equation}
\frac{d\sigma}{dQ^2}=\frac{A^2\sigma_0}{\mu^2_{\chi N}}\frac{m_{\rm eff} ^4}{\left(Q^2+m_{\rm eff} ^2\right)^2}G^2\left(Q^2\right)\label{lit}
\end{equation}
where 
\begin{equation}
\sigma_0=\frac{\alpha^2\mu^2_{\chi N}}{\pi m_{\rm eff} ^4}
\end{equation}
is the cross section at zero momentum transfer and $G\left(Q^2\right)$ is a form factor.
Literature take $G\left(Q^2\right)$ as Helm form factor $F(q) = 3 \frac{j_1(q R_1)}{q R_1} e^{-(q s)^2 / 2}$ for nuclei scattering 
and dipole form factor $G_P(Q^2)=1/(1+Q^2/\Lambda^2)^2$  for nucleon scattering with $\Lambda=770$ MeV.

\begin{figure}[h]
\centering
\includegraphics[width=.6\linewidth]
{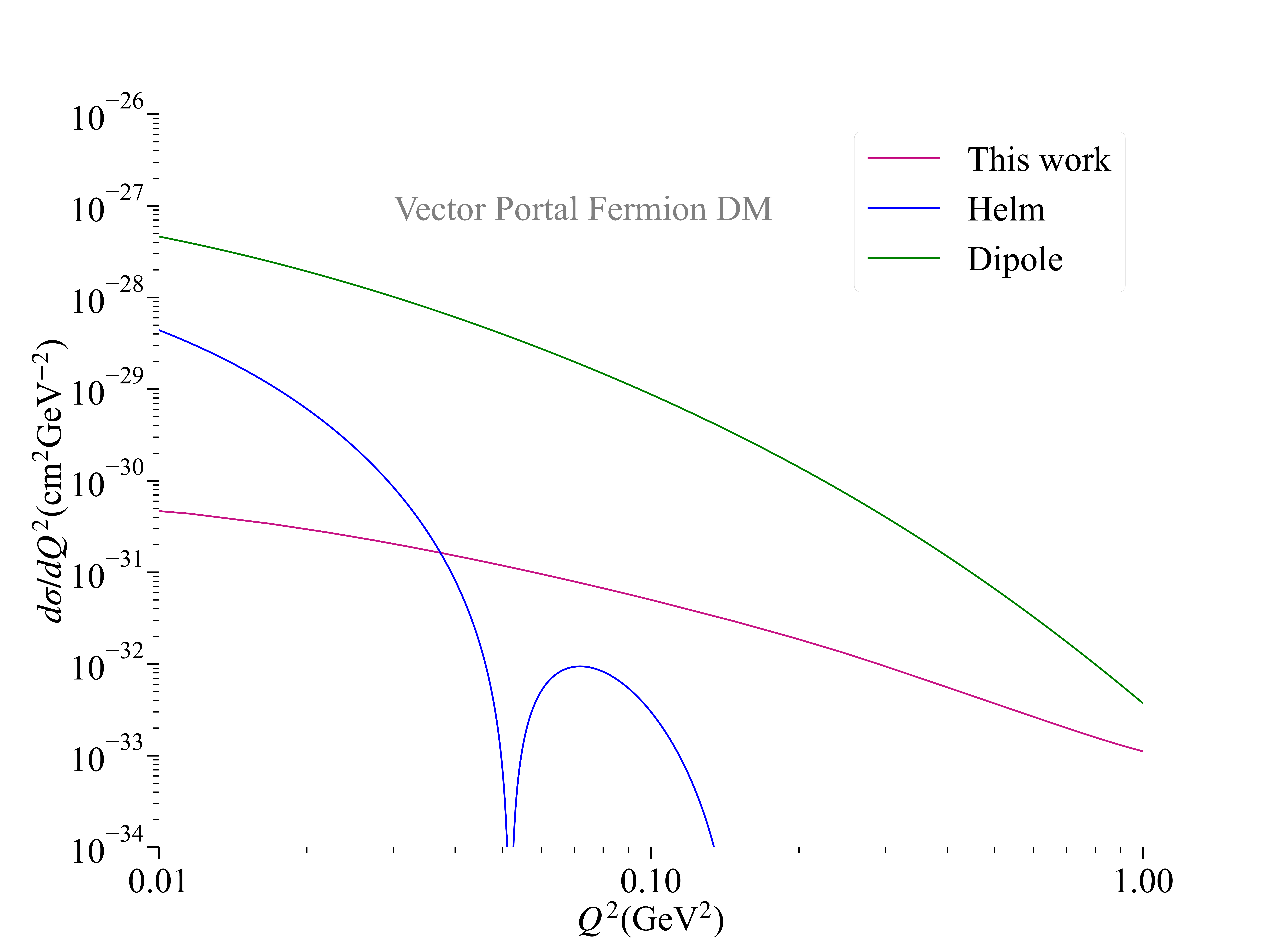}
\caption{Our nuclear treatment (magenta) is compared with dipole (green) form factor and Helm (blue) form factor treatment. \label{compare}}
\end{figure}

We compare our numerical result with nucleon-level quark operator finite momentum transfer information being convoluted to nuclei level with Fermi gas model for the same vector mediator model, setting dark matter mass $M=0.1$ GeV, kinetic energy $T=1$ GeV, $\alpha=0.001$, mediator mass $m_{\rm eff} =0.1$ GeV, with literature's results from Eq.~(\ref{lit}).
The differential cross sections are plotted in Fig.\ref{compare}.

Our result is distinctively different from either Helm or dipole form factor result.
The Helm form factor one has hills due to the spherical Bessel function  feature.
Ours is a smooth curve decreasing as momentum transfer increases, while it overlaps the Helm form factor result at low momentum transfer and yields same order of magnitude of area enclosed below the curve, indicating same order of magnitude of event rate at this regime but ours is higher at high momentum transfer regime.

The dipole form factor is usually taken for a single nucleon or light nuclei. Here for the Argon nuclei,  to apply the dipole form factor by force, the cross section is a result of simple addition of all nucleons' results.
It is a coherent result neglecting nucleon-nucleon interaction, therefore providing a much higher cross section than ours and the Helm form factor one's.

\section{Summary \label{sec:sum}}

We have formulated a general framework for computing cross sections for the detection of light-portal dark matter, which accounts for finite momentum transfer from microscopic effective quark and gluon operators at nucleon-level, then being convoluted to nucleus level. 
In this setup, dark matter interacts with Standard Model particles through a light mediator that also governs its self-interactions. 
The framework is particularly useful for scenarios involving light (semi-)relativistic dark matter, potentially boosted by cosmic rays\textcolor{red}{\footnote{Alternatively, this framework applies to dark matter reflected by hot solar nuclei~\cite{Emken:2017hnp}, which involves lower momentum transfers in the keV regime with a mediator mass also around the keV scale.}}, and is applicable to high-threshold fixed-target detectors such as large-volume neutrino experiments. 
By incorporating recent lattice QCD results for nucleon matrix elements and modeling nuclear effects via the relativistic Fermi gas model, we provide a systematic effective theory description. 
As a demonstration, we compute the momentum-transfer-dependent cross sections for spin-1 and spin-2 portal models capable of addressing the astrophysical core-cusp problem.

Specifically, we construct the effective interactions between dark matter and QCD bilinear operators by explicitly including the light mediator propagator. 
We consider QCD bilinear operators up to dimension 5, alongside scalar and fermion dark matter bilinear operators up to dimension 5. 
Excluding the dimension of the propagator, the resulting DM--QCD effective interactions span dimensions 6, 7, and 8. 
The nucleon matrix elements of these QCD operators are parametrized at finite momentum transfer. 
Furthermore, we extend the relativistic Fermi gas model to accommodate higher-rank hadronic tensors, thereby converting DM--nucleon interactions into DM--nucleus interactions. 
The detailed computational steps within this framework are provided for easy reference.

 We have truncated the spin of the light-portal to be no more than 2. In principle, one could also consider mediators with higher spins, which may arise from a scenario where the light-portal is, e.g., string-like and contains an infinite number of higher--spin states. However, in the absence of any enhancement of the effective coefficients, the higher spin contributions are suppressed by extra factors of the effective cutoff from the bilinears. This occurs because higher-spin propagators possess more Lorentz indices, necessitating the use of higher-dimensional DM/SM bilinear operators or additional derivatives for proper index contraction, thereby increasing the effective dimension of the operator.
 
This framework applies to differential cross section computation with finite momentum transfer of general light-portal dark matter direct detection. It can further be used to predict the event rate in the experiments by incorporating dark matter flux information.

\acknowledgments
QC gratefully acknowledges useful discussions with Yi Chen on hermitian operators, Gui-Jun Ding and Jiang-Hao Yu on cosmic ray boosted dark matter, Yang Li and Zhaofeng Liu on QCD operators and nucleon matrix elements, Keping Xie on nuclear effect.
QC is supported by National Natural Science Foundation of China under Grant No.\,12505118, Anhui Provincial Natural Science Foundation under Grant No.\,2508085QA034 and 
Scientific Research Foundation for High-level Talents of Anhui University of Science and Technology under Grant No.\,2024yjrc166.
SYZ acknowledges support from the National Natural Science Foundation of China under grant No.~12475074, No.~12075233 and No.~12247103.

\appendix

\section{Nucleon onshell identities \label{app:nid}}
The nucleon may be treated as a Dirac fermion, and from equation of motion, the following identities of operators in momentum space can be derived.
\ba
&&\bar{u}(p^\prime)\left(P^\nu-\sigma^{\mu\nu}iq_\mu\right)u(p)=2m_N\bar{u}(p^\prime)\gamma^\nu u(p)
\\
&&\bar{u}(p^\prime)\left(i\sigma^{\mu\nu}\gamma^5 P_\mu-q^\nu\gamma^5 \right)u(p)=2m_N\bar{u}(p^\prime)\gamma^5 \gamma^\nu u(p)
\\
&&\bar{u}(p^\prime)\left(\sigma^{\mu\nu} P_\mu+iq^\nu \right)u(p)=0
\\
&&\bar{u}(p^\prime)\left(i\gamma^5 P^\nu+\sigma^{\mu\nu}q_\mu \gamma^5 \right)u(p)=0
\\
&&\bar{u}(p^\prime)\left(\gamma^{[\alpha} P^{\nu]}+\frac{1}{2}\epsilon^{\mu\nu\alpha\beta}iq_\mu \gamma_\beta\gamma^5 \right)u(p)=0
\\
&&\bar{u}(p^\prime)\left(\gamma^{[\alpha} iq^{\nu]}-\frac{1}{2}\epsilon^{\mu\nu\alpha\beta}P_\mu \gamma_\beta\gamma^5 \right)u(p)=m_N\bar{u}(p^\prime)\sigma^{\nu\alpha} u(p)
\\
&&\bar{u}(p^\prime)\left(P^\nu i q^\alpha+q_\mu q^\alpha\sigma^{\mu\nu}\right)u(p)=2m_N\bar{u}(p^\prime)\gamma^\nu iq^\alpha u(p)\label{p1}
\\
&&\bar{u}(p^\prime)\left(P^\nu  P^\alpha-iq_\mu P^\alpha\sigma^{\mu\nu}\right)u(p)=2m_N\bar{u}(p^\prime)\gamma^\nu P^\alpha u(p)
\\
&&\bar{u}(p^\prime)\left(iq^\nu  P^\alpha+P_\mu P^\alpha\sigma^{\mu\nu}\right)u(p)=0
\\
&&\bar{u}(p^\prime)\left(q^\nu q^\alpha-iP_\mu q^\alpha\sigma^{\mu\nu}\right)u(p)=0\label{p2}
\\
&&\bar{u}(p^\prime)\left(P^\nu i q^\alpha+q_\mu q^\alpha\sigma^{\mu\nu}\right)i\gamma^5u(p)=0\label{p51}
\\
&&\bar{u}(p^\prime)\left(q^\nu q^\alpha-iP_\mu q^\alpha\sigma^{\mu\nu}\right)i\gamma^5u(p)=2m_N\bar{u}(p^\prime)\gamma^\nu \gamma^5 iq^\alpha u(p)\label{p52}
\\
&&\bar{u}(p^\prime)\left(P^\nu  P^\alpha-iq_\mu P^\alpha\sigma^{\mu\nu}\right)i\gamma^5u(p)=0\label{p53}
\\
&&\bar{u}(p^\prime)\left(q^\nu  P^\alpha-iP_\mu P^\alpha\sigma^{\mu\nu}\right)\gamma^5u(p)=2m_N\bar{u}(p^\prime)\gamma^\nu\gamma^5 P^\alpha u(p)\label{p54}
\ea

\section{Relativistic Fermi gas model details  \label{app:RFG}}
To treat the integral in Eq.~(\ref{cs_nuclear}), we first perform the integral over the nucleon three-momentum $\mathbf{p}$ by carrying out the angular ($\cos\theta$) integration using the one-dimensional delta function and converting the $|{\bf p}|$ integration to the $p^0$ integration,
\ba
&&\int\frac{d^3\mathbf{p}}{\left(2\pi\right)^3 2E_{p^\prime}}n_i(\mathbf{p})\left[1-n_f(\mathbf{p}^\prime)\right]
\frac{V}{4|k\cdot p|}(2\pi)\delta(p^0-p^{\prime\,0}+q^0) T^N_{ij}(p^0,\,\mathbf{p};\,q^0,\,\mathbf{q})\nn
\\
&\equiv&\int\,d^3\mathbf{p} f(\mathbf{p},\,q^0,\,\mathbf{q})T^N_{ij}(p^0,\,\mathbf{p};\,q^0,\,\mathbf{q})
\label{d3pintdef}
\\
& =&\int|\mathbf{p}|^2 d|\mathbf{p}| d \cos \theta d \phi \frac{V}{16 \pi^2} n_i(\mathbf{p})\left[1-n_f\left(\mathbf{p}^{\prime}\right)\right] 
T_{i j}^N\left(p^0, \mathbf{p} ; q^0, \mathbf{q}\right)
\\
&& \qquad\qquad\qquad\qquad\qquad\qquad\times \frac{\delta\left(p^0-\sqrt{m_N^2+\mathbf{p}^2+2|\mathbf{p}||\mathbf{q}| \cos \theta+\mathbf{q}^2}+q^0\right)}{2 E_{p^{\prime}} p^0 E_k}  \nn \\
&=&\frac{V}{16\pi}\int dp^0\frac{p^0+\epsilon_b}{p^0|\mathbf{q}|E_k}n_i(\mathbf{p})\left[1-n_f(\mathbf{p}^\prime)\right]T^N_{ij}(p^0,\,\mathbf{p};\,q^0,\,\mathbf{q})
\label{pintegral}
\ea
where $T^N_{ij}$ is the hadronic tensor given in Eq.~(\ref{hadronTensor}), $n_i(\mathbf{p})=\theta\left(p_F-\mathbf{p}\right)$, $n_f(\mathbf{p}^\prime)=\theta\left(p_F-\mathbf{p}^\prime\right)$ and we have defined
\ba
f(\mathbf{p},\,q^0,\,\mathbf{q})&=&\frac{V}{4|k\cdot p|8\pi^2E_{p^\prime}}n_i(\mathbf{p})\left[1-n_f(\mathbf{p}^\prime)\right]\delta(p^0-p^{\prime\,0}+q^0)
\nn\\
&=&\frac{m_TV}{32\pi^2|k\cdot p_T|}n_i(\mathbf{p})\left[1-n_f(\mathbf{p}^\prime)\right]\frac{\delta(p^0-p^{\prime\,0}+q^0)}{p^0E_{p^\prime}}\label{ffunc}
\ea
We approximate $k\cdot p$ by $E_k E_p$ by treating the initial nucleon almost at rest. We also introduce the target 4-momentum 
\be
 p_T^\mu=m_T\delta^\mu_0
\ee
at rest with mass $m_T=A(m_N-\epsilon_b)$ for $A$ nucleons. The integration over the dark matter three-momentum $\mathbf{k}^\prime$ will be performed later.

In performing the integration in Eq.~(\ref{pintegral}), we decompose the nucleon tensor and express it as combinations of tensors constructed solely from the nuclear momentum $p_T^\mu$ and the momentum transfer $q^\mu$. This allows us to convert nucleon-level physics to the nuclear level. Specifically, we choose the spatial direction of the momentum transfer to lie along the $3$-direction, writing it explicitly as 
\be
\label{qmuq0}
q_\mu = q_0 \delta_\mu^0 + |\mathbf{q}|  \delta_\mu^3 ~~~\Longrightarrow ~~~\delta_\mu^3=\frac{q_\mu}{|\mathbf{q}|}-\frac{q_0}{|\mathbf{q}|}\delta_\mu^0 .
\ee
This conversion works differently for different portals:
\begin{itemize}
\item For a spin-0 portal, the components $T^N_{ij}$ are scalars, so the conversion is straightforward. They depend on the Mandelstam variables $s=(k+p)^2=M^2+2(k\cdot p_T) p^0/m_T+p_0^2-(p_0+\epsilon_b)^2+m_N^2$ and $t=q^2$, as well as on the dark matter mass $M$ and the target mass $m_T$.

\item For a spin-1 portal, the components $T^N_{ij}$ are rank-2 tensors. In this case, we encounter terms involving products of up to two nucleon momenta. Specifically, we need to convert expressions such as $p^\mu p^\nu$ and $p^\mu q^\nu$ into rank-2 tensors constructed from $p_T^\mu$ and $q^\mu$.

\item For a spin-2 portal, the components $T^N_{ij}$ are rank-4 tensors. Here, terms may involve products of up to four nucleon momenta, such as $p^\mu q^\nu q^\alpha q^\beta$, $p^\mu p^\nu q^\alpha q^\beta$, $p^\mu p^\nu p^\alpha q^\beta$, and $p^\mu p^\nu p^\alpha p^\beta$. These must be converted into rank-4 tensors built from $p_T^\mu$ and $q^\mu$.
\end{itemize}

In performing the conversion of these tensors, a useful simplification is that $T^N_{ij}$ appears in the integration of $d^3\bfp$ and $n_i(\bfp)=n_i(\bfp)_{p_1\to -p_1~{\rm or }~p_2\to -p_2}$, which allows us to use the symmetry under $p_1\to -p_1~{\rm or }~p_2\to -p_2$. All odd functions of $p_1$ or $p_2$ vanish after integration $d^3\bfp$. Explicitly, we can make use of the following integrals (further details are provided in Appendix~\ref{app:trans}):
\vskip 5pt

\noindent (1) One-nucleon-momentum integral\,\footnote{This is in the context of Eq.~(\ref{d3pintdef}), with other factors neglected; a similar understanding is assumed for the following integrals.}: 
\ba
 &&\int d^3\bfp\, p_\mu =
\int d^3\bfp\left(\delta_\mu^0 p_0+\delta_\mu^i p_i\right)
=\int d^3\bfp\,\left[\left(p_0-p_3\frac{q_0}{|\mathbf{q}|}\right)\delta_\mu^0+p_3\frac{q_\mu}{|\mathbf{q}|}\right]\,,
\label{1pterm}
\ea
where we have used Eq.~(\ref{qmuq0}) and the fact that $\int d^3\bfp \,p_1=\int d^3\bfp\, p_2=0$
\ba
\int d^3\bfp \,\delta_\mu^i p_i
=\int d^3\bfp \,\delta_\mu^3 p_3=\int d^3\bfp\,\left[p_3\frac{q_\mu}{|\mathbf{q}|}-p_3\frac{q_0}{|\mathbf{q}|}\delta_\mu^0\right]\,.
\label{1pint}
\ea
In the final results, we will replace $\delta_\mu^0$ by $p^T_\mu/m_T$; however, for notational simplicity we keep $\delta_\mu^0$ in these intermediate expressions and the same for integrals below.
All the nucleon momentum $p$ integrals will be sorted into the format as in Eq.~(\ref{Afunc}) and we deal with them all together until the end of this section.
\vskip 5pt

\noindent (2) Two-nucleon-momentum integral
\ba
 \int d^3\bfp\, p_\mu p_\nu=\int d^3\bfp\left[\delta_\mu^0\delta_\nu^0 p_0^2+\delta_\mu^i\delta_\nu^j p_i p_j+\delta_\mu^i\delta_\nu^0 p_i p_0+\delta_\mu^0\delta_\nu^i p_0 p_i\right]\,,
\label{pmupnu}
\ea
where 
\ba
\int d^3\bfp\,\delta_\mu^i\delta_\nu^j p_i p_j
=\int d^3\bfp\,\left[\frac{1}{2}\Delta_{\mu\nu} \left(|\mathbf{p}|^2-p_3^2\right)+\delta_\mu^3\delta_\nu^3 p_3^2\right]\,,
\label{2pipj}
\ea
is the  two-nucleon-momentum spatial integral Eq.~(\ref{pipj}) and we have defined
\ba
\Delta_{\mu\nu}\equiv\delta_\mu^1\delta_\nu^1+\delta_\mu^2\delta_\nu^2=-g_{\mu\nu}+\delta_\mu^0\delta_\nu^0-\delta_\mu^3\delta_\nu^3\,.
\label{2pint}
\ea
Again, making use of Eq.~(\ref{qmuq0}) and Eq.~(\ref{1pterm}), we can obtain
\begin{align}
\label{2pterm}
&~~~~\int d^3\bfp\, p_\mu p_\nu
\\
&=\int d^3\bfp\,\bigg\{-g_{\mu\nu}\frac{1}{2}\left( |\mathbf{p}|^2-p_3^2\right)+
\delta_\mu^0\delta_\nu^0\bigg[p_0^2+\frac{1}{2}\left(1-\frac{q_0^2}{|\mathbf{q}|^2}\right)
\left( |\mathbf{p}|^2-p_3^2\right)+\frac{q_0^2}{|\mathbf{q}|^2}p_3^2
\nn\\
&-2\frac{q_0}{|\mathbf{q}|}p_0p_3\bigg]+\left(\frac{q_\mu \delta^0_\nu}{|\mathbf{q}|}+\frac{\delta^0_\mu q_\nu }{|\mathbf{q}|}\right)\left[\frac{1}{2}\frac{q_0}{|\mathbf{q}|}\left( |\mathbf{p}|^2-3p_3^2\right)+p_0p_3\right]
-\frac{q_\mu q_\nu}{2|\mathbf{q}|^2}\left( |\mathbf{p}|^2-3p_3^2\right)\bigg\}\,.\nn
\end{align}

\noindent (3) Three-nucleon-momentum integral
\begin{align}
&\int d^3\bfp\, p_\mu p_\nu p_\alpha
=\int d^3\bfp\,p_0^3\delta_\mu^0\delta_\nu^0\delta_\alpha^0+\int d^3\bfp\,p_0^2p_3\left(\delta_\mu^0\delta_\nu^0 \delta^3_\alpha
+\delta_\mu^0\delta_\alpha^0\delta^3_\nu+\delta_\alpha^0\delta_\nu^0\delta^3_\mu\right)
\nn\\
&+\int d^3\bfp\,p_0\left(\delta_\mu^0p_j p_k\delta^j_\nu \delta^k_\alpha+\delta_\nu^0 p_i p_k \delta^i_\mu \delta^k_\alpha+ \delta_\alpha^0 p_i p_j\delta^i_\mu \delta^j_\nu\right)
+\int d^3\bfp\,p_i p_j p_k \delta^i_\mu \delta^j_\nu \delta^k_\alpha\,,
\label{3greek}
\end{align}
where we need the three-nucleon-momentum spatial integral Eq.~(\ref{pipjpk})
\be
\!\!\int\!\! d^3\bfp\,\delta_\mu^i\delta_\nu^j \delta^k_\alpha p_i p_j p_k
=\!\int\!\! d^3\bfp\!\left\{\!\left[\Delta_{\mu\nu}\delta^3_\alpha
+\Delta_{\mu\alpha}\delta^3_\nu
+\Delta_{\nu\alpha}\delta^3_\mu\right]\!
\left(|\mathbf{p}|^2\!-p_3^2\right)\frac{p_3}{2}
+\delta_\mu^3\delta_\nu^3 \delta^3_\alpha p_3^3\right\},
\label{3pint}
\ee
as well as Eq.~(\ref{1pint}) and Eq.~(\ref{2pipj}). Here, $\delta_\mu^3$ should be understood as $=\frac{q_\mu}{|\mathbf{q}|}-\frac{q_0}{|\mathbf{q}|}\delta_\mu^0$.

\noindent (4) Four-nucleon-momentum integral
\begin{align}
&~~~~\int d^3\bfp\, p_\mu p_\nu p_\alpha p_\beta \label{4greek}
\\
&=\int d^3\bfp\, p_0^4 \delta_\mu^0\delta_\nu^0\delta_\alpha^0\delta_\beta^0+\int d^3\bfp\,p_0^3p_3\left(\delta_\mu^0\delta_\nu^0\delta_\alpha^0\delta_\beta^3
+\delta_\mu^0\delta_\nu^0\delta_\beta^0\delta_\alpha^3+\delta_\mu^0\delta_\alpha^0\delta_\beta^0\delta_\nu^3+\delta_\alpha^0\delta_\nu^0\delta_\beta^0\delta_\mu^3\right)
\nn\\
&+\int d^3\bfp\, p_0^2\Big(\delta_\mu^0\delta_\nu^0 p_k p_m\delta^k_\alpha\delta^m_\beta+\delta_\mu^0\delta_\alpha^0 p_j p_m\delta^j_\nu\delta^m_\beta
+\delta_\mu^0\delta_\beta^0 p_j p_k\delta^j_\nu\delta^k_\alpha+\delta_\nu^0\delta_\alpha^0p_i p_m\delta^i_\mu \delta^m_\beta
\nn\\
&+\delta_\nu^0\delta_\beta^0p_i p_k\delta^i_\mu \delta^k_\alpha+\delta_\alpha^0\delta_\beta^0p_i p_j\delta^i_\mu \delta^j_\nu\Big)+\int d^3\bfp\, p_0\Big(\delta_\mu^0p_jp_k p_m\delta^j_\nu\delta^k_\alpha\delta^m_\beta
+\delta_\nu^0p_ip_k p_m\delta^i_\mu\delta^k_\alpha\delta^m_\beta
\nn\\
&+\delta_\alpha^0p_ip_j p_m\delta^i_\mu\delta^j_\nu\delta^m_\beta+\delta_\beta^0p_ip_j p_k\delta^i_\mu\delta^j_\nu\delta^k_\alpha\Big)+\int d^3\bfp\, p_ip_j p_k p_m\delta^i_\mu\delta^j_\nu\delta^k_\alpha\delta^m_\beta
\,,\nn
\end{align}
where the four-nucleon-momentum spatial integral Eq.~(\ref{pipjpkpl})
\ba
&&\int d^3\bfp\,\delta_\mu^i\delta_\nu^j \delta^k_\alpha \delta^m_\beta p_i p_j p_k p_m
=\int d^3\bfp\, \bigg[\frac{1}{8}\left(\Delta_{\mu\nu}\Delta_{\alpha\beta}+\Delta_{\mu\alpha}\Delta_{\nu\beta}+\Delta_{\mu\beta}\Delta_{\nu\alpha}\right)
\left(|\mathbf{p}|^2-p_3^2\right)^2
\nn\\
&&+\frac{1}{2}\left[\Delta_{\mu\nu}\delta^3_\alpha \delta^3_\beta
+\Delta_{\mu\alpha}\delta^3_\nu \delta^3_\beta
+\Delta_{\mu\beta}\delta^3_\alpha \delta^3_\nu
+\Delta_{\nu\alpha}\delta^3_\beta\delta^3_\mu
+\Delta_{\nu\beta}\delta^3_\alpha\delta^3_\mu
+\Delta_{\alpha\beta}\delta^3_\nu\delta^3_\mu\right]\left(|\mathbf{p}|^2-p_3^2\right)p_3^2
\nn\\
&&+\delta_\mu^3\delta_\nu^3\delta^3_\alpha \delta^3_\beta p_3^4
\bigg]\label{4pint}
\ea
is needed as well as Eq.~(\ref{1pint}), Eq.~(\ref{2pipj}) and Eq.~(\ref{3pint}).

After converting the hadronic tensor $T^N_{ij}$ in Eq.~(\ref{pintegral}) from the nucleon-level to the nucleus level, i.e., replacing all the tensors made of $p^\mu$ and $q^\mu$ by 
$p_T^\mu$ and $q^\mu$ with Eqs.~(\ref{1pterm}, \ref{2pterm}, \ref{3greek}, \ref{4greek}),
the final integral terms in Eq.~(\ref{pintegral}) can be cast into a generic form in terms of 
the nucleon momentum integral basis below
\ba
A_{nmk}=\int d^3\bfp\, \frac{p_0^n p_z^{m}|\mathbf{p}|^{2k}}{m_N^r}f(\mathbf{p},\,q^0,\,\mathbf{q})\,,
\label{Afunc}
\ea
where $p_z\equiv p_3$, the non-negative integers $n,\,m,\,k$ satisfy $n+m+2k=r$ and  $r=0,\,1,\,2,\,3,\,4$. They are related to the integrals
\ba
B_j=\frac{V}{16\pi E_k|\mathbf{q}|} \int dp_0 \frac{p_0+\epsilon_b}{p_0}\left(\frac{p_0+\epsilon_b}{m_N}\right)^j
\label{Bfunc}
\ea
with $B_0=A_{000}$. It can be derived that (see Appendix~\ref{app:ABrelation})
\ba
A_{nmk}&=&\int \frac{d^3\bfp}{m_N^r}\, \sum_{i=0}^n\sum_{s=0}^m \sum_{l=0}^k C_n^i\left(-\epsilon_b\right)^{n-i}C_m^s c^{m-s} c^{\prime s}C_k^l
\left(-m_N^2\right)^{\left(k-l\right)}\left(p_0+\epsilon_b\right)^{i+s+2l}f(\mathbf{p},\,q^0,\mathbf{q})
\nn\\
&=&\sum_{i=0}^n\sum_{s=0}^m \sum_{l=0}^kC_n^i C_m^s C_k^l \left(-\epsilon_b\right)^{n-i} c^{m-s} c^{\prime s}(-1)^{k-l}m_N^{i+s+2k-r}B_{i+s+2l}\,,
\label{Afunc}
\ea
where $C_n^i ={n!}/({i!(n-k)!})$ is the binomial coefficient, and we have defined two constants $c=({q^2+\epsilon_b^2-2\epsilon_b q^0})/({2|\mathbf{q}|})$ and $c^\prime=({q^0-\epsilon_b})/{|\mathbf{q}|}$.
 All relevant $B$ functions with explicit expressions are given in Appendix~\ref{app:Bfunc}.

\section{Spatial integrals in the relativistic Fermi gas model \label{app:trans}}
Details of the spatial integrals used to convert tensors from the nucleon-level to the nucleus level are given below.

\noindent Two-nucleon-momentum spatial integral:
\ba
&&\int d^3\bfp\,\delta_\mu^i\delta_\nu^j p_i p_j
\nn\\
&&=\int d^3\bfp\,\left(\delta_\mu^1\delta_\nu^1 p_1^2+\delta_\mu^2\delta_\nu^2 p_2^2+\delta_\mu^3\delta_\nu^3 p_3^2\right)
\nn\\
&&=\int d^3\bfp\,\left[\left(\delta_\mu^1\delta_\nu^1 \sin^2{\phi}+\delta_\mu^2\delta_\nu^2\cos^2{\phi}\right)\left(p_1^2+p_2^2\right)+\delta_\mu^3\delta_\nu^3 p_3^2\right]
\nn\\
&&=\int |\mathbf{p}|^2d|\mathbf{p}|d\cos{\theta}\textcolor{black}{d\phi} \left(\delta_\mu^1\delta_\nu^1 \textcolor{black}{\sin^2{\phi}}+\delta_\mu^2\delta_\nu^2\textcolor{black}{\cos^2{\phi}}\right)\left(p_1^2+p_2^2\right)+\int d^3\bfp\,\delta_\mu^3\delta_\nu^3 p_3^2
\nn\\
&&=\int d^3\bfp\,\left[\frac{1}{2}\left(\delta_\mu^1\delta_\nu^1+\delta_\mu^2\delta_\nu^2\right)\left(p_1^2+p_2^2\right)+\delta_\mu^3\delta_\nu^3 p_3^2\right]
\nn\\
&&=\int d^3\bfp\,\left[\frac{1}{2}\left(-g_{\mu\nu}+\delta_\mu^0\delta_\nu^0-\delta_\mu^3\delta_\nu^3\right)\left(p_1^2+p_2^2\right)+\delta_\mu^3\delta_\nu^3 p_3^2\right]
\nn\\
&&=\int d^3\bfp\,\left[\frac{1}{2}\Delta_{\mu\nu} \left(|\mathbf{p}|^2-p_3^2\right)+\delta_\mu^3\delta_\nu^3 p_3^2\right]\,.
\label{pipj}
\ea
\noindent Three-nucleon-momentum spatial integral:
\ba
&&\int d^3\bfp\,\delta_\mu^i\delta_\nu^j \delta^k_\alpha p_i p_j p_k
\nn\\
&&=\int d^3\bfp\,\left[\left(\delta_\mu^1\delta_\nu^1p_1^2+\delta_\mu^2\delta_\nu^2p_2^2\right)\delta^3_\alpha p_3+(\nu \leftrightarrow \alpha)+(\mu \leftrightarrow \alpha)\right]+\int d^3\bfp\,\delta_\mu^3\delta_\nu^3 \delta^3_\alpha p_3^3
\nn\\
&&=\int |\mathbf{p}|^2d|\mathbf{p}|d\cos{\theta}d\phi \left[\left(\delta_\mu^1\delta_\nu^1\sin^2{\phi}+\delta_\mu^2\delta_\nu^2\cos^2{\phi}\right)|\mathbf{p}|^2\sin^2{\theta}|\mathbf{p}|\cos{\theta}
+(\nu \leftrightarrow \alpha)+(\mu \leftrightarrow \alpha)\right]
\nn\\
&&+\int d^3\bfp\,\delta_\mu^3\delta_\nu^3 \delta^3_\alpha p_3^3
\nn\\
&&=\int d^3\bfp\,\frac{1}{2}\left[\left(\delta_\mu^1\delta_\nu^1+\delta_\mu^2\delta_\nu^2\right)\delta^3_\alpha+\left(\delta_\mu^1\delta_\alpha^1+\delta_\mu^2\delta_\alpha^2\right)\delta^3_\nu+\left(\delta_\alpha^1\delta_\nu^1+\delta_\alpha^2\delta_\nu^2\right)\delta^3_\mu\right]
\left(|\mathbf{p}|^2-p_3^2\right)p_3
\nn\\
&&+\int d^3\bfp\,\delta_\mu^3\delta_\nu^3 \delta^3_\alpha p_3^3
\nn\\
&&=\int d^3\bfp\,\frac{1}{2}\Big[\left(-g_{\mu\nu}+\delta_\mu^0\delta_\nu^0-\delta_\mu^3\delta_\nu^3\right)\delta^3_\alpha
+\left(-g_{\mu\alpha}+\delta_\mu^0\delta_\alpha^0-\delta_\mu^3\delta_\alpha^3\right)\delta^3_\nu
\nn\\
&&+\left(-g_{\nu\alpha}+\delta_\nu^0\delta_\alpha^0-\delta_\nu^3\delta_\alpha^3\right)\delta^3_\mu\Big]\left(|\mathbf{p}|^2-p_3^2\right)p_3+\int d^3\bfp\,\delta_\mu^3\delta_\nu^3 \delta^3_\alpha p_3^3
\nn\\
&&=\int d^3\bfp\,\frac{1}{2}\left[\Delta_{\mu\nu}\delta^3_\alpha
+\Delta_{\mu\alpha}\delta^3_\nu
+\Delta_{\nu\alpha}\delta^3_\mu\right]
\left(|\mathbf{p}|^2-p_3^2\right)p_3
+\delta_\mu^3\delta_\nu^3 \delta^3_\alpha p_3^3
\label{pipjpk}
\ea
\noindent Four-nucleon-momentum spatial integral
\ba
&&\int d^3\bfp\,\delta_\mu^i\delta_\nu^j \delta^k_\alpha \delta^m_\beta p_i p_j p_k p_m
\nn\\
&&=\int d^3\bfp\, \Bigg[\frac{1}{8}\bigg[\left(\delta_\mu^1\delta_\nu^1+\delta_\mu^2\delta_\nu^2\right)\left(\delta_\alpha^1\delta_\beta^1+\delta^2_\alpha \delta^2_\beta\right)
+\left(\delta_\mu^1\delta_\alpha^1+\delta_\mu^2\delta_\alpha^2\right)\left(\delta_\nu^1\delta_\beta^1+\delta^2_\nu \delta^2_\beta\right)
\nn\\
&&+\left(\delta_\mu^1\delta_\beta^1+\delta_\mu^2\delta_\beta^2\right)\left(\delta_\nu^1\delta_\alpha^1+\delta^2_\nu \delta^2_\alpha\right)\bigg]
\left(|\mathbf{p}|^2-p_3^2\right)^2
\nn\\
&&+\frac{1}{2}\bigg[\left(\delta_\mu^1\delta_\nu^1+\delta_\mu^2\delta_\nu^2\right)\delta^3_\alpha \delta^3_\beta
+\left(\delta_\mu^1\delta_\alpha ^1+\delta_\mu^2\delta_\alpha ^2\right)\delta^3_\nu \delta^3_\beta
+\left(\delta_\mu^1\delta_\beta ^1+\delta_\mu^2\delta_\beta ^2\right)\delta^3_\alpha \delta^3_\nu
+\left(\delta_\nu^1\delta_\alpha ^1+\delta_\nu^2\delta_\alpha ^2\right)\delta^3_\beta\delta^3_\mu
\nn\\
&&+\left(\delta_\nu^1\delta_\beta ^1+\delta_\nu^2\delta_\beta ^2\right)\delta^3_\alpha\delta^3_\mu
+\left(\delta_\alpha^1\delta_\beta ^1+\delta_\alpha^2\delta_\beta ^2\right)\delta^3_\nu\delta^3_\mu\bigg]\left(|\mathbf{p}|^2-p_3^2\right)p_3^2
+\delta_\mu^3\delta_\nu^3\delta^3_\alpha \delta^3_\beta p_3^4
\Bigg]
\nn\\
&&=\int d^3\bfp\, \bigg[\frac{1}{8}\left(\Delta_{\mu\nu}\Delta_{\alpha\beta}+\Delta_{\mu\alpha}\Delta_{\nu\beta}+\Delta_{\mu\beta}\Delta_{\nu\alpha}\right)
\left(|\mathbf{p}|^2-p_3^2\right)^2
\nn\\
&&+\frac{1}{2}\left[\Delta_{\mu\nu}\delta^3_\alpha \delta^3_\beta
+\Delta_{\mu\alpha}\delta^3_\nu \delta^3_\beta
+\Delta_{\mu\beta}\delta^3_\alpha \delta^3_\nu
+\Delta_{\nu\alpha}\delta^3_\beta\delta^3_\mu
+\Delta_{\nu\beta}\delta^3_\alpha\delta^3_\mu
+\Delta_{\alpha\beta}\delta^3_\nu\delta^3_\mu\right]\left(|\mathbf{p}|^2-p_3^2\right)p_3^2
\nn\\
&&+\delta_\mu^3\delta_\nu^3\delta^3_\alpha \delta^3_\beta p_3^4
\bigg]\label{pipjpkpl}
\ea

\section{ Relations between $A$ and $B$ functions \label{app:ABrelation}}

The $A$ functions (Eq.~(\ref{Afunc})) can be converted to the simpler $B$ functions (Eq.~(\ref{Bfunc})) by the following integrals
\ba
&&\int d^3\bfp\,p_0^n f(\mathbf{p},\,q^0,\,\mathbf{q})
=\int d^3\bfp\,\left(p_0+\epsilon_b-\epsilon_b\right)^n f(\mathbf{p},\,q^0,\,\mathbf{q})
\nn\\
&&=\int d^3\bfp\,\sum_i C_n^i\left(p_0+\epsilon_b\right)^i\left(-\epsilon_b\right)^{n-i} f(\mathbf{p},\,q^0,\,\mathbf{q})
\nn\\
&&=\sum_{i=0}^n C_n^i\left(-\epsilon_b\right)^{n-i}m_N^i B_i
\ea
\ba
&&\int d^3\bfp\,p_z^m f(\mathbf{p},\,q^0,\,\mathbf{q})
=\int d^3\bfp\,|\mathbf{p}|^m\cos^m{\theta_0} f(\mathbf{p},\,q^0,\,\mathbf{q})
\nn\\
&&=\int d^3\bfp\,\left[c+c^\prime\left(p_0+\epsilon_b\right)\right]^m f(\mathbf{p},\,q^0,\,\mathbf{q})
=\int d^3\bfp\,\sum_{s=0}^m C_m^s c^{m-s} c^{\prime s}\left(p_0+\epsilon_b\right)^s f(\mathbf{p},\,q^0,\,\mathbf{q})
\nn\\
&&=\sum_{s=0}^m C_m^s c^{m-s} c^{\prime s}m_N^s B_s
\ea
\ba
&&\int d^3\bfp\, |\mathbf{p}|^{2k}f(\mathbf{p},\,q^0,\,\mathbf{q})
=\int d^3\bfp\, \left[\left(p_0+\epsilon_b\right)^2-m_N^2\right]^k f(\mathbf{p},\,q^0,\,\mathbf{q})
\nn\\
&&=\int d^3\bfp\, \sum_{l=0}^k C_k^l \left(p_0+\epsilon_b\right)^{2l}\left(-m_N^2\right)^{\left(k-l\right)} f(\mathbf{p},\,q^0,\,\mathbf{q})
\nn\\
&&= \sum_{l=0}^k C_k^l\left(-1\right)^{\left(k-l\right)}\left(m_N\right)^{2k}B_{2l}
\ea
where the angle $\theta_0$ is given by the Dirac delta function in Eq.~(\ref{ffunc}). More explicitly, we have
\ba
\cos{\theta_0}=\frac{2p^0(q^0-\epsilon_b)+q^{2}_0-|\mathbf{q}|^2-\epsilon_b^2}{2\sqrt{(p^0+\epsilon_b)^2-m_N^2}|\mathbf{q}|}\,.
\ea

\section{Explicit forms of $B$ functions in the relativistic Fermi gas model \label{app:Bfunc}}
Here we list the explicit expressions of the $B$ functions that are useful in the relativistic Fermi gas model calculations, up to spin-2 interactions relevant to our discussion.

\ba
B_0&=&\frac{V}{16\pi E_k|\mathbf{q}|} \bigg[\epsilon_b\log{\left(E_H/E_L\right)}+E_H-E_L\bigg]
\nn\\
B_1&=&\frac{V}{16\pi m_N E_k|\mathbf{q}|} \left[\epsilon^2_b\log{\left(E_H/E_L\right)}+2\left(E_H-E_L\right)\epsilon_b+\frac{1}{2}\left(E^2_H-E^2_L\right)\right]
\nn\\
B_2&=&\frac{V}{16\pi m_N^2 E_k|\mathbf{q}|} \left[\epsilon^3_b\log{\left(E_H/E_L\right)}+3\left(E_H-E_L\right)\epsilon^2_b+\frac{3}{2}\epsilon_b\left(E^2_H-E^2_L\right)+\frac{1}{3}\left(E^3_H-E^3_L\right)\right]
\nn\\
B_3&=&\frac{V}{16\pi m_N^3 E_k|\mathbf{q}|} \bigg[\epsilon^4_b\log{\left(E_H/E_L\right)}+4\epsilon^3_b\left(E_H-E_L\right)+3\epsilon^2_b\left(E^2_H-E^2_L\right)
+\frac{4}{3}\epsilon_b\left(E^3_H-E^3_L\right)
\nn\\
&&\hspace{40pt}+\frac{1}{4}\left(E^4_H-E^4_L\right)\bigg]
\nn\\
B_4&=&\frac{V}{16\pi m_N^4 E_k|\mathbf{q}|} \bigg[\epsilon^5_b\log{\left(E_H/E_L\right)}+5\epsilon^4_b\left(E_H-E_L\right)+5\epsilon^3_b\left(E^2_H-E^2_L\right)
+\frac{10}{3}\epsilon^2_b\left(E^3_H-E^3_L\right)
\nn\\
&&\hspace{40pt}+\frac{5}{4}\epsilon_b\left(E^4_H-E^4_L\right)+\frac{1}{5}\left(E^5_H-E^5_L\right)\bigg]
\nn\\
B_5&=&\frac{V}{16\pi m_N^5 E_k|\mathbf{q}|} \bigg[\epsilon^6_b\log{\left(E_H/E_L\right)}+6\epsilon^5_b\left(E_H-E_L\right)+\frac{15}{2}\epsilon^4_b\left(E^2_H-E^2_L\right)
+\frac{20}{3}\epsilon^3_b\left(E^3_H-E^3_L\right)
\nn\\
&&\hspace{40pt}+\frac{15}{4}\epsilon_b^2\left(E^4_H-E^4_L\right)+\frac{6}{5}\epsilon_b\left(E^5_H-E^5_L\right)+\frac{1}{6}\left(E^6_H-E^6_L\right)\bigg]
\nn\\
B_6&=&\frac{V}{16\pi m_N^6 E_k|\mathbf{q}|} \bigg[\epsilon^7_b\log{\left(E_H/E_L\right)}+7\epsilon^6_b\left(E_H-E_L\right)+\frac{21}{2}\epsilon^5_b\left(E^2_H-E^2_L\right)
+\frac{35}{3}\epsilon^4_b\left(E^3_H-E^3_L\right)
\nn\\
&&\hspace{40pt}+\frac{35}{4}\epsilon_b^3\left(E^4_H-E^4_L\right)+\frac{21}{5}\epsilon_b^2\left(E^5_H-E^5_L\right)+\frac{7}{6}\epsilon_b\left(E^6_H-E^6_L\right)
+\frac{1}{7}\left(E^7_H-E^7_L\right)\bigg]
\ea
where $E_H=E_F-\epsilon_b=\sqrt{m_N^2+p_F^2}-\epsilon_b$, $E_L=\mathrm{Max}\,\left(E_F-q^0,\,m_N-\epsilon_b,\,E_r-\epsilon_b\right)$, 
with $E_r=m_N\left[c c^\prime/m_N+\sqrt{1-c^{\prime 2}+\left(c/m_N\right)^2}\right]/\left(1-c^{\prime 2}\right)$.

\section{Relevant Feynman rules of QCD in curved space time \label{app: fmrules}}
The action for a quark in curved spacetime is~\cite{Nieh:1981xk}
\ba
S=\int d^4x\, \mathrm{det}\left(e^a_\mu\right)\left[\frac{1}{2}\left(\bar{\psi} i \gamma^a e^\mu_a \overrightarrow{D}_\mu \psi-\bar{\psi}\overleftarrow{D}_\mu i \gamma^a e^\mu_a \psi\right)
-m\bar{\psi}{\psi}\right]\,,
\ea
where $\mathrm{det}(e^a_\mu)=\sqrt{|\mathrm{det}\, g|}$,
$\overrightarrow{D}_\mu=\partial_\mu-ig_s A_\mu-\frac{i}{4}\omega_\mu^{ab}\sigma_{ab}$,
$\overleftarrow{D}_\mu=\overleftarrow{\partial}_\mu+ig_s A_\mu+\frac{i}{4}\omega_\mu^{ab}\sigma_{ab}$,
$\sigma_{ab}=\frac{i}{2}[\gamma_a, \,\gamma_b]$ and $\omega_\mu^{ab}=e_\nu^a\nabla_{\mu} e^{\nu b}=e_\nu^a(\partial_\mu e^{\nu b}+\Gamma^\nu_{\rho\mu}e^{\rho b})$.
More explicitly, the spin connection can be written as
\be
\omega_{\mu ab}=\frac{1}{2}\left(\partial_\mu e_{b\nu}-\partial_\nu e_{b\mu}\right)e_a^\nu
-\frac{1}{2}\left(\partial_\mu e_{a\nu}-\partial_\nu e_{a\mu}\right)e_b^\nu
-\frac{1}{2}e^\rho_a e^\sigma_b\left(\partial_\rho e_{c\sigma}-\partial_\sigma e_{c\rho}\right)e^c_\mu\,.
\ee
Expanding the action around Minkowski space, at linear order of $h$, along with the Non-abelian gauge Lagrangian, we find the Feynman rules~\cite{Han:1998sg}

\vspace{1cm}
\parbox{35mm}{
\begin{fmfgraph*}(70,100)
\centering
\fmfleft{i}
\fmfright{o}
\fmftop{v}
\fmfv{label=$\mu\nu$}{v}
  \fmf{fermion, label=$p_1$}{i,b}
    \fmf{fermion, label=$p_2$}{b,o}
  \fmf{dbl_zigzag}{b,v}
\end{fmfgraph*}
} 
\parbox{35mm}{
$$=i \kappa\Bigg[-\frac{1}{8}\left[\left(p_1^\mu+p_2^\mu\right)\gamma^\nu+\left(p_1^\nu+p_2^\nu\right)\gamma^\mu\right]
+\frac{1}{4}\eta^{\mu\nu}\left(\slashed{p}_1+\slashed{p}_2\right)-\frac{1}{2}m\eta^{\mu\nu}
\Bigg]\,,$$
}

\vspace{1cm}
\parbox{35mm}{
\begin{fmfgraph*}(70,100)
\centering
\fmfleft{i}
\fmfright{o}
\fmftop{v}
\fmfv{label=$\mu\nu$}{v}
  \fmf{gluon, label=$A_\alpha^a(p_1)$,label.dist=5pt,label.side=left}{i,b}
    \fmf{gluon, label=$A_\beta^b(p_2)$,label.dist=5pt,label.side=left}{b,o}
  \fmf{dbl_zigzag}{b,v}
\end{fmfgraph*}
} 
\parbox{35mm}{
\ba
&&=-i\frac{\kappa}{2} \delta^{ab}\Bigg[p_1\cdot p_2\big(\eta_{\mu\alpha}\eta_{\nu\beta}+\eta_{\mu\beta}\eta_{\nu\alpha}-\eta_{\mu\nu}\eta_{\alpha\beta}\big)
+\eta_{\mu\nu}p_{1\beta}p_{2\alpha}
\nn\\
&&-\big(\eta_{\mu\beta}p_{1\nu}p_{2\alpha}+\eta_{\mu\alpha}p_{1\beta}p_{2\nu}-\eta_{\alpha\beta}p_{1\mu}p_{2\nu}
+\eta_{\nu\beta}p_{1\mu}p_{2\alpha}+\eta_{\nu\alpha}p_{1\beta}p_{2\mu}
\nn\\
&&-\eta_{\alpha\beta}p_{1\nu}p_{2\mu}\big)
\Bigg]\,,\nn
\ea
}

where we have used the linearized vierbein
\ba
e_\mu^a=\delta_\mu^a+\frac{\kappa}{2}h_\mu^a\,, 
\quad e^\mu_a=\delta^\mu_a-\frac{\kappa}{2}h^\mu_a\,.
\ea

\end{fmffile}

\bibliographystyle{JHEP}
\bibliography{light_portal}

@article{deRham:2010kj,
    author = "de Rham, Claudia and Gabadadze, Gregory and Tolley, Andrew J.",
    title = "{Resummation of Massive Gravity}",
    eprint = "1011.1232",
    archivePrefix = "arXiv",
    primaryClass = "hep-th",
    doi = "10.1103/PhysRevLett.106.231101",
    journal = "Phys. Rev. Lett.",
    volume = "106",
    pages = "231101",
    year = "2011"
}

@article{Oman:2015xda,
    author = "Oman, Kyle A. and others",
    title = "{The unexpected diversity of dwarf galaxy rotation curves}",
    eprint = "1504.01437",
    archivePrefix = "arXiv",
    primaryClass = "astro-ph.GA",
    doi = "10.1093/mnras/stv1504",
    journal = "Mon. Not. Roy. Astron. Soc.",
    volume = "452",
    number = "4",
    pages = "3650--3665",
    year = "2015"
}

@article{Kaplinghat:2015aga,
    author = "Kaplinghat, Manoj and Tulin, Sean and Yu, Hai-Bo",
    title = "{Dark Matter Halos as Particle Colliders: Unified Solution to Small-Scale Structure Puzzles from Dwarfs to Clusters}",
    eprint = "1508.03339",
    archivePrefix = "arXiv",
    primaryClass = "astro-ph.CO",
    doi = "10.1103/PhysRevLett.116.041302",
    journal = "Phys. Rev. Lett.",
    volume = "116",
    number = "4",
    pages = "041302",
    year = "2016"
}

@article{Bondarenko:2019vrb,
    author = "Bondarenko, Kyrylo and Boyarsky, Alexey and Bringmann, Torsten and Hufnagel, Marco and Schmidt-Hoberg, Kai and Sokolenko, Anastasia",
    title = "{Direct detection and complementary constraints for sub-GeV dark matter}",
    eprint = "1909.08632",
    archivePrefix = "arXiv",
    primaryClass = "hep-ph",
    reportNumber = "DESY-19-140",
    doi = "10.1007/JHEP03(2020)118",
    journal = "JHEP",
    volume = "03",
    pages = "118",
    year = "2020"
}

@article{Dent:2019krz,
    author = "Dent, James B. and Dutta, Bhaskar and Newstead, Jayden L. and Shoemaker, Ian M.",
    title = "{Bounds on Cosmic Ray-Boosted Dark Matter in Simplified Models and its Corresponding Neutrino-Floor}",
    eprint = "1907.03782",
    archivePrefix = "arXiv",
    primaryClass = "hep-ph",
    doi = "10.1103/PhysRevD.101.116007",
    journal = "Phys. Rev. D",
    volume = "101",
    number = "11",
    pages = "116007",
    year = "2020"
}

@article{Ge:2020yuf,
    author = "Ge, Shao-Feng and Liu, Jianglai and Yuan, Qiang and Zhou, Ning",
    title = "{Diurnal Effect of Sub-GeV Dark Matter Boosted by Cosmic Rays}",
    eprint = "2005.09480",
    archivePrefix = "arXiv",
    primaryClass = "hep-ph",
    doi = "10.1103/PhysRevLett.126.091804",
    journal = "Phys. Rev. Lett.",
    volume = "126",
    number = "9",
    pages = "091804",
    year = "2021"
}

@article{Bell:2021xff,
    author = "Bell, Nicole F. and Dent, James B. and Dutta, Bhaskar and Ghosh, Sumit and Kumar, Jason and Newstead, Jayden L. and Shoemaker, Ian M.",
    title = "{Cosmic-ray upscattered inelastic dark matter}",
    eprint = "2108.00583",
    archivePrefix = "arXiv",
    primaryClass = "hep-ph",
    reportNumber = "MI-HET-758",
    doi = "10.1103/PhysRevD.104.076020",
    journal = "Phys. Rev. D",
    volume = "104",
    pages = "076020",
    year = "2021"
}

@article{Alvey:2022pad,
    author = "Alvey, James and Bringmann, Torsten and Kolesova, Helena",
    title = "{No room to hide: implications of cosmic-ray upscattering for GeV-scale dark matter}",
    eprint = "2209.03360",
    archivePrefix = "arXiv",
    primaryClass = "hep-ph",
    reportNumber = "CERN-TH-2022-147",
    doi = "10.1007/JHEP01(2023)123",
    journal = "JHEP",
    volume = "01",
    pages = "123",
    year = "2023"
}

@article{Bell:2023sdq,
    author = "Bell, Nicole F. and Newstead, Jayden L. and Shaukat-Ali, Iman",
    title = "{Cosmic-ray boosted dark matter confronted by constraints on new light mediators}",
    eprint = "2309.11003",
    archivePrefix = "arXiv",
    primaryClass = "hep-ph",
    doi = "10.1103/PhysRevD.109.063034",
    journal = "Phys. Rev. D",
    volume = "109",
    number = "6",
    pages = "063034",
    year = "2024"
}

@article{Cappiello:2024acu,
    author = "Cappiello, Christopher V. and Liu, Qinrui and Mohlabeng, Gopolang and Vincent, Aaron C.",
    title = "{Cosmic ray-boosted dark matter at IceCube}",
    eprint = "2405.00086",
    archivePrefix = "arXiv",
    primaryClass = "hep-ph",
    doi = "10.1103/PhysRevD.110.095031",
    journal = "Phys. Rev. D",
    volume = "110",
    number = "9",
    pages = "095031",
    year = "2024"
}

@article{Diurba:2025lky,
    author = "Diurba, Richard and Kole{\v{s}}ov{\'a}, Helena",
    title = "{Prospects of detecting cosmic ray up-scattered dark matter with DUNE}",
    eprint = "2504.16996",
    archivePrefix = "arXiv",
    primaryClass = "hep-ph",
    doi = "10.1007/JHEP07(2025)202",
    journal = "JHEP",
    volume = "07",
    pages = "202",
    year = "2025"
}

@article{Emken:2018run,
    author = "Emken, Timon and Kouvaris, Chris",
    title = "{How blind are underground and surface detectors to strongly interacting Dark Matter?}",
    eprint = "1802.04764",
    archivePrefix = "arXiv",
    primaryClass = "hep-ph",
    reportNumber = "CP3-ORIGINS-2018-007, CP3-Origins-2018-007 DNRF90",
    doi = "10.1103/PhysRevD.97.115047",
    journal = "Phys. Rev. D",
    volume = "97",
    number = "11",
    pages = "115047",
    year = "2018"
}

@article{Nieves:2011pp,
    author = "Nieves, J. and Ruiz Simo, I. and Vicente Vacas, M. J.",
    title = "{Inclusive Charged--Current Neutrino--Nucleus Reactions}",
    eprint = "1102.2777",
    archivePrefix = "arXiv",
    primaryClass = "hep-ph",
    doi = "10.1103/PhysRevC.83.045501",
    journal = "Phys. Rev. C",
    volume = "83",
    pages = "045501",
    year = "2011"
}

@article{Ankowski:2014yfa,
    author = "Ankowski, Artur M. and Benhar, Omar and Sakuda, Makoto",
    title = "{Improving the accuracy of neutrino energy reconstruction in charged-current quasielastic scattering off nuclear targets}",
    eprint = "1404.5687",
    archivePrefix = "arXiv",
    primaryClass = "nucl-th",
    doi = "10.1103/PhysRevD.91.033005",
    journal = "Phys. Rev. D",
    volume = "91",
    number = "3",
    pages = "033005",
    year = "2015"
}

@article{Emken:2017hnp,
    author = "Emken, Timon and Kouvaris, Chris and Nielsen, Niklas Gr{\o}nlund",
    title = "{The Sun as a sub-GeV Dark Matter Accelerator}",
    eprint = "1709.06573",
    archivePrefix = "arXiv",
    primaryClass = "hep-ph",
    reportNumber = "CP3-ORIGINS-2017-035",
    doi = "10.1103/PhysRevD.97.063007",
    journal = "Phys. Rev. D",
    volume = "97",
    number = "6",
    pages = "063007",
    year = "2018"
}

@article{Elor:2021swj,
    author = "Elor, Gilly and McGehee, Robert and Pierce, Aaron",
    title = "{Maximizing Direct Detection with Highly Interactive Particle Relic Dark Matter}",
    eprint = "2112.03920",
    archivePrefix = "arXiv",
    primaryClass = "hep-ph",
    reportNumber = "LCTP-21-26, MITP-21-044",
    doi = "10.1103/PhysRevLett.130.031803",
    journal = "Phys. Rev. Lett.",
    volume = "130",
    number = "3",
    pages = "031803",
    year = "2023"
}

@article{Dror:2019dib,
    author = "Dror, Jeff A. and Elor, Gilly and Mcgehee, Robert",
    title = "{Absorption of Fermionic Dark Matter by Nuclear Targets}",
    eprint = "1908.10861",
    archivePrefix = "arXiv",
    primaryClass = "hep-ph",
    doi = "10.1007/JHEP02(2020)134",
    journal = "JHEP",
    volume = "02",
    pages = "134",
    year = "2020"
}

@article{Su:2022wpj,
    author = "Su, Liangliang and Wu, Lei and Zhou, Ning and Zhu, Bin",
    title = "{Accelerated-light-dark-matter{\textendash}Earth inelastic scattering in direct detection}",
    eprint = "2212.02286",
    archivePrefix = "arXiv",
    primaryClass = "hep-ph",
    doi = "10.1103/PhysRevD.108.035004",
    journal = "Phys. Rev. D",
    volume = "108",
    number = "3",
    pages = "035004",
    year = "2023"
}

@article{Nieh:1981xk,
    author = "Nieh, H. T. and Yan, M. L.",
    title = "{Quantized Dirac Field in Curved Riemann-cartan Background. 1. Symmetry Properties, Green's Function}",
    reportNumber = "ITP-SB-81-18",
    doi = "10.1016/0003-4916(82)90186-5",
    journal = "Annals Phys.",
    volume = "138",
    pages = "237",
    year = "1982"
}

@article{Han:1998sg,
    author = "Han, Tao and Lykken, Joseph D. and Zhang, Ren-Jie",
    title = "{On Kaluza-Klein states from large extra dimensions}",
    eprint = "hep-ph/9811350",
    archivePrefix = "arXiv",
    reportNumber = "MADPH-98-1092, FERMILAB-PUB-98-364",
    doi = "10.1103/PhysRevD.59.105006",
    journal = "Phys. Rev. D",
    volume = "59",
    pages = "105006",
    year = "1999"
}

@article{Hackett:2023rif,
    author = "Hackett, Daniel C. and Pefkou, Dimitra A. and Shanahan, Phiala E.",
    title = "{Gravitational Form Factors of the Proton from Lattice QCD}",
    eprint = "2310.08484",
    archivePrefix = "arXiv",
    primaryClass = "hep-lat",
    reportNumber = "MIT-CTP/5630, FERMILAB-PUB-23-592-T",
    doi = "10.1103/PhysRevLett.132.251904",
    journal = "Phys. Rev. Lett.",
    volume = "132",
    number = "25",
    pages = "251904",
    year = "2024"
}

@article{Song:2023jqm,
    author = "Song, Huayang and Sun, Hao and Yu, Jiang-Hao",
    title = "{Complete EFT operator bases for dark matter and weakly-interacting light particle}",
    eprint = "2306.05999",
    archivePrefix = "arXiv",
    primaryClass = "hep-ph",
    doi = "10.1007/JHEP05(2024)103",
    journal = "JHEP",
    volume = "05",
    pages = "103",
    year = "2024"
}

@article{Nair:2025sfr,
    author = "Nair, Sreeraj and Mondal, Chandan and Xu, Siqi and Zhao, Xingbo and Vary, James P.",
    collaboration = "BLFQ",
    title = "{Proton gravitational structure and mass decomposition on the light front}",
    eprint = "2506.07554",
    archivePrefix = "arXiv",
    primaryClass = "hep-ph",
    doi = "10.1103/79fs-5ghd",
    journal = "Phys. Rev. D",
    volume = "112",
    number = "11",
    pages = "114001",
    year = "2025"
}

@article{Bhattacharya:2025blb,
    author = "Bhattacharya, Shohini and Fuyuto, Kaori and Mereghetti, Emanuele and Richardson, Thomas R.",
    title = "{Toward the determination of CP-odd pion-nucleon couplings}",
    eprint = "2504.01105",
    archivePrefix = "arXiv",
    primaryClass = "hep-ph",
    reportNumber = "LA-UR-25-23006",
    doi = "10.1103/fhm8-flsg",
    journal = "Phys. Rev. C",
    volume = "112",
    number = "2",
    pages = "025501",
    year = "2025"
}

@article{Jungman:1995df,
    author = "Jungman, Gerard and Kamionkowski, Marc and Griest, Kim",
    title = "{Supersymmetric dark matter}",
    eprint = "hep-ph/9506380",
    archivePrefix = "arXiv",
    reportNumber = "SU-4240-605, UCSD-PTH-95-02, IASSNS-HEP-95-14, CU-TP-677",
    doi = "10.1016/0370-1573(95)00058-5",
    journal = "Phys. Rept.",
    volume = "267",
    pages = "195--373",
    year = "1996"
}

@article{Bertone:2004pz,
    author = "Bertone, Gianfranco and Hooper, Dan and Silk, Joseph",
    title = "{Particle dark matter: Evidence, candidates and constraints}",
    eprint = "hep-ph/0404175",
    archivePrefix = "arXiv",
    reportNumber = "FERMILAB-PUB-04-047-A",
    doi = "10.1016/j.physrep.2004.08.031",
    journal = "Phys. Rept.",
    volume = "405",
    pages = "279--390",
    year = "2005"
}

@article{XENON:2025vwd,
    author = "Aprile, E. and others",
    collaboration = "XENON",
    title = "{WIMP Dark Matter Search using a 3.1 tonne $\times$ year Exposure of the XENONnT Experiment}",
    eprint = "2502.18005",
    archivePrefix = "arXiv",
    primaryClass = "hep-ex",
    month = "2",
    year = "2025"
}

@article{PandaX:2024qfu,
    author = "Bo, Zihao and others",
    collaboration = "PandaX",
    title = "{Dark Matter Search Results from 1.54{\,}{\,}Tonne{\textperiodcentered}Year Exposure of PandaX-4T}",
    eprint = "2408.00664",
    archivePrefix = "arXiv",
    primaryClass = "hep-ex",
    doi = "10.1103/PhysRevLett.134.011805",
    journal = "Phys. Rev. Lett.",
    volume = "134",
    number = "1",
    pages = "011805",
    year = "2025"
}

@article{LZ:2022lsv,
    author = "Aalbers, J. and others",
    collaboration = "LZ",
    title = "{First Dark Matter Search Results from the LUX-ZEPLIN (LZ) Experiment}",
    eprint = "2207.03764",
    archivePrefix = "arXiv",
    primaryClass = "hep-ex",
    doi = "10.1103/PhysRevLett.131.041002",
    journal = "Phys. Rev. Lett.",
    volume = "131",
    number = "4",
    pages = "041002",
    year = "2023"
}

@article{Chen:2019gtm,
    author = "Chen, Qing and Hill, Richard J.",
    title = "{Direct detection rate of heavy Higgsino-like and Wino-like dark matter}",
    eprint = "1912.07795",
    archivePrefix = "arXiv",
    primaryClass = "hep-ph",
    reportNumber = "FERMILAB-PUB-19-564-T",
    doi = "10.1016/j.physletb.2020.135364",
    journal = "Phys. Lett. B",
    volume = "804",
    pages = "135364",
    year = "2020"
}

@article{Chen:2023bwg,
    author = "Chen, Qing and Ding, Gui-Jun and Hill, Richard J.",
    title = "{General heavy WIMP nucleon elastic scattering}",
    eprint = "2309.02715",
    archivePrefix = "arXiv",
    primaryClass = "hep-ph",
    reportNumber = "USTC-ICTS/PCFT-23-26, FERMILAB-PUB-23-423-T",
    doi = "10.1103/PhysRevD.108.116023",
    journal = "Phys. Rev. D",
    volume = "108",
    number = "11",
    pages = "116023",
    year = "2023"
}

@article{Chen:2018uqz,
    author = "Chen, Chien-Yi and Hill, Richard J. and Solon, Mikhail P. and Wijangco, Alexander M.",
    title = "{Power Corrections to the Universal Heavy WIMP-Nucleon Cross Section}",
    eprint = "1801.08551",
    archivePrefix = "arXiv",
    primaryClass = "hep-ph",
    reportNumber = "CAL-TH-2017-063, FERMILAB-PUB-17-400-T",
    doi = "10.1016/j.physletb.2018.04.021",
    journal = "Phys. Lett. B",
    volume = "781",
    pages = "473--479",
    year = "2018"
}

@article{Bringmann:2018cvk,
    author = "Bringmann, Torsten and Pospelov, Maxim",
    title = "{Novel direct detection constraints on light dark matter}",
    eprint = "1810.10543",
    archivePrefix = "arXiv",
    primaryClass = "hep-ph",
    doi = "10.1103/PhysRevLett.122.171801",
    journal = "Phys. Rev. Lett.",
    volume = "122",
    number = "17",
    pages = "171801",
    year = "2019"
}

@article{Super-Kamiokande:2017dch,
    author = "Kachulis, C. and others",
    collaboration = "Super-Kamiokande",
    title = "{Search for Boosted Dark Matter Interacting With Electrons in Super-Kamiokande}",
    eprint = "1711.05278",
    archivePrefix = "arXiv",
    primaryClass = "hep-ex",
    doi = "10.1103/PhysRevLett.120.221301",
    journal = "Phys. Rev. Lett.",
    volume = "120",
    number = "22",
    pages = "221301",
    year = "2018"
}

@article{PandaX-II:2021kai,
    author = "Cui, Xiangyi and others",
    collaboration = "PandaX-II",
    title = "{Search for Cosmic-Ray Boosted Sub-GeV Dark Matter at the PandaX-II Experiment}",
    eprint = "2112.08957",
    archivePrefix = "arXiv",
    primaryClass = "hep-ex",
    doi = "10.1103/PhysRevLett.128.171801",
    journal = "Phys. Rev. Lett.",
    volume = "128",
    number = "17",
    pages = "171801",
    year = "2022"
}

@article{Super-Kamiokande:2022ncz,
    author = "Abe, K. and others",
    collaboration = "Super-Kamiokande",
    title = "{Search for Cosmic-Ray Boosted Sub-GeV Dark Matter Using Recoil Protons at Super-Kamiokande}",
    eprint = "2209.14968",
    archivePrefix = "arXiv",
    primaryClass = "hep-ex",
    doi = "10.1103/PhysRevLett.130.031802",
    journal = "Phys. Rev. Lett.",
    volume = "130",
    number = "3",
    pages = "031802",
    year = "2023",
    note = "[Erratum: Phys.Rev.Lett. 131, 159903 (2023)]"
}

@article{LZ:2025iaw,
    author = "Aalbers, J. and others",
    collaboration = "LZ",
    title = "{New Constraints on Cosmic Ray-Boosted Dark Matter from the LUX-ZEPLIN Experiment}",
    eprint = "2503.18158",
    archivePrefix = "arXiv",
    primaryClass = "hep-ex",
    reportNumber = "FERMILAB-PUB-25-0191-V",
    doi = "10.1103/nr92-jvt3",
    journal = "Phys. Rev. Lett.",
    volume = "134",
    number = "24",
    pages = "241801",
    year = "2025"
}

@article{PandaX:2024syk,
    author = "Shen, Guofang and others",
    collaboration = "PandaX, (PandaX Collaboration)25,26",
    title = "{Search for Solar Boosted Dark Matter Particles at the PandaX-4T Experiment}",
    eprint = "2412.19970",
    archivePrefix = "arXiv",
    primaryClass = "hep-ex",
    doi = "10.1103/PhysRevLett.134.161003",
    journal = "Phys. Rev. Lett.",
    volume = "134",
    number = "16",
    pages = "161003",
    year = "2025"
}

@article{Essig:2011nj,
    author = "Essig, Rouven and Mardon, Jeremy and Volansky, Tomer",
    title = "{Direct Detection of Sub-GeV Dark Matter}",
    eprint = "1108.5383",
    archivePrefix = "arXiv",
    primaryClass = "hep-ph",
    reportNumber = "SLAC-PUB-14538",
    doi = "10.1103/PhysRevD.85.076007",
    journal = "Phys. Rev. D",
    volume = "85",
    pages = "076007",
    year = "2012"
}

@article{Liang:2024ecw,
    author = "Liang, Jin-Han and Liao, Yi and Ma, Xiao-Dong and Wang, Hao-Lin",
    title = "{A systematic investigation on dark matter-electron scattering in effective field theories}",
    eprint = "2406.10912",
    archivePrefix = "arXiv",
    primaryClass = "hep-ph",
    doi = "10.1007/JHEP07(2024)279",
    journal = "JHEP",
    volume = "07",
    pages = "279",
    year = "2024"
}

@article{Krnjaic:2024bdd,
    author = "Krnjaic, Gordan and Rocha, Duncan and Trickle, Tanner",
    title = "{The non-relativistic effective field theory of dark matter-electron interactions}",
    eprint = "2407.14598",
    archivePrefix = "arXiv",
    primaryClass = "hep-ph",
    reportNumber = "FERMILAB-PUB-24-0356-T",
    doi = "10.1007/JHEP03(2025)165",
    journal = "JHEP",
    volume = "03",
    pages = "165",
    year = "2025"
}

@article{Hill:2013hoa,
    author = "Hill, Richard J. and Solon, Mikhail P.",
    title = "{WIMP-nucleon scattering with heavy WIMP effective theory}",
    eprint = "1309.4092",
    archivePrefix = "arXiv",
    primaryClass = "hep-ph",
    reportNumber = "EFI-13-23",
    doi = "10.1103/PhysRevLett.112.211602",
    journal = "Phys. Rev. Lett.",
    volume = "112",
    pages = "211602",
    year = "2014"
}

@article{Moore:1994yx,
    author = "Moore, B.",
    title = "{Evidence against dissipationless dark matter from observations of galaxy haloes}",
    doi = "10.1038/370629a0",
    journal = "Nature",
    volume = "370",
    pages = "629",
    year = "1994"
}

@article{Navarro:1995iw,
    author = "Navarro, Julio F. and Frenk, Carlos S. and White, Simon D. M.",
    title = "{The Structure of cold dark matter halos}",
    eprint = "astro-ph/9508025",
    archivePrefix = "arXiv",
    doi = "10.1086/177173",
    journal = "Astrophys. J.",
    volume = "462",
    pages = "563--575",
    year = "1996"
}

@article{Navarro:1996gj,
    author = "Navarro, Julio F. and Frenk, Carlos S. and White, Simon D. M.",
    title = "{A Universal density profile from hierarchical clustering}",
    eprint = "astro-ph/9611107",
    archivePrefix = "arXiv",
    doi = "10.1086/304888",
    journal = "Astrophys. J.",
    volume = "490",
    pages = "493--508",
    year = "1997"
}

@article{Flores:1994gz,
    author = "Flores, Ricardo A. and Primack, Joel R.",
    title = "{Observational and theoretical constraints on singular dark matter halos}",
    eprint = "astro-ph/9402004",
    archivePrefix = "arXiv",
    reportNumber = "SCIPP-93-01-REV, SCIPP-93-01",
    doi = "10.1086/187350",
    journal = "Astrophys. J. Lett.",
    volume = "427",
    pages = "L1--4",
    year = "1994"
}

@article{Tulin:2017ara,
    author = "Tulin, Sean and Yu, Hai-Bo",
    title = "{Dark Matter Self-interactions and Small Scale Structure}",
    eprint = "1705.02358",
    archivePrefix = "arXiv",
    primaryClass = "hep-ph",
    doi = "10.1016/j.physrep.2017.11.004",
    journal = "Phys. Rept.",
    volume = "730",
    pages = "1--57",
    year = "2018"
}

@article{Spergel:1999mh,
    author = "Spergel, David N. and Steinhardt, Paul J.",
    title = "{Observational evidence for selfinteracting cold dark matter}",
    eprint = "astro-ph/9909386",
    archivePrefix = "arXiv",
    doi = "10.1103/PhysRevLett.84.3760",
    journal = "Phys. Rev. Lett.",
    volume = "84",
    pages = "3760--3763",
    year = "2000"
}

@inproceedings{Battaglieri:2017aum,
    author = "Battaglieri, Marco and others",
    title = "{US Cosmic Visions: New Ideas in Dark Matter 2017: Community Report}",
    booktitle = "{U.S. Cosmic Visions: New Ideas in Dark Matter}",
    eprint = "1707.04591",
    archivePrefix = "arXiv",
    primaryClass = "hep-ph",
    reportNumber = "FERMILAB-CONF-17-282-AE-PPD-T",
    month = "7",
    year = "2017"
}

@article{Kang:2020afi,
    author = "Kang, Yoo-Jin and Lee, Hyun Min",
    title = "{Effective theory for self-interacting dark matter and massive spin-2 mediators}",
    eprint = "2003.09290",
    archivePrefix = "arXiv",
    primaryClass = "hep-ph",
    doi = "10.1088/1361-6471/abe529",
    journal = "J. Phys. G",
    volume = "48",
    number = "4",
    pages = "045002",
    year = "2021"
}

@article{Chen:2023oip,
    author = "Chen, Qing and Zhou, Shuang-Yong",
    title = "{Bigravity portal dark matter}",
    eprint = "2310.03079",
    archivePrefix = "arXiv",
    primaryClass = "hep-ph",
    reportNumber = "USTC-ICTS/PCFT-23-30",
    doi = "10.1103/PhysRevD.109.095035",
    journal = "Phys. Rev. D",
    volume = "109",
    number = "9",
    pages = "095035",
    year = "2024"
}

@article{Bhattacharya:2025yba,
    author = "Bhattacharya, Shohini and Cichy, Krzysztof and Constantinou, Martha and Metz, Andreas and Miller, Joshua and Petreczky, Peter and Steffens, Fernanda",
    title = "{Generalized Parton Distributions from Lattice QCD with Asymmetric Momentum Transfer: Tensor Case}",
    eprint = "2505.11288",
    archivePrefix = "arXiv",
    primaryClass = "hep-lat",
    month = "5",
    year = "2025"
}

@article{Alexandrou:2019ali,
    author = "Alexandrou, C. and others",
    title = "{Moments of nucleon generalized parton distributions from lattice QCD simulations at physical pion mass}",
    eprint = "1908.10706",
    archivePrefix = "arXiv",
    primaryClass = "hep-lat",
    doi = "10.1103/PhysRevD.101.034519",
    journal = "Phys. Rev. D",
    volume = "101",
    number = "3",
    pages = "034519",
    year = "2020"
}

@article{Bhattacharya:2024wtg,
    author = "Bhattacharya, Shohini and Cichy, Krzysztof and Constantinou, Martha and Gao, Xiang and Metz, Andreas and Miller, Joshua and Mukherjee, Swagato and Petreczky, Peter and Steffens, Fernanda and Zhao, Yong",
    title = "{Moments of axial-vector GPD from lattice QCD: quark helicity, orbital angular momentum, and spin-orbit correlation}",
    eprint = "2410.03539",
    archivePrefix = "arXiv",
    primaryClass = "hep-lat",
    reportNumber = "LA-UR-24-29020",
    doi = "10.1007/JHEP01(2025)146",
    journal = "JHEP",
    volume = "01",
    pages = "146",
    year = "2025"
}

@article{Diehl:2005jf,
    author = "Diehl, M. and Hagler, Ph.",
    title = "{Spin densities in the transverse plane and generalized transversity distributions}",
    eprint = "hep-ph/0504175",
    archivePrefix = "arXiv",
    reportNumber = "DESY-05-061",
    doi = "10.1140/epjc/s2005-02342-6",
    journal = "Eur. Phys. J. C",
    volume = "44",
    pages = "87--101",
    year = "2005"
}

@article{Alexandrou:2022dtc,
    author = "Alexandrou, C. and others",
    title = "{Moments of the nucleon transverse quark spin densities using lattice QCD}",
    eprint = "2202.09871",
    archivePrefix = "arXiv",
    primaryClass = "hep-lat",
    doi = "10.1103/PhysRevD.107.054504",
    journal = "Phys. Rev. D",
    volume = "107",
    number = "5",
    pages = "054504",
    year = "2023"
}

@article{Wang:2024lrm,
    author = "Wang, Bigeng and He, Fangcheng and Wang, Gen and Draper, Terrence and Liang, Jian and Liu, Keh-Fei and Yang, Yi-Bo",
    collaboration = "\ensuremath{\chi}QCD",
    title = "{Trace anomaly form factors from lattice QCD}",
    eprint = "2401.05496",
    archivePrefix = "arXiv",
    primaryClass = "hep-lat",
    doi = "10.1103/PhysRevD.109.094504",
    journal = "Phys. Rev. D",
    volume = "109",
    number = "9",
    pages = "094504",
    year = "2024"
}

@article{Liu:2024rdm,
    author = "Liu, Wei-Yang and Shuryak, Edward and Zahed, Ismail",
    title = "{Glue in hadrons at medium resolution and the QCD instanton vacuum}",
    eprint = "2404.03047",
    archivePrefix = "arXiv",
    primaryClass = "hep-ph",
    doi = "10.1103/PhysRevD.110.054005",
    journal = "Phys. Rev. D",
    volume = "110",
    number = "5",
    pages = "054005",
    year = "2024"
}

@article{Hoferichter:2012wf,
    author = "Hoferichter, M. and Ditsche, C. and Kubis, B. and Meissner, U. G.",
    title = "{Dispersive analysis of the scalar form factor of the nucleon}",
    eprint = "1204.6251",
    archivePrefix = "arXiv",
    primaryClass = "hep-ph",
    doi = "10.1007/JHEP06(2012)063",
    journal = "JHEP",
    volume = "06",
    pages = "063",
    year = "2012"
}

@article{Cao:2024zlf,
    author = "Cao, Xiong-Hui and Guo, Feng-Kun and Li, Qu-Zhi and Yao, De-Liang",
    title = "{Dispersive determination of nucleon gravitational form factors}",
    eprint = "2411.13398",
    archivePrefix = "arXiv",
    primaryClass = "hep-ph",
    doi = "10.1038/s41467-025-62278-9",
    journal = "Nature Commun.",
    volume = "16",
    number = "1",
    pages = "6979",
    year = "2025"
}

@article{Cao:2025dkv,
    author = "Cao, Xiong-Hui and Guo, Feng-Kun and Li, Qu-Zhi and Wu, Bo-Wen and Yao, De-Liang",
    title = "{Gravitational form factors of pions, kaons and nucleons from dispersion relations}",
    eprint = "2507.05375",
    archivePrefix = "arXiv",
    primaryClass = "hep-ph",
    month = "7",
    year = "2025"
}

@article{Haxton:2024lyc,
    author = "Haxton, Wick and McElvain, Kenneth and Menzo, Tony and Rule, Evan and Zupan, Jure",
    title = "{Effective theory tower for {\ensuremath{\mu}} {\textrightarrow} e conversion}",
    eprint = "2406.13818",
    archivePrefix = "arXiv",
    primaryClass = "hep-ph",
    reportNumber = "LA-UR-24-24937, N3AS-24-023",
    doi = "10.1007/JHEP11(2024)076",
    journal = "JHEP",
    volume = "11",
    pages = "076",
    year = "2024"
}

@article{Dutta:2024kuj,
    author = "Dutta, Bhaskar and Huang, Wei-Chih and Kim, Doojin and Newstead, Jayden L. and Park, Jong-Chul and Ali, Iman Shaukat",
    title = "{Prospects for Light Dark Matter Searches at Large-Volume Neutrino Detectors}",
    eprint = "2402.04184",
    archivePrefix = "arXiv",
    primaryClass = "hep-ph",
    doi = "10.1103/PhysRevLett.133.161801",
    journal = "Phys. Rev. Lett.",
    volume = "133",
    number = "16",
    pages = "161801",
    year = "2024"
}

@article{Ema:2018bih,
    author = "Ema, Yohei and Sala, Filippo and Sato, Ryosuke",
    title = "{Light Dark Matter at Neutrino Experiments}",
    eprint = "1811.00520",
    archivePrefix = "arXiv",
    primaryClass = "hep-ph",
    reportNumber = "DESY-18-194",
    doi = "10.1103/PhysRevLett.122.181802",
    journal = "Phys. Rev. Lett.",
    volume = "122",
    number = "18",
    pages = "181802",
    year = "2019"
}

@article{Super-Kamiokande:2002weg,
    author = "Fukuda, Y. and others",
    editor = "Ilyin, V. A. and Korenkov, V. V. and Perret-Gallix, D.",
    collaboration = "Super-Kamiokande",
    title = "{The Super-Kamiokande detector}",
    doi = "10.1016/S0168-9002(03)00425-X",
    journal = "Nucl. Instrum. Meth. A",
    volume = "501",
    pages = "418--462",
    year = "2003"
}

@article{Hyper-Kamiokande:2018ofw,
    author = "Abe, K. and others",
    collaboration = "Hyper-Kamiokande",
    title = "{Hyper-Kamiokande Design Report}",
    eprint = "1805.04163",
    archivePrefix = "arXiv",
    primaryClass = "physics.ins-det",
    month = "5",
    year = "2018"
}

@article{DUNE:2020lwj,
    author = "Abi, Babak and others",
    collaboration = "DUNE",
    title = "{Deep Underground Neutrino Experiment (DUNE), Far Detector Technical Design Report, Volume I Introduction to DUNE}",
    eprint = "2002.02967",
    archivePrefix = "arXiv",
    primaryClass = "physics.ins-det",
    reportNumber = "FERMILAB-PUB-20-024-ND, FERMILAB-DESIGN-2020-01",
    doi = "10.1088/1748-0221/15/08/T08008",
    journal = "JINST",
    volume = "15",
    number = "08",
    pages = "T08008",
    year = "2020"
}

@article{JUNO:2021vlw,
    author = "Abusleme, Angel and others",
    collaboration = "JUNO",
    title = "{JUNO physics and detector}",
    eprint = "2104.02565",
    archivePrefix = "arXiv",
    primaryClass = "hep-ex",
    doi = "10.1016/j.ppnp.2021.103927",
    journal = "Prog. Part. Nucl. Phys.",
    volume = "123",
    pages = "103927",
    year = "2022"
}

@article{Pate:2024acz,
    author = "Pate, S. F. and Papavassiliou, V. and Schaub, J. P. and Trujillo, D. P. and Ivanov, M. V. and Barbaro, M. B. and Giusti, C.",
    title = "{Global fit of electron and neutrino elastic scattering data to determine the strange quark contribution to the vector and axial form factors of the nucleon}",
    eprint = "2402.10854",
    archivePrefix = "arXiv",
    primaryClass = "hep-ph",
    reportNumber = "FERMILAB-PUB-24-0074-V",
    doi = "10.1103/PhysRevD.109.093001",
    journal = "Phys. Rev. D",
    volume = "109",
    number = "9",
    pages = "093001",
    year = "2024"
}

@article{Alexandrou:2021wzv,
    author = "Alexandrou, C. and Bacchio, S. and Constantinou, M. and Hadjiyiannakou, K. and Jansen, K. and Koutsou, G.",
    title = "{Quark flavor decomposition of the nucleon axial form factors}",
    eprint = "2106.13468",
    archivePrefix = "arXiv",
    primaryClass = "hep-lat",
    doi = "10.1103/PhysRevD.104.074503",
    journal = "Phys. Rev. D",
    volume = "104",
    pages = "074503",
    year = "2021"
}

@article{Alexandrou:2023qbg,
    author = "Alexandrou, Constantia and Bacchio, Simone and Constantinou, Martha and Finkenrath, Jacob and Frezzotti, Roberto and Kostrzewa, Bartosz and Koutsou, Giannis and Spanoudes, Gregoris and Urbach, Carsten",
    collaboration = "Extended Twisted Mass",
    title = "{Nucleon axial and pseudoscalar form factors using twisted-mass fermion ensembles at the physical point}",
    eprint = "2309.05774",
    archivePrefix = "arXiv",
    primaryClass = "hep-lat",
    doi = "10.1103/PhysRevD.109.034503",
    journal = "Phys. Rev. D",
    volume = "109",
    number = "3",
    pages = "034503",
    year = "2024"
}

@article{Smith:1972xh,
    author = "Smith, R. A. and Moniz, E. J.",
    title = "{NEUTRINO REACTIONS ON NUCLEAR TARGETS}",
    doi = "10.1016/0550-3213(75)90612-4",
    journal = "Nucl. Phys. B",
    volume = "43",
    pages = "605",
    year = "1972",
    note = "[Erratum: Nucl.Phys.B 101, 547 (1975)]"
}

@article{Moniz:1969sr,
    author = "Moniz, E. J.",
    title = "{Pion electroproduction from nuclei}",
    doi = "10.1103/PhysRev.184.1154",
    journal = "Phys. Rev.",
    volume = "184",
    pages = "1154--1161",
    year = "1969"
}

@article{Bhattacharya:2011ah,
    author = "Bhattacharya, Bhubanjyoti and Hill, Richard J. and Paz, Gil",
    title = "{Model independent determination of the axial mass parameter in quasielastic neutrino-nucleon scattering}",
    eprint = "1108.0423",
    archivePrefix = "arXiv",
    primaryClass = "hep-ph",
    reportNumber = "EFI-11-23",
    doi = "10.1103/PhysRevD.84.073006",
    journal = "Phys. Rev. D",
    volume = "84",
    pages = "073006",
    year = "2011"
}

@article{Bhattacharya:2024win,
    author = "Bhattacharya, Bhubanjyoti and Carey, Sam and Cohen, Erez O. and Paz, Gil",
    title = "{Quasielastic lepton-nucleus scattering and the correlated Fermi gas model}",
    eprint = "2405.05342",
    archivePrefix = "arXiv",
    primaryClass = "hep-ph",
    reportNumber = "WSU-HEP-2402",
    doi = "10.1103/PhysRevD.111.096021",
    journal = "Phys. Rev. D",
    volume = "111",
    number = "9",
    pages = "096021",
    year = "2025"
}

@article{Hill:2010yb,
    author = "Hill, Richard J. and Paz, Gil",
    title = "{Model independent extraction of the proton charge radius from electron scattering}",
    eprint = "1008.4619",
    archivePrefix = "arXiv",
    primaryClass = "hep-ph",
    reportNumber = "EFI-PREPRINT-10-21",
    doi = "10.1103/PhysRevD.82.113005",
    journal = "Phys. Rev. D",
    volume = "82",
    pages = "113005",
    year = "2010"
}

@article{Hill:2014yxa,
    author = "Hill, Richard J. and Solon, Mikhail P.",
    title = "{Standard Model anatomy of WIMP dark matter direct detection II: QCD analysis and hadronic matrix elements}",
    eprint = "1409.8290",
    archivePrefix = "arXiv",
    primaryClass = "hep-ph",
    reportNumber = "EFI-PREPRINT-14-25",
    doi = "10.1103/PhysRevD.91.043505",
    journal = "Phys. Rev. D",
    volume = "91",
    pages = "043505",
    year = "2015"
}

@article{Fierz:1939ix,
    author = "Fierz, M. and Pauli, W.",
    title = "{On relativistic wave equations for particles of arbitrary spin in an electromagnetic field}",
    doi = "10.1098/rspa.1939.0140",
    journal = "Proc. Roy. Soc. Lond. A",
    volume = "173",
    pages = "211--232",
    year = "1939"
}

@article{deRham:2014zqa,
    author = "de Rham, Claudia",
    title = "{Massive Gravity}",
    eprint = "1401.4173",
    archivePrefix = "arXiv",
    primaryClass = "hep-th",
    doi = "10.12942/lrr-2014-7",
    journal = "Living Rev. Rel.",
    volume = "17",
    pages = "7",
    year = "2014"
}

@article{Tulin:2013teo,
    author = "Tulin, Sean and Yu, Hai-Bo and Zurek, Kathryn M.",
    title = "{Beyond Collisionless Dark Matter: Particle Physics Dynamics for Dark Matter Halo Structure}",
    eprint = "1302.3898",
    archivePrefix = "arXiv",
    primaryClass = "hep-ph",
    doi = "10.1103/PhysRevD.87.115007",
    journal = "Phys. Rev. D",
    volume = "87",
    number = "11",
    pages = "115007",
    year = "2013"
}

@article{Huang:2015yga,
    author = "Huang, Qing-Guo and Ribeiro, Raquel H. and Xing, Yu-Hang and Zhang, Ke-Chao and Zhou, Shuang-Yong",
    title = "{On the uniqueness of the non-minimal matter coupling in massive gravity and bigravity}",
    eprint = "1505.02616",
    archivePrefix = "arXiv",
    primaryClass = "hep-th",
    doi = "10.1016/j.physletb.2015.07.003",
    journal = "Phys. Lett. B",
    volume = "748",
    pages = "356--360",
    year = "2015"
}

@article{Cates:2011pz,
    author = "Cates, G. D. and de Jager, C. W. and Riordan, S. and Wojtsekhowski, B.",
    title = "{Flavor decomposition of the elastic nucleon electromagnetic form factors}",
    eprint = "1103.1808",
    archivePrefix = "arXiv",
    primaryClass = "nucl-ex",
    reportNumber = "JLAB-PHY-11-1325",
    doi = "10.1103/PhysRevLett.106.252003",
    journal = "Phys. Rev. Lett.",
    volume = "106",
    pages = "252003",
    year = "2011"
}

@article{Qattan:2011ke,
    author = "Qattan, I. A. and Alsaad, A. and Arrington, J.",
    title = "{Reexamination of phenomenological two-photon exchange corrections to the proton form factors and $e^{+-}p$ scattering}",
    eprint = "1109.1441",
    archivePrefix = "arXiv",
    primaryClass = "nucl-ex",
    doi = "10.1103/PhysRevC.84.054317",
    journal = "Phys. Rev. C",
    volume = "84",
    pages = "054317",
    year = "2011"
}

\end{document}